\documentclass[fleqn,usenatbib]{mnras}

\usepackage{newtxtext,newtxmath}

\usepackage[T1]{fontenc}

\DeclareRobustCommand{\VAN}[3]{#2}
\let\VANthebibliography\thebibliography
\def\thebibliography{\DeclareRobustCommand{\VAN}[3]{##3}\VANthebibliography}

\usepackage{graphicx}	
\usepackage{amsmath}	
\usepackage{supertabular}
\usepackage[normalem]{ulem}
\usepackage{longtable}
\usepackage{pdflscape}
\usepackage{afterpage}
\usepackage{rotating}
\usepackage{lscape}
\usepackage{caption, subcaption}

\usepackage{siunitx}
\newcommand\mc[1]{\multicolumn{1}{c}{#1}}
\newcommand\mcn[1]{\multicolumn{1}{c|}{#1}}

\newcolumntype{d}[1]{D{.}{.}{#1}}
\robustify\bfseries

\title[The subpulse modulation of 1198 pulsars]{The Thousand-Pulsar-Array programme on MeerKAT - VIII. The subpulse modulation of 1198 pulsars}
  
\author[Song et al.]{
X.~Song,$^{1}$\thanks{E-mail: xsong@pulsarastronomy.net (XS)}
P.~Weltevrede,$^{1}$
A.~Szary,$^{2,3}$
G.~Wright,$^{1}$
M.~J.~Keith,$^{1}$
A.~Basu,$^{1}$
S.~Johnston,$^{4}$\newauthor
A.~Karastergiou,$^{5}$
R.~A.~Main,$^{6}$
L.~S.~Oswald,$^{5,7}$
A.~Parthasarathy,$^{6}$
B.~Posselt,$^{5,8}$\newauthor
M.~Bailes,$^{9,10}$
S.~Buchner,$^{11}$
B.~Hugo,$^{11,12}$ 
M.~Serylak,$^{13,14}$ \\
$^{1}$Jodrell Bank Centre for Astrophysics, Department of Physics and Astronomy, University of Manchester, Manchester M13 9PL, UK\\
$^{2}$Janusz Gil Institute of Astronomy, University of Zielona G{\'o}ra, Licealna 9, 65-417 Zielona G{\'o}ra, Poland\\
$^{3}$ASTRON, Netherlands Institute for Radio Astronomy, Oude Hoogeveensedijk 4, 7991 PD, Dwingeloo, The Netherlands\\
$^{4}$Australia Telescope National Facility, CSIRO Space and Astronomy, PO~Box~76, Epping NSW~1710, Australia\\
$^{5}$ Department of Astrophysics, University of Oxford, Denys Wilkinson Building, Keble Road, Oxford OX1 3RH, UK\\
$^{6}$Max-Plank-Institut f\"{u}r Radioastronomie, Auf dem H\"{u}gel 69, D-53121 Bonn, Germany\\
$^{7}$ Magdalen College, University of Oxford, Oxford OX1 4AU, UK\\
$^{8}$Department of Astronomy \& Astrophysics, Pennsylvania State University, 525 Davey Lab, University Park, PA 16802, USA \\
$^{9}$Centre for Astrophysics and Supercomputing, Swinburne University of Technology, PO~Box~218, Hawthorn, Vic~3122, Australia\\
$^{10}$ARC Centre of Excellence for
Gravitational Wave Discovery (OzGrav)\\
$^{11}$South African Radio Astronomy Observatory (SARAO). 2 Fir Street, Black River Park Observatory, Cape Town, 7925\\
$^{12}$Department of Physics and Electronics, Rhodes University, Artillery Road, Grahamstown, 6139, South Africa\\
$^{13}$SKA Observatory, Jodrell Bank, Lower Withington, Macclesfield, SK11 9FT, United Kingdom\\
$^{14}$Department of Physics and Astronomy, University of the Western Cape, Bellville, Cape Town, 7535, South Africa\\
}

\date{Accepted XXX. Received YYY; in original form ZZZ}

\pubyear{2022}

\begin{document}
\label{firstpage}
\pagerange{\pageref{firstpage}--\pageref{lastpage}}
\maketitle

\newcommand{\numsubarray}{690} 
\newcommand{\driftnum}{418} 
\newcommand{\driftfrac}{35\%} 
\newcommand{\lononlynum}{115} 
\newcommand{\mededotdrift}{7.34e+31} 
\newcommand{\mededotnofeature}{9.74e+32} 
\newcommand{\highedotdrift}{111} 
\newcommand{\censuspsr}{1198} 
\newcommand{\totalpulses}{1,646,896} 
\newcommand{\longobs}{887} 
\newcommand{\thousandpulses}{763} 
\newcommand{\subarrayobs}{26} 
\newcommand{\fitsnrf}{0.62} 
\newcommand{\fitsnrs}{400} 
\newcommand{\fitsnrallf}{0.72} 
\newcommand{\fitsnralls}{280} 
\newcommand{\fitsnrerrf}{0.08} 
\newcommand{\fitsnrerrs}{100} 
\newcommand{\fitsnrallerrf}{0.06} 
\newcommand{\fitsnrallerrs}{50} 
\newcommand{\medage}{$4.6\times 10^6$} 
\newcommand{\numdomdrift}{403} 
\newcommand{\numdomlon}{130} 
\newcommand{\driftingpositive}{189} 
\newcommand{\driftingnegative}{214} 
\newcommand{\numseverescatter}{30} 
\newcommand{\numscatter}{105} 
\newcommand{\numsnragree}{762} 
\newcommand{\numsnrall}{1060} 

\newcommand{\tpawes}{116} 
\newcommand{\tpaweshasp}{40} 
\newcommand{\tpawesconsist}{30} 
\newcommand{\tpanowes}{40} 
\newcommand{\tpawespconsist}{33} 
\newcommand{\tpabl}{49} 
\newcommand{\tpablconsist}{46} 
\newcommand{\tpanoblall}{43} 
\newcommand{\tpanobl}{30} 
\newcommand{\tpanobldrift}{21} 
\newcommand{\tpabe}{10} 
\newcommand{\tpabepconsist}{7} 
\newcommand{\tpabepdrift}{5} 
\newcommand{\inbenotpa}{2} 
\newcommand{\intpanoblinbe}{8} 
\newcommand{\tpawesptwo}{29} 
\newcommand{\nrpulsarswithIPtotal}{34} 
\newcommand{\nrpulsarswithIPautodetect}{33} 
\newcommand{\nrpulsarswithIPmanual}{1} 
\newcommand{\nrpulsarswithnoIP}{7} 
\newcommand{\nrMPdoublePlusNrIPdouble}{376} 
\newcommand{\nrMPorIPManualDouble}{134} 
\newcommand{\nrMPorIPManualOnpulseNoDouble}{118} 
\newcommand{\nrPulsarsManualOffpulse}{90} 
\newcommand{\nrPulsarsNoOffSubtrTwodfs}{7} 
\newcommand{\NoOffSubtrTwodfspsrs}{PSRs J0828$-$3417, J1019$-$5749, J1107$-$5907, J1316$-$6232, J1406$-$5806, J1746$-$2856 and J1801$-$2304} 
\newcommand{\nrPulsarsNskipOrNread}{55} 
\newcommand{\nrPulsarsAutoClean}{79} 
\newcommand{\nrPulsarsManualFFTsize}{48} 
\newcommand{\nrPulsarsManualBins}{16} 
\newcommand{\manualbinspsrs}{PSRs J0111$-$7131, J0133$-$6957, J0729$-$1448, J0905$-$5127, J1331$-$5245, J1632$-$1013, J1700$-$4422, J1702$-$4128, J1737$-$3102, J1837$-$0822, J1851+0418, J1856+0245, J1901+0234, J1901+1306, J1908+0916 and J2048+2255} 
\newcommand{\psrcatversion}{1.67} 
\newcommand{\nrbootstrapitts}{1000} 
\newcommand{\nrcatOffpulse}{13} 
\newcommand{\psrcatoffpulsepsrs}{PSRs J1015$-$5719, J1133$-$6250, J1302$-$6350, J1622$-$4950, J1627$-$5936, J1750$-$3503, J1826$-$1334, J1827$-$0958, J1831$-$0952, J1842+1332, J1843$-$0355, J1851+0418 and J1930+1852} 
\newcommand{\nrfoldharmonics}{12} 
\newcommand{\nrManualVrange}{57} 
\newcommand{\nrComponentWidth}{4} 
\newcommand{\ComponentWidthpsrs}{PSRs J1107$-$5907, J1406$-$5806, J1550$-$5418 and J1828$-$1101} 
\newcommand{\nrNoBaselineSlopeSub}{7} 
\newcommand{\NoBaselineSlopeSubpsrs}{PSRs J0828$-$3417, J1019$-$5749, J1107$-$5907, J1406$-$5806, J1622$-$4950, J1630$-$4733 and J1721$-$3532} 
\newcommand{\numrficlean}{131} 
\newcommand{\percentzap}{3\%} 
\newcommand{\zapspec}{17} 
\newcommand{\zapspecpsrs}{PSRs J0038$-$2501, J0514$-$4407, J0733$-$2345, J0932$-$5327, J1016$-$5857, J1151$-$6108, J1244$-$6359, J1535$-$4114, J1538+2345, J1629$-$3825, J1755$-$2550, J1828+1359, J1844$-$0452, J1850+0423, J1859+1526, J1917+1353 and J2136$-$1606} 
\newcommand{\noshuffleapp}{PSRs J1016$-$5857, J1535$-$4114, J1538+2345, J1914+0631, J1916$-$2939, J1936+1536 and J1936+1536} 

\newcommand{\Powersellng}{-1.8} 
\newcommand{\Powersellngerr}{0.4} 
\newcommand{\Powersellnan}{-0.13} 
\newcommand{\Powersellnanerr}{0.02} 
\newcommand{\Powersellnbn}{0.80} 
\newcommand{\Powersellnbnerr}{0.02} 
\newcommand{\Powersellnsig}{0.35} 
\newcommand{\Powersellnsigerr}{0.01} 
\newcommand{\corrPowerselln}{0.30} 
\newcommand{\correrrPowerselln}{0.03} 
\newcommand{\driftratedriftlng}{-0.2} 
\newcommand{\driftratedriftlngerr}{0.1} 
\newcommand{\driftratedriftlnan}{-0.4} 
\newcommand{\driftratedriftlnanerr}{0.1} 
\newcommand{\driftratedriftlnbn}{0.49} 
\newcommand{\driftratedriftlnbnerr}{0.03} 
\newcommand{\driftratedriftlnsig}{0.45} 
\newcommand{\driftratedriftlnsigerr}{0.02} 
\newcommand{\corrdriftratedriftln}{0.20} 
\newcommand{\correrrdriftratedriftln}{0.04} 
\newcommand{\Phorabsdriftlng}{-5.5} 
\newcommand{\Phorabsdriftlngerr}{2.0} 
\newcommand{\Phorabsdriftlnan}{0.14} 
\newcommand{\Phorabsdriftlnanerr}{0.03} 
\newcommand{\Phorabsdriftlnbn}{1.32} 
\newcommand{\Phorabsdriftlnbnerr}{0.03} 
\newcommand{\Phorabsdriftlnsig}{0.42} 
\newcommand{\Phorabsdriftlnsigerr}{0.02} 
\newcommand{\corrPhorabsdriftln}{0.46} 
\newcommand{\correrrPhorabsdriftln}{0.04} 
\newcommand{\driftfracdriftlng}{-1.1} 
\newcommand{\driftfracdriftlngerr}{0.5} 
\newcommand{\driftfracdriftlnan}{0.21} 
\newcommand{\driftfracdriftlnanerr}{0.04} 
\newcommand{\driftfracdriftlnbn}{-0.08} 
\newcommand{\driftfracdriftlnbnerr}{0.03} 
\newcommand{\driftfracdriftlnsig}{0.49} 
\newcommand{\driftfracdriftlnsigerr}{0.02} 
\newcommand{\corrdriftfracdriftln}{0.32} 
\newcommand{\correrrdriftfracdriftln}{0.05} 
\newcommand{\AvModsnrlng}{0.11} 
\newcommand{\AvModsnrlngerr}{0.07} 
\newcommand{\AvModsnrlnan}{0.13} 
\newcommand{\AvModsnrlnanerr}{0.02} 
\newcommand{\AvModsnrlnbn}{-0.076} 
\newcommand{\AvModsnrlnbnerr}{0.009} 
\newcommand{\AvModsnrlnsig}{0.199} 
\newcommand{\AvModsnrlnsigerr}{0.005} 
\newcommand{\corrAvModsnrln}{0.24} 
\newcommand{\correrrAvModsnrln}{0.05} 
\newcommand{\Phorabsfracdriftlng}{-2.2} 
\newcommand{\Phorabsfracdriftlngerr}{0.4} 
\newcommand{\Phorabsfracdriftlnan}{0.26} 
\newcommand{\Phorabsfracdriftlnanerr}{0.03} 
\newcommand{\Phorabsfracdriftlnbn}{0.45} 
\newcommand{\Phorabsfracdriftlnbnerr}{0.03} 
\newcommand{\Phorabsfracdriftlnsig}{0.42} 
\newcommand{\Phorabsfracdriftlnsigerr}{0.02} 
\newcommand{\corrPhorabsfracdriftln}{0.49} 
\newcommand{\correrrPhorabsfracdriftln}{0.04} 

\newcommand{\Pverdriftparag}{-1.7} 
\newcommand{\Pverdriftparagerr}{0.3} 
\newcommand{\Pverdriftparaan}{0.09} 
\newcommand{\Pverdriftparaanerr}{0.02} 
\newcommand{\Pverdriftparabn}{-0.00} 
\newcommand{\Pverdriftparabnerr}{0.02} 
\newcommand{\Pverdriftparacn}{0.77} 
\newcommand{\Pverdriftparacnerr}{0.02} 
\newcommand{\Pverdriftparasig}{0.35} 
\newcommand{\Pverdriftparasigerr}{0.01} 
\newcommand{\corrPverdriftpara}{0.34} 
\newcommand{\correrrPverdriftpara}{0.05} 
\newcommand{\Pverstddriftparag}{-2.6} 
\newcommand{\Pverstddriftparagerr}{0.4} 
\newcommand{\Pverstddriftparaan}{-0.04} 
\newcommand{\Pverstddriftparaanerr}{0.01} 
\newcommand{\Pverstddriftparabn}{0.17} 
\newcommand{\Pverstddriftparabnerr}{0.03} 
\newcommand{\Pverstddriftparacn}{-1.77} 
\newcommand{\Pverstddriftparacnerr}{0.02} 
\newcommand{\Pverstddriftparasig}{0.34} 
\newcommand{\Pverstddriftparasigerr}{0.02} 
\newcommand{\corrPverstddriftpara}{0.36} 
\newcommand{\correrrPverstddriftpara}{0.05} 
\newcommand{\AvModsnrparag}{0.03} 
\newcommand{\AvModsnrparagerr}{0.06} 
\newcommand{\AvModsnrparaan}{0.12} 
\newcommand{\AvModsnrparaanerr}{0.05} 
\newcommand{\AvModsnrparabn}{0.18} 
\newcommand{\AvModsnrparabnerr}{0.03} 
\newcommand{\AvModsnrparacn}{-0.084} 
\newcommand{\AvModsnrparacnerr}{0.009} 
\newcommand{\AvModsnrparasig}{0.198} 
\newcommand{\AvModsnrparasigerr}{0.005} 
\newcommand{\corrAvModsnrpara}{0.26} 
\newcommand{\correrrAvModsnrpara}{0.04} 

\newcommand{\corrlogpstdavm}{0.03} 
\newcommand{\correrrlogpstdavm}{0.05} 
\newcommand{\corrlogpstdavmpower}{0.06} 
\newcommand{\correrrlogpstdavmpower}{0.08} 
\newcommand{\corrlogpstdavmsig}{-0.10} 
\newcommand{\correrrlogpstdavmsig}{0.09} 
\newcommand{\corrlogpoweravm}{-0.04} 
\newcommand{\correrrlogpoweravm}{0.03} 
\newcommand{\corrlogpstdminm}{0.09} 
\newcommand{\correrrlogpstdminm}{0.05} 
\newcommand{\corrlogpstdminmpower}{0.08} 
\newcommand{\correrrlogpstdminmpower}{0.09} 
\newcommand{\corrlogpstdminmsig}{-0.08} 
\newcommand{\correrrlogpstdminmsig}{0.08} 
\newcommand{\corrlogpowerminm}{-0.03} 
\newcommand{\correrrlogpowerminm}{0.04} 
\newcommand{\corrlogavmaone}{0.09} 
\newcommand{\errlogavmaone}{0.04} 
\newcommand{\corrlogavmatwo}{-0.11} 
\newcommand{\errlogavmatwo}{0.04} 
\newcommand{\corrlogavmathree}{0.17} 
\newcommand{\errlogavmathree}{0.04} 
\newcommand{\corrlogavmafour}{-0.17} 
\newcommand{\errlogavmafour}{0.04} 
\newcommand{\corrlogavmafit}{-0.22} 
\newcommand{\errlogavmafit}{0.04} 
\newcommand{\corrlogminmaone}{0.08} 
\newcommand{\errlogminmaone}{0.04} 
\newcommand{\corrlogminmatwo}{-0.12} 
\newcommand{\errlogminmatwo}{0.03} 
\newcommand{\corrlogminmathree}{0.16} 
\newcommand{\errlogminmathree}{0.03} 
\newcommand{\corrlogminmafour}{-0.17} 
\newcommand{\errlogminmafour}{0.03} 
\newcommand{\corrlogminmafit}{-0.22} 
\newcommand{\errlogminmafit}{0.03} 

\newcommand{\secintro}{1} 
\newcommand{\secmethod}{2} 
\newcommand{\secsinglepulsepipeline}{2.1} 
\newcommand{\secprepsrsalsa}{2.2} 
\newcommand{\secpsrsalsa}{2.3} 
\newcommand{\secbaselinesub}{2.3.1} 
\newcommand{\seclrfsmod}{2.3.2} 
\newcommand{\sectwodfs}{2.3.3} 
\newcommand{\secshufflenorm}{2.3.4} 
\newcommand{\sectwodfsana}{2.3.5} 
\newcommand{\secmcmc}{2.4} 
\newcommand{\secdataquality}{3} 
\newcommand{\secsample}{3.1} 
\newcommand{\secchecklit}{3.2} 
\newcommand{\secfracdrfiters}{4} 
\newcommand{\secppdot}{5} 
\newcommand{\secresultpver}{5.1} 
\newcommand{\secresultpverstd}{5.2} 
\newcommand{\secresultpasym}{5.3} 
\newcommand{\secresultphor}{5.4} 
\newcommand{\secresultdriftrate}{5.5} 
\newcommand{\secresultmod}{5.6} 
\newcommand{\secdiscussion}{6} 
\newcommand{\secoverviewdis}{6.1} 
\newcommand{\secpversim}{6.2} 
\newcommand{\secphorevo}{6.3} 
\newcommand{\seccomptheo}{6.4} 
\newcommand{\appspectra}{B} 
\newcommand{\appppdothist}{C} 
\newcommand{\figspec}{1} 

\newcommand{\onpulsestack}{D1} 
\newcommand{\onmanualonpulse}{D2} 
\newcommand{\onrfi}{D3} 
\newcommand{\onautorebinning}{D4} 
\newcommand{\onautofft}{D5} 
\newcommand{\onoffpulse}{D6} 
\newcommand{\onbayesian}{D7} 
\newcommand{\onsample}{D8} 
\newcommand{\oncompsurvey}{D9} 
\newcommand{\oncompwes}{D9.1} 
\newcommand{\oncompbasu}{D9.2} 
\newcommand{\onmod}{E} 

\newcommand{\figpulsestack}{D1} 
\newcommand{\figperiodpulse}{D2} 
\newcommand{\figsnrcomp}{D3} 
\newcommand{\figsnrperiod}{D4} 

\newcommand{\p}{$P$}
\newcommand{\pdot}{$\dot{P}$}
\newcommand{\edot}{$\dot{E}$}
\newcommand{\absr}{$|r|$}
\newcommand{\bsurf}{$B$}
\newcommand{\age}{$\tau_{\mathrm{c}}$}
\newcommand{\pasym}{$P_\mathrm{asym}$}
\newcommand{\pstd}{$\sigma_{1/P_3}$}
\newcommand{\sig}{$\sigma_{e}$}
\newcommand{\drate}{$|P_2|/P_3$}

\begin{abstract}

We report on the subpulse modulation properties of \censuspsr\ pulsars using the Thousand-Pulsar-Array programme on MeerKAT. About \driftfrac\ of the analysed pulsars exhibit drifting subpulses which are more pronounced towards the deathline, consistent with previous studies. We estimate that this common phenomenon is detectable in 60\% of the overall pulsar population if high quality data were available for all.
This large study reveals the evolution of drifting subpulses across the pulsar population in unprecedented detail. In particular, we find that the modulation period $P_3$ follows a V-shaped evolution with respect to the characteristic age \age, such that the smallest $P_3$ values, corresponding to the Nyquist period $P_3\simeq2$, are found at $\tau_c \simeq 10^{7.5}$ yr. The V-shaped evolution can be interpreted and reproduced if young pulsars possess aliased fast intrinsic $P_3$, which monotonically increase, ultimately achieving a slow unaliased $P_3$. Enhancement of irregularities in intrinsic subpulse modulation by aliasing in small \age\ pulsars would explain their observed less well defined $P_3$'s and weaker spectral features. Modelling these results as rotating subbeams, their circulation must slow down as the pulsar evolves. This is the opposite to that expected if circulation is driven by $\mathbf{E}\times\mathbf{B}$ drift. This can be resolved if the observed $P_3$ periodicity is due to a beat between an $\mathbf{E}\times\mathbf{B}$ system and the pulsar period. As a by-product, we identified the correct periods and spin-down rates for 12 pulsars, for which harmonically related values were reported in the literature.

\end{abstract}

\begin{keywords}
catalogues -- pulsars: general
\end{keywords}



\section{Introduction}
\label{sec:intro}

Radio signals from pulsars have been observed for over five decades, with the highly magnetised and compact nature of pulsars making them unique laboratories to study physics under extreme conditions. The pulses from single stellar rotations have shown much larger variations in intensity and pulse phase compared to the average pulse shape, and often show subpulses with much narrower pulse widths. These single pulses always show a degree of apparent random pulse-to-pulse variability, but this is often accompanied by a high degree of organisation. In many pulsars subpulses `march' through the pulse profile window \citep{Drake1968}, now more commonly known as `drifting' subpulses. The single pulses form diagonal `drift bands' in the pulse stack (see e.g. Fig.~1 in \citealt{WES2006} and online material \ref{app:pulse_stack}), 
and the drift patterns can be quantified via $P_3$ and $P_2$ as the repeat separation between drift bands (in the pulse number direction expressed in a number of rotational periods $P$) and between successive subpulses (in the pulse phase direction), respectively. Unlike $P_3$, $P_2$ depends on observing frequency and pulse longitude. It is therefore in general different in different profile components (e.g. \citealt{WES2006,wse07}). This makes $P_2$ a somewhat ill defined quantity.

Subpulse drift has been found to be a complex phenomenon. For example, the observed $P_3$ and $P_2$ values can be complicated by the effect of aliasing, which occurs if the intrinsic periodicity of the observed pattern ($P_3$) is less than twice our sampling rate ($P_3<2$). This results in the observed $P_3$ being larger than its intrinsic value (e.g.\,\ PSR~B0943+10 by \citealt{Deshpande1999,Gil2003alias}). The $P_2$ values can be different for different pulse profile components. In some cases the drift bands are such that the subpulses are stationary in phase, and only amplitude modulations are seen (e.g.\,\citealt{WES2006,Basu2016}). In others, the subpulses associated with different pulse profile components are observed to drift in opposite directions at a given time (so-called `bi-drifting' pulsars, see e.g.\ \citealt{Qiao2004,Champion2005,wel16,Wright2017pulsar,Szary2017}). The drifting subpulse pattern has also been found to occasionally switch among different forms, referred to as drift mode changes \citep[e.g.][]{Huguenin1970}. In these cases their drift rates typically have a consistent $P_2$ but different $P_3$ values. In contrast, other pulsars continuously change their apparent drift direction as a function of time (e.g.\, PSRs~B0826$-$34, \citealt{Biggs1985,Gupta2004}; B0540+23, \citealt{Nowakowski1991}; J1750$-$3503, \citealt{Szary2022}).

More dramatic forms of mode changing can also occur, with one mode exhibiting drifting subpulses and the other having disorganised subpulse modulation (e.g.\ PSR~B0943+10 \citealt{Deshpande2001}). Both kinds of mode change can result in average pulse profiles switching between different stable shapes over long timescales. 
A further not uncommon phenomenon is `nulling', which is observed when emission ceases for a few pulses to hundreds of pulse periods \citep[e.g.][]{Taylor1971,Wang2007}, and can be thought of as an extreme case of mode changing. The mode changing phenomenon, as well as nulling (the data presented here will be analysed for nulling in Keith et al.\ in prep.), adds variability with a characteristic timescale to the observed emission, which can complicate the detection of periodicities associated with drifting subpulses. 

In addition to studying the single pulse phenomena of individual sources, large surveys of subpulse modulation help identify trends in the pulsar population and constrain the underlying emission processes. A single observational set-up, fixed frequency band, and systematic approach in the analysis provide additional benefits. Early surveys of subpulse modulation include \citet{Backus1981} and \citet{Ashworth1982}, where 20 and 52 sources were assessed at around 400~MHz and they found about half of the sample showing drifting subpulses. \citet{Rankin1986} collected statistics of pulsars with drifting subpulses and empirically linked this phenomenon to the pulse profile morphology. Single pulse observations dedicated to the systematic study of such pulsars were carried out centred at a wavelength of $\sim21$ and $\sim92$ cm by \citet{WES2006,wse07} using the Westerbork Synthesis Radio Telescope (WSRT). In particular, by analysing 187 pulsars, the 21 cm survey significantly increased the number of pulsars with known drifting subpulses by discovering this phenomenon in 42 pulsars. This unbiased search for pulsar subpulse modulation concluded, after taking the signal-to-noise ratio (S/N) of the data into account, that more than 50\% of pulsars exhibit drifting subpulses or periodic modulation features. It therefore represents a very common phenomenon throughout the pulsar population. In addition, these results confirmed the finding of \citet{Wolszczan1980,Ashworth1982,Rankin1986} that pulsars of large characteristic age are more likely to have detectable drifting subpulses with more precise drift features.
 
\citet{Basu2016,Basu2019} studied 123 pulsars using the Giant Metrewave Radio Telescope (GMRT) at low frequency (at 618 and 333~MHz), of which 61  show drifting subpulses. They found an anti-correlation between $P_3$ and the spin-down energy loss rate \edot, with typically larger $P_3$ values present at low \edot. In particular, pulsars with conal profile components, usually with $\dot{E} < 5\times10^{32}$ erg/s, are more likely to have drifting subpulses (see also \citealt{Rankin1986}). Pulsars with \edot\ above this boundary have core dominated pulse profiles, and amplitude modulation (without detectable drift) with large $P_3$. 

Theoretical explanations of drifting subpulses were established relatively early after their discovery. The most well developed model is the well known \citealt{RS1975} model. These authors attribute drifting subpulses to a rotating `carousel' of discrete sparks (discharges) that are formed in a gap (charge depleted region) above the neutron star surface. The electron-positron pairs produced in the discharges are responsible for the observed radio emission. A circulation of the carousel of radio beamlets around the magnetic axis, due to the $\bf E \times B$ drift, is predicted. The circulation timescale of the carousel together with the number of sparks define the $P_3$ of the observed pattern of drifting subpulses. Varying drift rates can then be thought of as changes in the number of beamlets or rotation speed of the carousel. 

The sparking gap model was further developed by e.g.\,\citet{Gil2003model} allowing for an additional screening of the electric field in the vacuum gap. In contrast to the classical picture, \citet{Basu2016} discussed the possibility that for a pulsar with a magnetic dipole axis inclined to the rotation axis, the sparks circulate around the rotation axis rather than the magnetic axis. Very different geometrical interpretations of drifting subpulses have been proposed. For example, \citet{Rosen2008} proposed a non-radial oscillation model without invoking circulations in the magnetosphere. \citet{Gogoberidze2005} suggested that drifting subpulses are due to modulation in the emission region generated by drift waves, in some form of magnetospheric oscillations. 

Here we aim to establish evolutionary trends of drifting subpulses, and subpulse modulation in general. This requires great sensitivity and sufficiently long pulse sequences. The MeerKAT Thousand-Pulsar-Array (TPA) project is designed to observe over a thousand southern hemisphere pulsars, a large fraction of the total known pulsar population, with a uniform observing setup. This provides an unprecedented view on the pulsar population. The MeerKAT telescope consists of 64 dishes, with an effective diameter of around 13 meters each. The full array has a total system equivalent flux density of 7.5~mJy, a receiver temperature of 18~K, and a wide bandwidth of around 642~MHz at L-band for single pulse studies \citep{Johnston2020}. The adopted observing strategy resulted in at least one long observation with high instantaneous sensitivity, as well as regular observations utilising subarrays (see \citealt{Song2021} for a detailed description of the TPA legacy datasets). 

We report on the findings of a systematic analysis of subpulse modulations for \censuspsr\ pulsars in the TPA legacy data. This is the largest subpulse modulation survey to date, with in total 1.6 million stellar rotations being analysed. A catalogue of subpulse modulation properties for a significant fraction of the known pulsars is provided, and various trends across the pulsar population are identified. 

The structure of the paper is as follows. Sec.\,\ref{sec:method} describes how drifting subpulses are identified in the spectral domain. The used sample is defined in Sec.\,\ref{sec:dataquality}, and the results from the identification of drifting subpulses are compared with results from other major drifting subpulse surveys. In Sec.\,\ref{sec:fracdriters} the fraction of pulsars with periodic subpulse modulation in the pulsar population is quantified, and correlations with $P$ and \pdot\ (spin-down rate) are identified in Sec.\,\ref{sec:ppdot}. Implications of our findings are discussed in Sec.\,\ref{sec:discussion}, followed by conclusions. Online materials\footnote{Available on zenodo via: \url{10.5281/zenodo.6900582}} include a table summarising all our measured subpulse modulation properties, and figures presenting the fluctuation spectra for all sources in the analysed sample, as well as additional descriptions related to the data analysis adopted pipeline and complementary statistical analysis.

\section{Observations and data processing}
\label{sec:method}

\subsection{Recording and generation of single pulse data}
\label{sec:singlepulsepipeline}

The data were observed with the L-band receivers (with a centre frequency which is precisely 1283.58203125\,MHz) of the MeerKAT array, and recorded by the PTUSE\footnote{Pulsar Timing User Supplied Equipment} instrument in single-pulse observing mode \citep{Bailes2020}. This produced filterbank data with a channel width of 0.8359375\,MHz and $\sim38.28\;\mu {\rm s}$ time resolution. The recorded bandwidth is 642\,MHz for data recorded between May 2019 and 10 February 2020, and  856\,MHz otherwise. 

After recording, the data are further processed using the methodology in Keith et al., (in prep.). In this process the filterbank data from the telescope are `folded' using {\sc dspsr} \citep{vanStraten2011}, producing an archive containing a sequence of pulses for each pulse period. Where possible, it is ensured that the boundaries between different pulse numbers do not coincide with observable emission. Polarisation calibration is applied, and the most by RFI affected frequency channels are removed. A set of fully frequency-averaged single-pulse archives  are produced for each observation.
The science ready data are analysed for subpulse modulation, which is described next.

\subsection{Profile alignment and on-pulse identification}
\label{sec:prepsrsalsa}

A python script generates and executes a sequence of shell commands which perform spectral analysis in an automated way using {\sc{psrsalsa}}\footnote{https://github.com/weltevrede/psrsalsa} (see Sec.\,\ref{sec:psrsalsa}) and extract statistical information from the generated output.

The process starts by combining the single-pulse data for a given observation as produced by the process described in Sect.\,\ref{sec:singlepulsepipeline} to form a continuous pulse-stack suitable for further processing. The data are aligned with a smooth representation of the pulse profile which we will refer to as the `template' (see \citealt{Posselt2021} for details about how these are generated). These steps are performed using  {\sc{psrchive}} tools \citep{hvm04}.

Before spectral analysis can take place, the pulse phase ranges corresponding to the on-pulse and off-pulse (no detectable signal from the pulsar) regions need to be identified. This is determined automatically where possible, but is defined manually where needed. 

The algorithm identifies a main pulse (MP) and possibly an additional interpulse (IP) interval based on the template. As described in Sec.~\ref{sec:twodfs}, some
of the further analysis benefits from splitting the on-pulse
regions. Up to two pulse longitude intervals of
interest can be specified, which we will refer to as `components'. As an example, columns (a) and (b) in Fig.~\ref{fig:allspectra}
correspond to pulsars without an IP detected (no MP/IP next to the pulsar name), while the other pulsars have an IP (the corresponding IP figures can be found in the online material). Panel a i) of Fig.~\ref{fig:allspectra}
shows the pulse profile of PSR J1539$-$6332 (solid line). This double-peaked profile has been split into two components, indicated by the different shaded regions underneath the profile. In contrast, the profile of PSR J1539$-$4828 (panel b i) is treated as a single component.

Appendix~\ref{app:spectra} has further information related to the definition of on- and off-pulse regions. The off-pulse  regions are required for baseline subtraction (see Sect.\,\ref{sec:baselinesub}) and analysis of the spectra (see Sect.\,\ref{sec:lrfsmod} and \ref{sec:twodfs}). For {\nrPulsarsNoOffSubtrTwodfs} pulsars the profile is too wide so that no suitable regions can be identified for off-pulse subtraction, which affects some of the further processing (see section \ref{sec:psrsalsa}). These are excluded from the analysis where relevant as explained in Sec.\,\ref{sec:ppdot}.

\subsection{Spectral analysis}
\label{sec:psrsalsa}

All further processing is done using {\sc{psrsalsa}}, and we refer to \citet{wel16} and references therein for additional descriptions of the techniques used. The procedure is summarised below, with an emphasis on extensions to the methodology developed for this survey.

\subsubsection{Baseline subtraction and RFI rejection}
\label{sec:baselinesub}
By default the `baseline' is subtracted for each individual pulse by fitting a first order polynomial to the off-pulse regions. This deals with slow (compared to $P$) variations in the baseline. This option is set for most pulsars, except for {\nrNoBaselineSlopeSub} pulsars (see online material {\onmanualonpulse}) with small off-pulse regions, where the slope subtraction would otherwise introduce variability in the baseline.

The process described in Sec.~\ref{sec:singlepulsepipeline} identifies, and removes, the frequency channels worst affected by RFI. See online material \ref{sec:app_RFI} 
for details of further RFI rejection, with typically a modest {\percentzap} of analysed pulses affected.

\subsubsection{LRFS and modulation index}
\label{sec:lrfsmod}

The longitude-resolved fluctuation spectrum (LRFS; \citealt{Backer1970}) is computed by calculating the Fast Fourier Transform (FFT) of sequences of flux densities for successive pulse longitudes. The LRFS helps to identify periodic emission and reveals the pulse longitudes for which the periodicity occurs. It can also be used to compute the longitude-resolved modulation index, which is a measure of variability defined as the standard deviation of the signal divided by the mean. An advantage of calculating it via the spectral domain is that the off-pulse spectral response is subtracted from the LRFS (reducing the power in the power spectrum). As a consequence, the bias on the modulation index introduced by the white-noise and possible remaining variations in the baseline is suppressed. Periodic RFI, affecting a narrow range of fluctuation frequencies, can be effectively excluded from the analysis in the spectral domain. This has been done for {\zapspec} pulsars (see online material {\onrfi}). 

The optimum pulse longitude resolution is determined automatically by considering the fraction of pulse longitude bins for which a significant modulation index could be detected (see online material \ref{sec:AutoRebinning}). 
The FFT length is automatically determined by considering the number of spectral bins in the LRFS with a significant detection of spectral power for most pulsars (see Appendix~\ref{app:spectra} for details).

\begin{landscape}

\begin{figure}
    \centering
    \begin{minipage}[t]{0.19\linewidth}
        \includegraphics[width=0.99\linewidth]{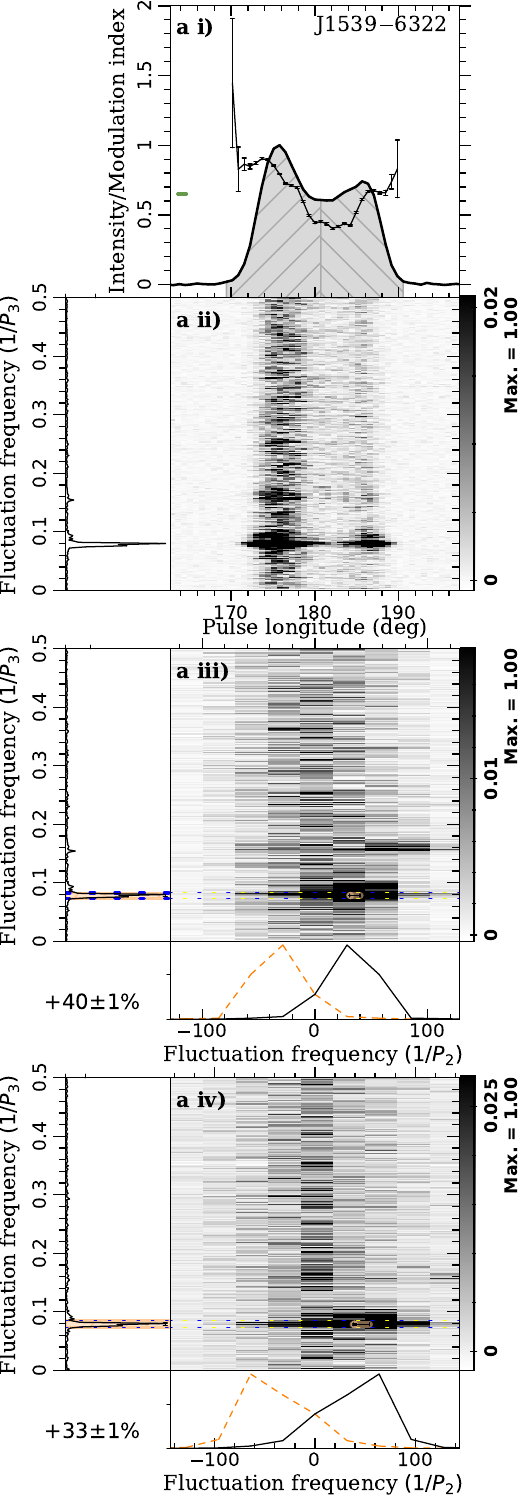}
    \end{minipage}
    \begin{minipage}[t]{0.19\linewidth}
    \includegraphics[width=0.99\linewidth,trim=0 -44mm 0 0, clip]{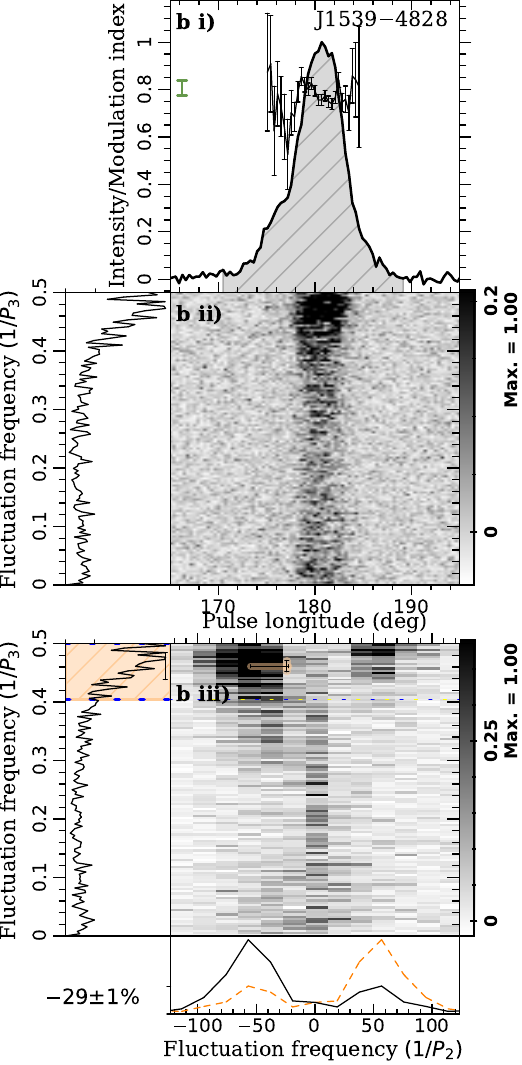}
    \end{minipage}
    \begin{minipage}[t]{0.19\linewidth}
        \includegraphics[width=0.99\linewidth]{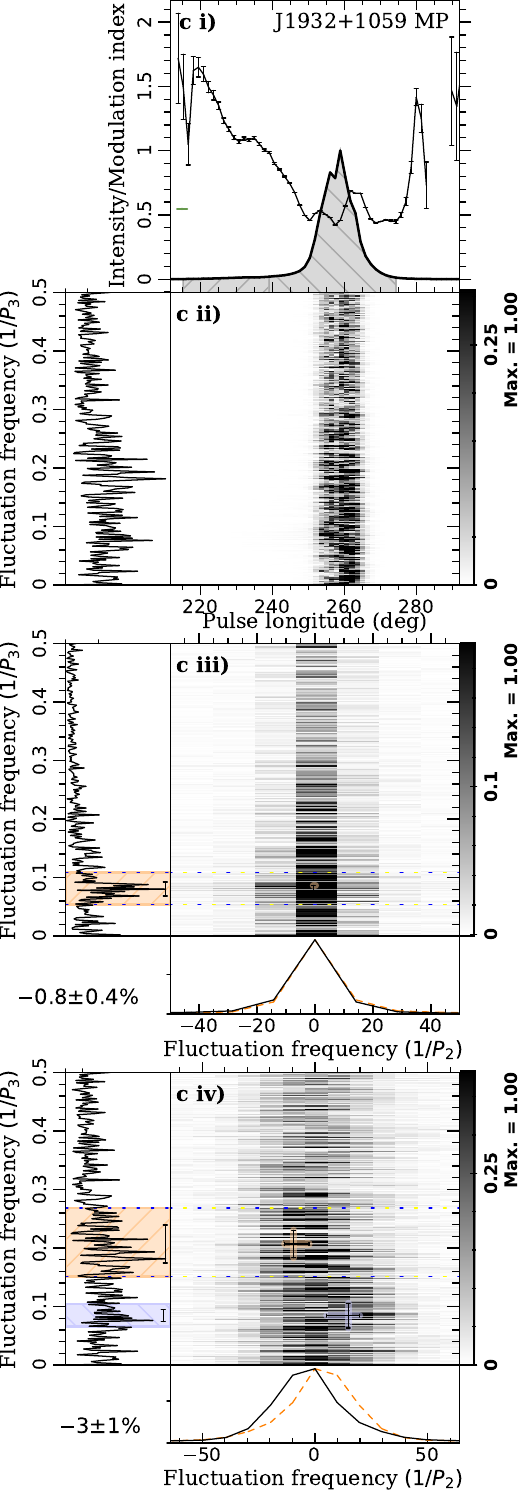}
    \end{minipage}
    \begin{minipage}[t]{0.19\linewidth}
      \includegraphics[width=0.99\linewidth]{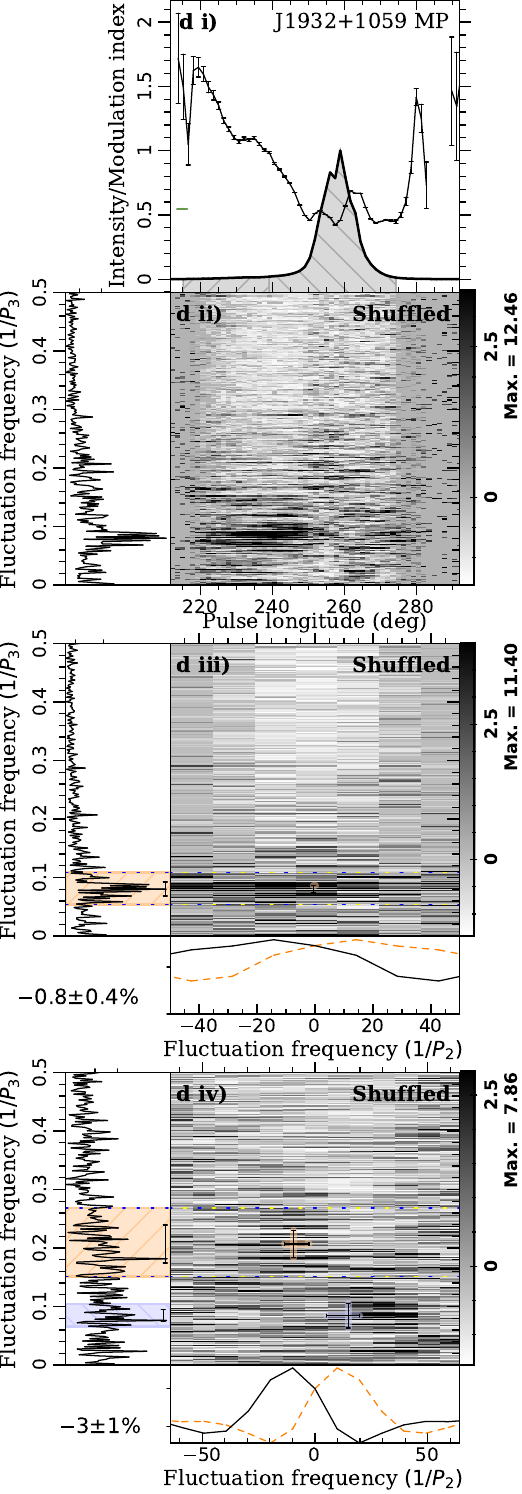}
      \end{minipage}
     \begin{minipage}[t]{0.19\linewidth}
        \includegraphics[width=0.99\linewidth,trim=0 -44mm 0 0, clip]{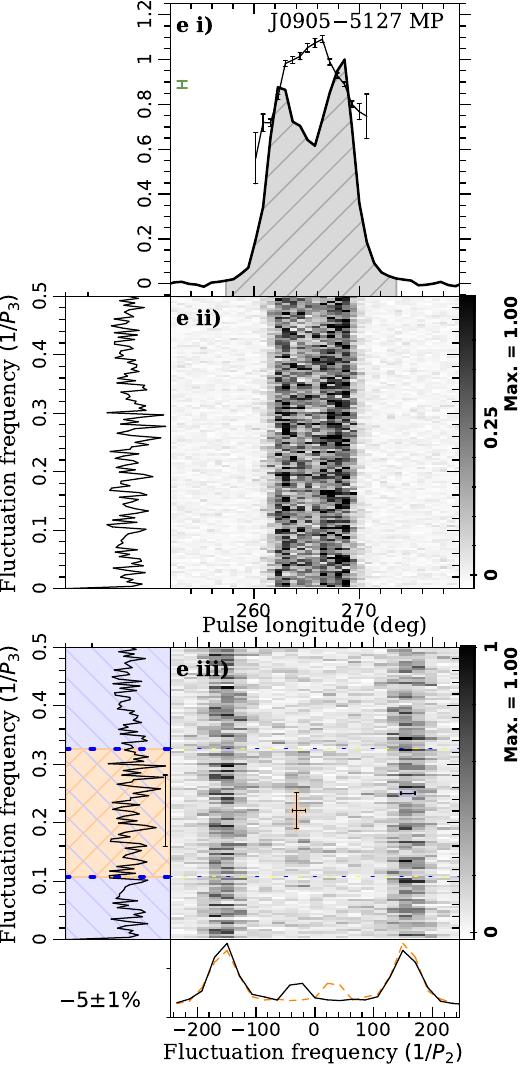}
    \end{minipage}
    \caption{{\em Column (a)}: a sharp spectral drifting subpulse feature and its harmonic. {\em Column (b)}: a spectral feature near the $P_3=2$ alias border. 
    {\em Column (c)}: the spectral results of PSR~J1932+1059. {\em Column (d)}: the  shuffle-normalised spectra for the same pulsar as column (c). A $1/P_3\simeq0.08$~cpp spectral feature dominates the appearance of the shuffle-normalised LRFS at a pulse longitude range between $220^{\circ}$ and $250^{\circ}$, which is invisible in column (c). Similarly, the spectral features corresponding to drifting subpulses are also much clearer in the shuffle-normalised 2DFS in panels (d iii and d iv). {\em Column (e)}: a $P_2$-only spectral feature (vertical bands of power extending the full $1/P_3$ range in panel e iii), as well as a weaker feature associated with drifting subpulses as indicated with a cross in the 2DFS.
    {\em Panels (i)} show the normalised average pulse profile (solid line) and modulation indices (points with errorbars, with ${\bar{m}}$ shown at the far left). The LRFS {\em (panels ii)} cover the same pulse longitude range. 
    For each pulsar there is at least one ({\em panels iii}), and sometimes a second ({\em panels iv}) 2DFS, corresponding to the shaded hatched regions under the pulse profile.
All spectra have a colour bar corresponding to spectral power, which can have a maximum value less than that indicated as `Max.' next to it. If so, the brighter features are clipped such that samples exceeding this threshold value are set to the darkest colour, thereby putting more emphasis to weaker features.
The horizontally integrated power in all spectra are shown in the left side panels.
The power in the 2DFS is vertically integrated between the dotted lines, or the full range when no dotted lines are shown in the 2DFS. This is shown in the bottom panels (solid line) attached to the 2DFS. The dashed line is mirrored about $1/P_2=0$ to highlight asymmetries.
For each spectral feature with a $P_2$ and $P_3$ measurement, their values are shown as small black error bars on the 2DFS and the shaded region in the left side panels indicate the spectral range included in the measurement. When multiple spectral features are analysed in the same 2DFS, different colours are used. The errorbars in the left side panels correspond to $\sigma_{1/P_3}$ (not visible for PSR~J1539$-$6322) and 
the dotted lines highlight for each 2DFS the dominant spectral feature with a $P_3$ value measured. The thicker dotted lines indicate the dominant feature of the pulsar.
The percentage indicated for each 2DFS corresponds to $P_\mathrm{asym}$. See Sec.~\ref{sec:psrsalsa} for more details.}
\label{fig:allspectra}
\end{figure}

\end{landscape}

In the figures, the dynamic range in the spectra are optimised to emphasise the weaker spectral features (see Appendix~\ref{app:spectra} for details).
An example LRFS is shown as a grey-scale map in panel (a ii) of Fig.~\ref{fig:allspectra} for PSR J1539$-$6322\footnote{Single pulses of this pulsar are shown in the online material \ref{app:pulse_stack}.}.
The horizontal axis of the LRFS corresponds to the pulse longitude in degrees, which is aligned with the pulse profile shown above. The vertical axis corresponds to the $1/P_3$ fluctuation frequency in cycles per period (cpp), where $P_3$ is the vertical separation between drift bands. The horizontally integrated power is shown in the side panel. The LRFS for PSR~J1539$-$6322 shows a sharp spectral feature seen across the pulse at $1/P_3 \simeq 0.08$ cpp, corresponding to $P_3\simeq 12$. Weaker spectral features (lighter colours) are seen at other fluctuation frequencies. Some are associated with stochastic subpulse modulation, although in this case the $1/P_3 \simeq 0.16$ cpp feature is the first harmonic of the fundamental frequency. No drifting subpulses were reported before for this pulsar. 

\cite{WES2006,wse07} analysed the minimum in the longitude resolved modulation index. An alternative way to extract a single value for the modulation index is to compute the flux density weighted and pulse-longitude averaged modulation index (for both the MP and IP), we define as
\begin{align}
\label{EqAvMod}
  {\bar{m}} = \frac{\sum_im_iI_i}{\sum_iI_i} = \frac{\sum_i\sigma_i}{\sum_iI_i}.
\end{align}
Here for on-pulse bin $i$ the longitude resolved modulation index and standard deviation\footnote{Since the off-pulse spectrum has been subtracted from the LRFS, the spectral power derived variance $\sigma_i^2$ can be negative. In those cases $\sigma_i$ is taken to be
$-(-\sigma_i^2)^{1/2}$.} are $m_i$ and $\sigma_i$ respectively, and $I_i$ is the flux density of the profile. In the last step in Eq.~\ref{EqAvMod}, we used the definition $m_i\equiv\sigma_i/I_i$. The error on $m_i$ and ${\bar{m}}$ are derived using bootstrapping (see \citealt{wel16}). In Eq.~\ref{EqAvMod}, $m_i$ for all on-pulse pulse longitude bins are included, regardless of their significance.
The longitude-resolved modulation indices are shown as solid points with errorbars in panels (i) in Fig.\,\ref{fig:allspectra}, and that of ${\bar{m}}$ is indicated at the far left in the same panels.

\cite{WES2006} suggested the use of the minimum in the longitude resolved modulation index, because the modulation index tends to be the largest where the pulse profile is weak. So the minimum should be relatively free from a S/N bias. In this work we prefer the use of ${\bar{m}}$, as it incorporates the emission from all pulse longitudes for which emission is detected.

To check the S/N bias, simulations have been done by adding white noise to pulsar data. This shows that there is still a S/N bias in the  minimum $m_i$ arising from the fact that a smaller $m_i$ requires a higher S/N to become significant. The simulations show that ${\bar{m}}$ is less susceptible to such a bias. However, there is a bias for S/N $\lesssim100$ such that the measured $\bar{m}$ tends to be too low. The analysis in Sec.\,\ref{sec:resultmod} takes this into account.

\subsubsection{2DFS evaluation}
\label{sec:twodfs}

The LRFS, when computed as a power spectrum (Sec.~\ref{sec:lrfsmod}) and as shown in for example Fig.~\ref{fig:allspectra}, does not reveal whether periodic subpulse modulation corresponds to subpulse drifting (phase modulation) or longitude stationary (amplitude) modulation. Phase drift can be revealed using the two-dimensional fluctuation spectrum (2DFS). The 2DFS is the power spectrum obtained by computing FFTs in both the pulse longitude and the pulse number direction \citep{ES2002}. The used pulse longitude resolution and the FFT sizes are the same as those used to compute the LFRS. The FFTs in the pulse longitude direction are restricted to the profile components (indicated by the shaded regions of the pulse profiles in panels i) of Fig.~\ref{fig:allspectra}, and up to two 2DFSs are calculated per MP and IP. As for the LRFSs, the dynamic range is adjusted to ensure weak features are visible (see Appendix~\ref{app:spectra}) and an off-pulse spectrum is subtracted. Details can be found in online material \ref{sec:app_offpulse_sub_2dfs}, 
including an explanation of differences compared to what is described in \citet{wel16}.

Examples of 2DFSs are shown in panels (iii) and (iv) of Fig.\,\ref{fig:allspectra}. The vertical axis has the same unit as that of the LRFS, while the horizontal axis corresponds to the fluctuation frequency $1/P_2$ in cpp. For drifting subpulses, $P_2$ is the separation between drift bands in the pulse longitude direction. A visual aid to assess the asymmetry of a spectral feature in the 2DFS with respect to $1/P_2=0$ is the bottom panel attached to the 2DFS, which shows the vertically integrated power of the 2DFS between the dotted lines indicating the most relevant feature to assess the asymmetry for. When no such feature is identified, there are no dotted lines shown in the 2DFS, and the full spectrum is integrated over. To highlight any asymmetry, a mirrored curve is shown as the dashed line. In the case of PSR J1539$-$6322 column (a) of Fig.~\ref{fig:allspectra},
a 2DFS is produced for each of the two profile components. The subpulses for both components have positive drifting directions (subpulses drift towards the trailing side of the profile), corresponding to the power in the 2DFS peaking at a positive $1/P_2$ frequency\footnote{Following \cite{WES2006,wse07}, the 2DFS is mirrored about the vertical axis to ensure that positive drift corresponds to a positive $1/P_2$ frequency.}. The spectral features peak at $1/P_2\simeq30$ and $\simeq60$ cpp for the leading and trailing components respectively, corresponding to $P_2\simeq360\degr{}/30=12\degr{}$ and $6\degr{}$. The first profile component shows a weaker spectral feature with the $1/P_3$ and $1/P_2$ frequencies doubled, as expected for the first harmonic of the drifting subpulse feature.

Column (b) of Fig.~\ref{fig:allspectra} (for PSR J1539$-$4828) shows an example of a broad spectral feature close to the alias boarder at $1/P_3=0.5$ cpp. The vertical extension of the spectral feature indicates that the modulation periodicity $P_3$ is variable. When $P_3<2$, the sampling of the pattern once per pulse period is not enough to resolve the periodicity unaliased. As a result, a periodicity slightly faster than 2 will be observed slightly slower than 2 with a drift direction which is opposite. Indeed the 2DFS shows spectral features at $\pm60$ cpp suggestive of opposite apparent detected drift directions. In this example, negative drift dominates over positive drift.

\subsubsection{Shuffle-normalised spectra}
\label{sec:shuffle_norm}

There are two contributions to the uncertainty in the found spectral features. One is attributed to the (possibly white) system noise. However, for most analysed pulsars, the significance of features is limited by the finite number of pulses. This can make a spectral feature spurious rather than reproducible in longer observations. What we will refer to as ``shuffle-normalised spectra'' (both the LRFS and 2DFS) are constructed to assess whether a spectral feature is significant or not. 

Spectra are computed for {\nrbootstrapitts} pulse sequences where the order of pulses is randomised, resulting in an average spectral power and root-mean-square (rms) for each spectral bin. No significant structures in $1/P_3$ are expected in the spectra of the randomised pulse sequences. The shuffle-normalised spectra are calculated such that for each bin in the original spectrum the average power is subtracted, before dividing by the rms. In addition, any spectral bins for which the power in the original spectrum does not exceed two times the off-pulse spectral rms are set to zero. The resulting shuffle-normalised spectra highlight spectral features associated with organised subpulse modulation and suppress stochastic subpulse variability.

Examples of shuffle-normalised spectra are shown in column (d) of Fig.~\ref{fig:allspectra}. These have identical markings of the $P_2$ and $P_3$ measurements, and $1/P_3$ ranges associated to each spectral feature, compared to the regular spectra (column c).
The regular LRFS (panel c ii) is dominated by variability associated with the main part of the pulse profile between pulse longitude $\sim250^{\circ}$ and $\sim270^{\circ}$, with a hint of a broad (diffuse) excess of spectral power between 0.1 and 0.3 cpp. In contrast, the shuffle-normalised LRFS (panel d ii) reveals a much sharper spectral feature at $\sim0.08$ cpp associated with the very low intensity leading part of the pulse profile (between pulse longitude $220^{\circ}$ and $250^{\circ}$). This is a clear demonstration of the power of the shuffle-normalised spectra. The weak spectral features, corresponding to organised subpulse modulation, stand out more clearly by suppressing spectral power associated with stochastic pulse-to-pulse variability. Also in the first 2DFS (corresponding the leading part of the profile) the spectral feature is clearer in the shuffle-normalised spectrum.
This spectral feature corresponds to the subpulse modulation reported by \citet{Kou2021}. Appendix~\ref{app:spectra} has more details related to shuffle normalised spectra. All the regular and shuffle-normalised spectra can be found in the online material.

\subsubsection{2DFS analysis}
\label{sec:twodfsana}

The spectra of all pulsars were investigated for evidence of significant spectral features. Identified features in the 2DFS are classified to be a drifting subpulse feature if there is a significant offset of the feature with respect to the $1/P_2$ axis. If no such offset is found, but there is a spectral feature which peaks at a specific non-zero $1/P_3$, this is defined as a $P_3$-only feature. For example, the spectral feature in panel d iii of Fig.~\ref{fig:allspectra} is classified as a $P_3$-only feature, while that in panel d iv is classed as a drifting subpulse feature. Finally, there are three pulsars (PSRs J0905$-$5127, J1835$-$1106 and J1856+0113) where subpulse modulation is identified with a specific $P_2$ separation between subpulses, but without any phase relation of subpulses between successive pulses. For example, panel e iii in  Fig.~\ref{fig:allspectra} shows two symmetrical vertical bands of power centered at $1/P_2=\pm150$~cpp, attributed to subpulses appearing with a typical separation within the pulse of $\sim360\degr/150\simeq2.4\degr$. There is no significant structure in the $1/P_3$ direction as the pulsar has no memory of at which pulse longitude the subpulses appeared in the previous pulse. As a consequence, these spectral features disappear in the shuffle-normalised spectra (see online material). Besides this main spectral feature which is classified as a $P_2$-only feature, there is a weak drifting subpulse feature with $1/P_2$ at around $-50$~cpp with a $1/P_3$ around 0.2 to 0.3 cpp 

The spectra were inspected by eye for features of interest and for each pulsar the dominant feature was identified. The dominant feature is what is used in the statistical analysis when spectral properties of different pulsars are compared. See Appendix~\ref{app:spectra} for details, including how the significance of spectral features is assessed.

\cite{WES2006,wse07} classified drift subpulse features as either `coherent' or `diffuse' depending on the width of the spectral feature in $1/P_3$. Rather than judging this by eye, in this work the width of the feature is measured from the horizontally integrated spectral power of the 2DFS over the selected $1/P_2$ range for the feature. To suppress any excess spectral power not associated with the feature of interest, a similar approach to produce the shuffle-normalised spectra is used. However, here the average power of the spectra arising from the shuffled pulse sequences is subtracted, but the spectra are not normalised. This step helps suppressing the `baseline' of the resulting integrated spectrum. 
An example of a shuffle-subtracted spectrum can be found in Appendix~\ref{app:spectra}, as well as further details related to this baseline subtraction.
The resulting spectral power as a function of spectral bin $i$ is $\mathcal{P}_i$. The variance associated with the width of the spectral feature is defined as
\begin{align}
\label{EqSpectralWidth}
\sigma_{1/P_3}^2 &= \frac{\sum_i\mathcal{P}_i\times((1/P_3)_i-\mu_{1/P_3})^2}{\sum_i\mathcal{P}_i},
\end{align}
where $(1/P_3)_i$ is the frequency (in cpp) associated with spectral power $\mathcal{P}_i$, and the mean spectral frequency of the feature is defined as
\begin{align}
\label{eq:pstd}
\mu_{1/P_3} &= \frac{\sum_i\mathcal{P}_i\times(1/P_3)_i}{\sum_i\mathcal{P}_i}.
\end{align}
In Eqs.~\ref{EqSpectralWidth} and \ref{eq:pstd} the summation is only over the $1/P_3$ region covering the feature.
The quoted values of $\sigma_{1/P_3}$ are set to be at least the spectral resolution if the feature is unresolved. An error on $\sigma_{1/P_3}$ is derived by repeating the computation many times after adding random Gaussian noise to the 2DFS with a rms equal to that measured in the off-pulse spectrum. 

The spectral shape of a drift feature can be distinctly non-Gaussian. An advantage of the used method to measure the spectral width is that no fitting is involved, and that no specific spectral shape has to be assumed. When the spectral shape is non-Gaussian, the reported standard deviation does not fully quantify the shape of the spectral feature, but it is indicative of its width. Similarly, this procedure of subtracting the baseline from $\mathcal{P}_i$ is a relatively simple, but in general an effective way to isolate the spectral feature of interest. The main advantage of the methodology used is that no detailed modelling of the 2DFS is required, which is difficult to automate for a large number of pulsars. 

Also the strength of the spectral power associated with drifting subpulses is quantified with a procedure that does not rely on fitting the spectral shapes of features in the 2DFS. This is done by quantifying the overall asymmetry of spectral power $\mathrm{2DFS}_{i,j}$ in the 2DFS, defined as
\begin{align}
  \label{eq:AsymDriftPower}
P_\mathrm{asym} &= \frac{|\sum_j\sum_i (\mathrm{2DFS}_{i,j}-\mathrm{2DFS}_{i,j'})|}{\sum_i\sum_j (\mathrm{2DFS}_{i,j} + \delta_{j,j'}\mathrm{2DFS}_{i,j'})},
\end{align}
Here index $i$ covers the full spectral range $1/P_3$ (so between 0 and 0.5 cpp), while $j$ covers half of the range corresponding to $1/P_2$ (e.g. only positive frequencies). The index $j'$ is defined to correspond to a frequency $1/P_2$ with the same magnitude, but the opposite sign compared to that of index $j$. The Kronecker delta $\delta_{j,j'}$ ensures that the spectral bin corresponding to $1/P_2=0$ (for which $j=j'$) is not double counted. The $1/P_2$ Nyquist frequency, which per definition is not associated with a drift direction, does not contribute to the numerator of Eq.~\ref{eq:AsymDriftPower}, but is included in the denominator.

Eq.~\ref{eq:AsymDriftPower} defines the fraction of subpulse modulation that is associated with an excess of drift in one particular direction. It is defined as a positive quantity, given the numerator contains the absolute value of the difference in spectral power between the two halves of the 2DFS. The sign of the numerator of Eq.~\ref{eq:AsymDriftPower}, without taking the absolute value, defines which drift direction dominates in the spectrum. The presence of noise (system noise as well as stochastic pulse-shape variability) implies that always some asymmetry can be expected in the 2DFS. The significance of the asymmetry is quantified by randomising (shuffling) the order of the pulses many times and re-computing $P_\mathrm{asym}$. After shuffling, no significant drift can be expected, hence the standard deviation of the obtained values of $P_\mathrm{asym}$ quantifies its uncertainty.

A S/N bias is introduced in Eq.~\ref{eq:AsymDriftPower} because the absolute value of the numerator is used. To quantify the bias, a few pulsars with drifting subpulses for which high S/N observations were available were analysed further. Different amounts of white noise were added to the pulse stack and \pasym\ was recalculated. If the $\mathrm{S/N}\gtrsim100$, no significant bias is observed. For the statistical analysis related to \pasym\ only observations with a S/N above 100 are considered to minimise this bias (see Sec.\,\ref{sec:resultpasym}). 

\subsection{$P$-\pdot\ correlation analysis}

\label{sec:mcmc}

How the quantities derived from the spectral analysis vary as a function of $P$ and \pdot\ is investigated in Sect.~\ref{sec:ppdot}. The correlations between a measured quantity $F$ (e.g. $P_3$) and a combination of $P$ and \pdot\ are parameterised as
\begin{equation}
    \log_{10} F = c (\log_{10} (P^{a}\dot{P}^{b}))^2 + d \log_{10}(P^{a} \dot{P}^{b}) + e.
    \label{eq:mcmcfit}
\end{equation}
The constants $a$ and $b$ define the direction in the $P$-\pdot\ diagram, and the quadratic form parameterised with the constants $c, d, e$ allows a non-linear variation to be described. Where a linear relation is observed in the $P$-\pdot\ diagram, $c$ can be fixed at zero. The ratio $a/b$, or $b/a$, defines a gradient in $\log_{10} (P)$-$\log_{10} (\dot{P})$ space, hence $a$ and $b$ are not independent parameters.
Therefore, when fitting Eq.\,\ref{eq:mcmcfit}, either $a$ or $b$ is set to 1. Unless the correlation is dominated by a dependence on $P$ or \pdot, the choice of which parameter to keep as a free parameter is arbitrary, but is informed by which resulted in more Gaussian-like posterior distributions or a larger correlation coefficient (see below).
Under the assumption of energy loss via magnetic dipole radiation, characteristic age \age, kinetic energy loss rate {\edot} and the dipolar component of the surface magnetic field strength {\bsurf} correspond to $a/b=-1$, $-3$ and $+1$, respectively. The true age and surface  magnetic field strength can be significantly different from the derived \age\ and \bsurf, e.g.\ in the presence of other energy-loss mechanisms.

The best-fitted parameters are determined by a Bayesian Markov Chain Monte Carlo (MCMC) approach. In addition to the parameters in Eq.\,\ref{eq:mcmcfit}, an extra noise term {\sig} is included. This accounts for a non-perfect correlation, even after accounting for the measurement uncertainties. Details of the fitting process can be found in online material \ref{sec:app_Bayesian}. 

To assess the significance of the found correlations, a weighted correlation coefficient between the measured $\log_{10} F$ values and the model (Eq.\,\ref{eq:mcmcfit}) is calculated. The data are weighted down by a variance equal to the quadrature sum of the measurement error and \sig.
The uncertainty on the correlation coefficient is estimated using bootstrapping, whereby the correlation coefficient is repeatedly re-calculated using randomly selected samples with replacement from the whole dataset, after applying offsets drawn from a Gaussian distribution with a standard deviation set to the measurement error. The standard deviation of the distribution of correlation coefficients is taken to be the uncertainty on the measured correlation coefficient.

\section{The Sample and a Survey Comparisons}
\label{sec:dataquality}

\subsection{The sample}
\label{sec:sample}

The observing strategy of the TPA project is defined in \citet{Song2021}. This strategy ensures high fidelity observations suitable for the detection of profile variability by taking into account the time required to achieve the desired profile stability. A pulsar monitoring campaign is currently being carried out, which adopts a strategy which exploits the ability of MeerKAT to form two subarrays. In addition, for the brightest half of the sample a longer pulse sequence (1024 or more pulses) was recorded using all available dishes from the MeerKAT array to maximise the sensitivity to detect individual pulses. Also for many pulsars with a lower flux density more than 1024 pulses were recorded to reach the required sensitivity and profile stability \citep{Song2021}. 

We here present the analysis of data for \censuspsr\ pulsars. For the majority of pulsars (\longobs) an observation of at least 1024 pulses is used. See online material \ref{app:sample} 
for a more detailed description of the distribution of observation lengths.

A comparison between the measured signal-to-noise ratio (S/N) of the integrated pulse profiles and that aimed for using the methodology of \citet{Song2021} shows that for many pulsars (for \numsnragree\ out of \numsnrall) the S/N of the data is consistent with the expected S/N within a factor of two (see online material Fig.~\ref{fig:snrcompdata}). 
Reasons why the S/N is lower than expected include that for some pulsars their actual flux densities (see \citealt{Posselt2022} for our TPA measurements) are different compared to the catalogued values assumed (see \citealt{Song2021} for a discussion), radio-frequency interference (RFI), and the use of the less sensitive subarray observations in \subarrayobs\ cases. Subarray observations were used for various reasons, including technical faults, less favourable interstellar scintillation or nulling conditions. The level of deviation is consistent with the findings of \citet{Song2021}.

A consequence of the spread in S/N is that not all observations are equally sensitive in detecting drifting subpulses, and this will be further discussed in Sec.\,\ref{sec:fracdriters}. Furthermore, the S/N is correlated with other pulsar parameters, in particular, with pulse period $P$ such that long period pulsars typically have a somewhat higher S/N (see online material Fig.~\ref{fig:snrvsp}). 
This is because long period pulsars were observed with longer observing length $t_{\mathrm{obs}}$, since the majority of pulsars in the sample (\thousandpulses\ out of \censuspsr) were observed for $\sim1000$ pulses. So higher S/Ns ($\propto \sqrt{t_{\mathrm{obs}}}$) are observed for longer period pulsars. Also the duty cycle $W/P$, where $W$ is the pulse width, plays a role and is correlated with the spin parameters \citep{Posselt2021}.
The S/N bias is further discussed in Sec.\,\ref{sec:fracdriters} and \ref{sec:ppdot}.

Some of the statistical analysis relies on information (in particular the spin parameters) from the pulsar catalogue (ATNF catalogue\footnote{\url{http://www.atnf.csiro.au/research/pulsar/psrcat}} version {\psrcatversion}, \citealt{atnf}), except for PSR~J1357$-$62 for which we used parameters supplied by S. Johnston (priv. comm.). 
For two pulsars we used the name 
from the pulsar catalogue (PSRs~J0514$-$4408 and J1402$-$5021) rather than the name used when the data was recorded (PSRs~J0514$-$4407 and J1402$-$5124).

In addition,  {\nrfoldharmonics} pulsars were identified to be folded at a harmonic of the true pulse period, implying that the pulse period in the catalogue was incorrect. This resulted in very sharp spectral features in our analysis. The data (after correcting both $P$ and \pdot) was refolded, as summarised in Table\,\ref{tab:foldharmonics}.

Within the sample of pulsars, there are \numseverescatter\ showing evidence of severe scattering with a timescale comparable to or longer than the pulse widths. These pulsars were visually identified, and included a sub-sample of those studied in \citet{Oswald2021}. For some of them, the large duty cycles are problematic (see Sect.\,\ref{sec:lrfsmod} and \ref{sec:twodfs}). Therefore, these pulsars are excluded from analysis where relevant (see Sect.\,\ref{sec:ppdot}). 

\begin{table}
    \centering
    \begin{tabular}{cclc}
\hline \hline
Pulsar & Harmonic & $P$ & $\dot{P}$\\ 
& & [s] & [s/s]\\ \hline
J0211$-$8159 & 3 & 3.2 & 8.7$\times10^{-16}$\\
J0711+0931 & 2 & 2.4 & $8.0\times10^{-16}$\\
J0836$-$4233 & 2 & 1.5 & $2.6\times10^{-15}$\\
J1427$-$4158 & 2 & 1.2 & $1.2\times10^{-15}$\\
J1618$-$4723 & 2 & 0.41 & $4.0\times10^{-15}$\\ 
J1639$-$4604 & 2 & 0.53 & $5.8\times10^{-15}$\\
J1714$-$1054 & 3 & 2.1 & $1.8\times10^{-16}$\\
J1739$-$3951 & 2 & 0.7 & $4.0\times10^{-17}$\\
J1802+0128 & 2 & 1.1 & $4.2\times10^{-15}$\\
J1830$-$0131 & 3 & 0.46 & $6.3\times10^{-15}$\\
J1848$-$1150 & 2 & 2.6 & $2.9\times10^{-15}$\\
J1946$-$1312 & 2 & 1.0 & $4.0\times10^{-15}$\\ \hline
    \end{tabular}
    \caption{Pulsars for which $P$ and {\pdot} were corrected (as shown) to a harmonic.}
    \label{tab:foldharmonics}
\end{table}

\subsection{Comparison with other surveys}
\label{sec:checklit}

\label{sec:online}

\begin{table*}
\sisetup{detect-weight=true,detect-inline-weight=math}
\caption{\label{tab:p3s} The first few entries of the table that can be found in the online material summarising the spectral subpulse modulation measurements. For each pulsar $P$ and \pdot, as well as the from these derived $\tau_c$ and $\dot{E}$ are reported. This is followed by the number of pulses used for the modulation index analysis and the corresponding S/N. The parameters $\bar{m}$, $P_{\mathrm{asym}}$, $P_2$, $P_3$ and $\sigma_{1/P_3}$ are defined in Sec.~\ref{sec:lrfsmod} and \ref{sec:twodfsana}. Each pulsar can have multiple profile components, each of which can have multiple spectral features. Each feature is assigned a class (drift/$P_3$-only/$P_2$-only correspond to drifting subpulses, $P_3$-only or $P_2$-only features as defined in the main text). A more in-depth description of the parameters is provided with the full online table.}
\resizebox{\linewidth}{!}{
\begin{tabular}{llS[table-format=1.4]SS[table-format=5.3]S[table-format=3.3]|rrS[table-format=1.5]S|ccS[table-format=2.3]S[table-format=1.3]r@{\;}l}
\hline\hline
\hspace{5mm}PSRJ & \hspace{3mm}PSRB & \mc{$P$} & \mc{$\dot P$} & \mc{$\tau_c$} & \mcn{$\dot{E}$} & Pulses & S/N & \mc{$\bar{m}$} & \mcn{$P_{\mathrm{asym}}$} & Class & Comp & \mc{$P_3$} & \mc{\hspace{4mm}$\sigma_{1/P_3}$} & \multicolumn{2}{c}{$P_2$} \\ 
 & & \mc{(s)} & \mc{($10^{-15}$ s/s)} & \mc{($10^{5}$ yr)} & \mcn{($10^{33}$ erg/s)} & & & & \mcn{(\%)} & & & & \mc{\hspace{4mm}(cpp)} & \multicolumn{2}{c}{(deg)} \\ \hline 
J0034$-$0721 & B0031$$-$$07 & 0.943 & 0.408 & 366.0 & 0.019 & 1040 & 1985 & 1.458(1) & -15.0(7) & drift & MP C1 & \bfseries 6.6(2) & 0.010(4) & $-23\!$ & $^{+3}_{-11}$  \\ 
 
 & & & & & & & & & & drift & MP C1 & 9.9(7) & 0.015(6) & $-23\!$ & $^{+2}_{-64}$  \\ 
 
J0038$-$2501 &  & 0.2569 & 0.001 & 53600.0 & 0.002 & 2336 & 73 & 2.01(9) & 38(16)  & & & & & &  \\ 
 
J0045$-$7042 &  & 0.6323 & 2.49 & 40.2 & 0.389 & 1632 & 36 & 1.6(2) & -13(15)  & & & & & &  \\  
... & & & & & & & & & & & & & & & \\ 
\hline
\end{tabular}
}
\end{table*}

The main table summarising our measurements can be found in the online material, of which the first entries are also in Table~\ref{tab:p3s}. The spectral figures are in the online material as well, with Fig.~\ref{fig:allspectra} being an example. Before analysing their statistical properties, the performance of the employed analysis techniques  (Sect.\,\ref{sec:method}) is assessed first. This is done by comparing our results for individual sources with those obtained by \citet{WES2006}, \citet{Basu2016} and \citet{Basu2019}. This comparison gives an indication of the completeness and accuracy of our analysis.

\citet{WES2006} analysed 187 pulsars using WSRT data at an observing frequency comparable to the data presented here and \tpawes\ pulsars were in both surveys. In addition, a comparison is made with the results of \citet{Basu2016} and \citet{Basu2019}, who analysed a combined dataset of 123 pulsars observed with GMRT at a lower frequency of 618 MHz and 333 MHz. \citet{Basu2019} analysed a larger sample compared to \citet{Basu2016}, but with a focus on phase modulations only. They reported 61 pulsars with drifting subpulses, of which \tpabl\ pulsars are covered in our sample. In online material \ref{sec:app_survey_comparison} 
we report on the small fraction of the sample for which our results do not confirm the reported results, or for which we detect not-reported subpulse modulation. In addition we comment on the minority of pulsars in the overlapping samples for which the $P_3$ measurements are inconsistent.

\section{The fraction of pulsars with drifting subpulses}

\label{sec:fracdriters}
\begin{figure}
    \centering
    \includegraphics[width=0.9\linewidth]{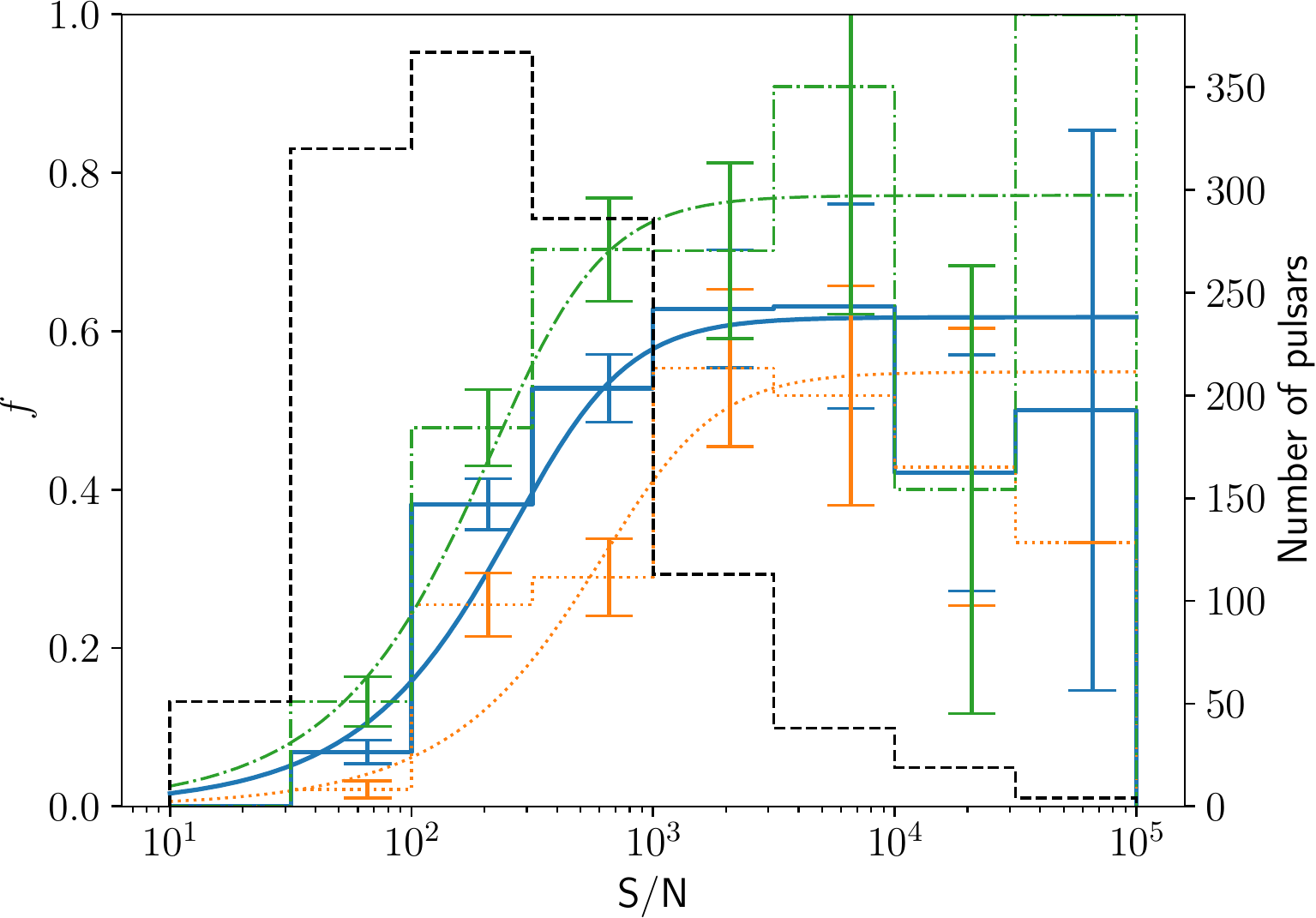}
    \caption{Distribution of the fraction of pulsars with drifting subpulses $f$ (solid histogram), with errorbars corresponding to the square-root of the number of pulsars with drifting subpulses divided by the total number of pulsars in a given S/N bin (dashed histogram). The solid curve is the fit of the $f$ with Eq.~\ref{eq:funcsnr}. The dotted and dash-dotted $f$ distributions (fitted with the same functional form) correspond to pulsars with \age\ below and above the median \age\ of the sample. }
    \label{fig:snrpdmp}
\end{figure}

Of the total \censuspsr\ pulsars in our sample, we found evidence for drifting subpulses in \driftnum\ pulsars, which corresponds to \driftfrac. As pointed out by for example \citet{WES2006}, the detectability of drifting subpulses depends on the S/N of the observation, so the intrinsic fraction of pulsars with drifting subpulses must be higher. 
This is demonstrated in Fig.\,\ref{fig:snrpdmp}, which shows the fraction of pulsars with drifting subpulses (solid histogram) as a function of the S/N of the integrated pulse profile of the observation. It is evident that at low S/N, drifting subpulses are less likely to be detected. The fraction of pulsars with drifting subpulses $f$ increases strongly towards higher S/N, before the distribution flattens at around $f=0.6$ when the S/N exceeds $\sim1000$. 

The observed distribution $f$ has a functional shape comparable to what can be used to describe high-pass filters,
and is fitted as
\begin{equation}
    f=f_\infty \frac{(\mathrm{S/N})/\mathrm{(S/N)}_0}{\sqrt{1 + (\mathrm{(S/N})/\mathrm{(S/N)}_0)^2}},
    \label{eq:funcsnr}
\end{equation}
where $f_\infty$ is the fraction of pulsars with detectable drifting subpulses in the limit of high S/N, and $\mathrm{(S/N)}_0$ quantifies the location of the turn-over in the distribution where $f=f_\infty/\sqrt{2}$.
The observed distribution is well described with this function with the fitted parameters $f_\infty=\fitsnrf \pm \fitsnrerrf$ and $\mathrm{(S/N)}_0=\fitsnrs\pm \fitsnrerrs$ (solid blue curve in Fig.~\ref{fig:snrpdmp}).
For many observations $\mathrm{S/N}<\mathrm{(S/N)}_0$ (see the dashed black distribution in Fig.~\ref{fig:snrpdmp}). This highlights that the \driftfrac\ of pulsars with detectable drifting subpulses severely underestimates the intrinsic fraction of pulsars with drifting subpulses, which is better estimated as $f_\infty$. This is consistent with the statement by \citet{WES2006} that over half of all pulsars would have detectable drifting subpulses in high enough S/N data, although the sample studied here is large enough to show that even for observations with very high S/N drifting subpulses are not detectable in all pulsars. When $P_3$-only pulsars are included in $f$, the plateau would be slightly higher with $f_\infty=\fitsnrallf \pm \fitsnrallerrf$.

One should be careful with interpreting the dependence of $f$ with S/N, as the S/N of the observations is correlated with $P$ (see Sec.\,\ref{sec:dataquality}). Not only is high S/N data beneficial for the detectability of drifting subpulses, long period pulsars (with higher S/N) are intrinsically more likely to show drifting subpulses since they are associated with large \age\ (e.g.\,\citealt{Rankin1986,WES2006}, but see also Sec.~\ref{sec:resultpasym}).
To disentangle these two effects, $f$ as a function of S/N was considered for pulsars with \age\ below and above the median value of \medage\ years (shown as the dotted orange and dash-dotted green distributions in Fig.\,\ref{fig:snrpdmp}).
For both age groups $f$ increases with S/N, and older pulsars are more likely to show drifting subpulses. Fitting of Eq.\,\ref{eq:funcsnr} reveals that the turnover $\mathrm{(S/N)_0}$ is similar for all distributions, suggesting that the dominant aspect to the S/N dependence of $f$ is the observational bias making drifting subpulses easier to detect when the S/N is high.
The pulsar luminosity is observed to be uncorrelated with $f$, demonstrating that the physical parameters governing the radio luminosity do not contribute to the detectability of drifting subpulses in pulsars. 

Apart from S/N, there are other factors which affect the detectability of drifting subpulses. For some pulsars drifting subpulses can be highly regular, making them easier to detect. For others the drifting subpulses patterns are highly irregular, potentially interrupted by nulling or intervals of time corresponding to an emission mode where there are no detectable drifting subpulses. 
The length of the observations analysed here are in general sufficiently long to allow the detection of drifting subpulses, and the length is found to be uncorrelated with $f$ (not shown).

Our results confirm that drifting subpulses are a very common phenomenon for radio pulsars, hence the physical conditions required for the production of drifting subpulses cannot be very different compared to the required conditions for the radio emission mechanism itself. Drifting subpulses might well be a fundamental property of the emission mechanism, although they are not detectable for all pulsars. The origin of the dependence of $f$ with the spin parameters is explored in more detail in Sect.~\ref{sec:ppdot}.

\section{Subpulse modulation in $P$-\pdot\ space}
\label{sec:ppdot}

In this section we quantify the evolution of subpulse modulation in the $P$-\pdot\ diagram, with a focus on properties related to drifting subpulses. The best fitted $P$ and \pdot\ relations (using the MCMC method described in Sec.\,\ref{sec:mcmc}) are summarised in Table~\ref{tab:relation}, including the strength of the correlation. Different properties are discussed in turn, aimed at building a phenomenological understanding of how subpulse modulation evolves in the $P$-\pdot\ diagram. The found correlations are compared with an evolution with \age\ or \edot\ (see Table~\ref{tab:relation}). It should be stressed that $P$, \pdot, \age\ and \edot\ are not independent parameters. However, these are useful parameters to relate the measurements to emission models. The qualitative interpretations are described geometrically in terms of the carousel model. Nevertheless, they could apply to any model capable of producing periodic subpulse modulation. The properties presented here will be further explored and compared to predictions in Sec.\,\ref{sec:discussion}.

\begin{table}
    \centering
    \setlength\tabcolsep{2.5pt}
    \begin{tabular}{clccl}
    \hline \hline
      Property & \multicolumn{1}{c}{Fit} & $c=0$ & Correlation & Consistent \\ \hline 
    $P_3$ & $a/b=\Pverdriftparag\pm\Pverdriftparagerr$ & N & $\corrPverdriftpara\pm\correrrPverdriftpara$ & \age\ \\
    $\sigma_{1/P_3}$ & $a/b=\Pverstddriftparag\pm\Pverstddriftparagerr$ & N & $\corrPverstddriftpara\pm\correrrPverstddriftpara$ & \edot\ \\
    \pasym\ & $a/b=\Powersellng\pm\Powersellngerr$ & Y & $\corrPowerselln\pm\correrrPowerselln$ & \age, \edot\ \\
    $|P_2|$ & $a/b=\Phorabsdriftlng\pm\Phorabsdriftlngerr$ & Y & $\corrPhorabsdriftln\pm\correrrPhorabsdriftln$ & \edot, \age, \pdot\ \\
    $|D|$ & $b/a=\driftratedriftlng\pm\driftratedriftlngerr$ & Y & $\corrdriftratedriftln\pm\correrrdriftratedriftln$ & \edot, $P$ \\
    $|D|/(W_{50}/2)$ & $a/b=\driftfracdriftlng\pm\driftfracdriftlngerr$ & Y & $\corrdriftfracdriftln\pm\correrrdriftfracdriftln$ & \age, \pdot\ \\
    $\bar{m}$ & $b/a=\AvModsnrparag\pm\AvModsnrparagerr$ & N & $\corrAvModsnrpara\pm\correrrAvModsnrpara$ & $P$ \\ \hline
    \end{tabular}
    \caption{A summary of the correlations found (see Sec.~\ref{sec:ppdot}) for the subpulse modulation properties (first column) with $P$ and \pdot. The next two columns explains if $a/b$ ($b$ is fixed to 1) or $b/a$ ($a$ is fixed to 1) is fitted for and the result, and if $c$ is fixed at zero in Eq.~\ref{eq:mcmcfit}. The final two columns show the strength of correlation quantified with the correlation coefficient, and if it is consistent with evolution in the \age\ ($a/b=-1$), \edot\ ($a/b=-3$), $P$ ($b=0$) or \pdot\ ($a=0$) direction.}
    \label{tab:relation}
\end{table}

\subsection{$P_3$ evolution}
\label{sec:resultp3}

Before exploring how pulsars with drifting subpulses are distributed in $P$-$\dot{P}$ space, it is instructive to consider the evolution of $P_3$ first. Many pulsars have multiple measured $P_3$ values. This could be because there are multiple profile components with separate $P_3$ measurements (this is discussed in Sec.~\ref{sec:comptheo}), or multiple periodicities were identified in a given component. Here we analyse a single $P_3$ for each pulsar, identified to be the dominant drifting subpulse feature by virtue of being the strongest and most striking. Preference is given to more well-defined (narrow) features. The used $P_3$ values are bold in Table~\ref{tab:p3s} and the corresponding online table. Pulsars severely affected by scattering were excluded.

\begin{figure}
    \centering
    \includegraphics[width=\linewidth]{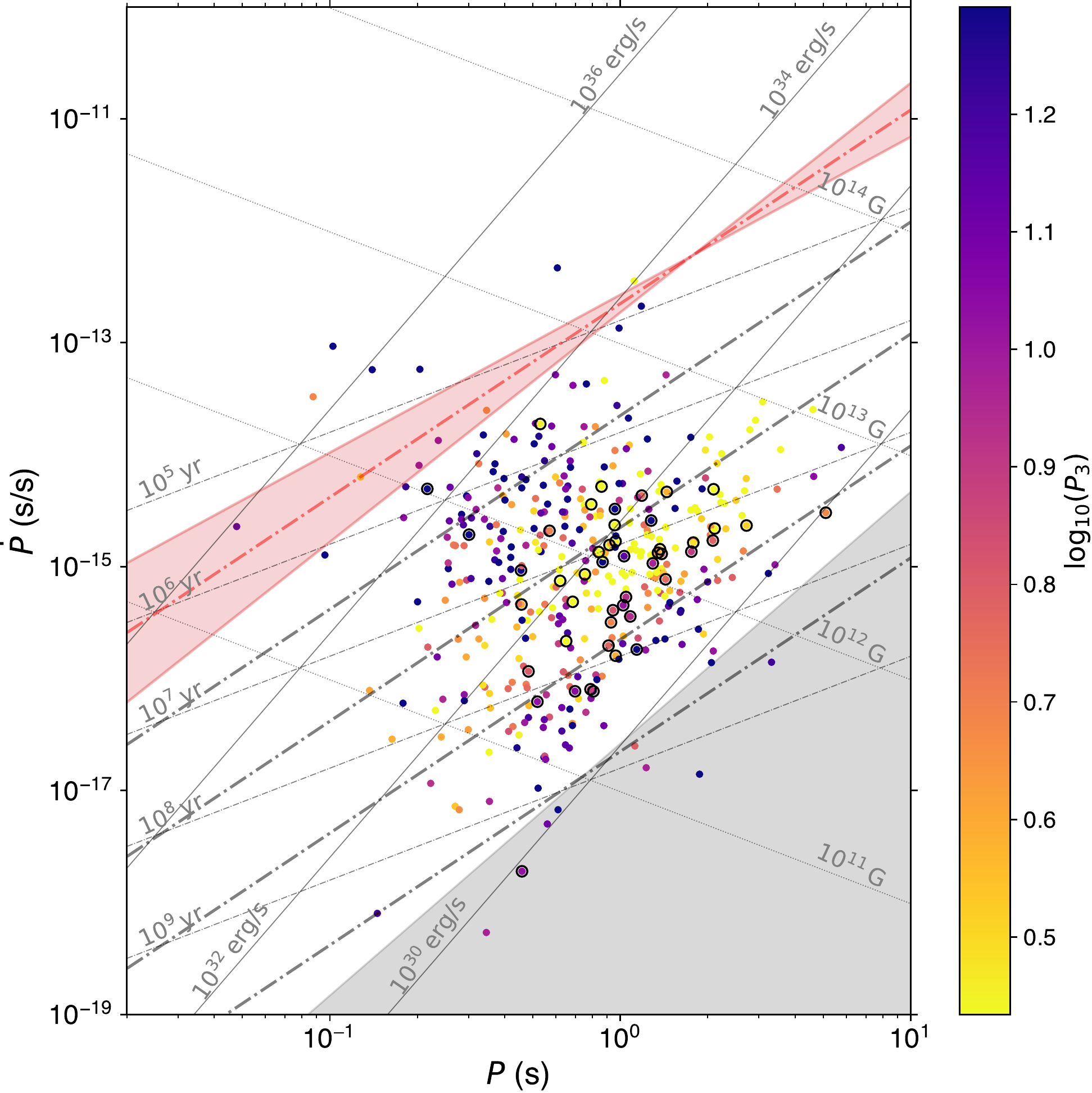}
    \caption{$P$-$\dot{P}$ diagram where the colour of the dots represent $\log_{10}(P_3)$ for each pulsar with a dominant drifting subpulse feature identified. Pulsars which have multiple distinct $P_3$ measurements for a single profile component are marked with black circles. The dot-dashed lines correspond to constant $P_3$ values according to a fit of Eq.\,\ref{eq:mcmcfit} (with $b=1$). The top shaded region indicates the $1\sigma$ uncertainty on the slope. Lines of constant \age, \bsurf\ and \edot, 
    are shown as the double dot-dashed, dotted and solid lines, respectively. The region below the death line (Eq.\,4 of \citealt{Zhang2000}) is shaded in the bottom-right corner.}
    \label{fig:parappdotp3}
\end{figure}

\begin{figure}
    \centering
    \includegraphics[width=\linewidth,trim= 0 0 0 1.3cm, clip]{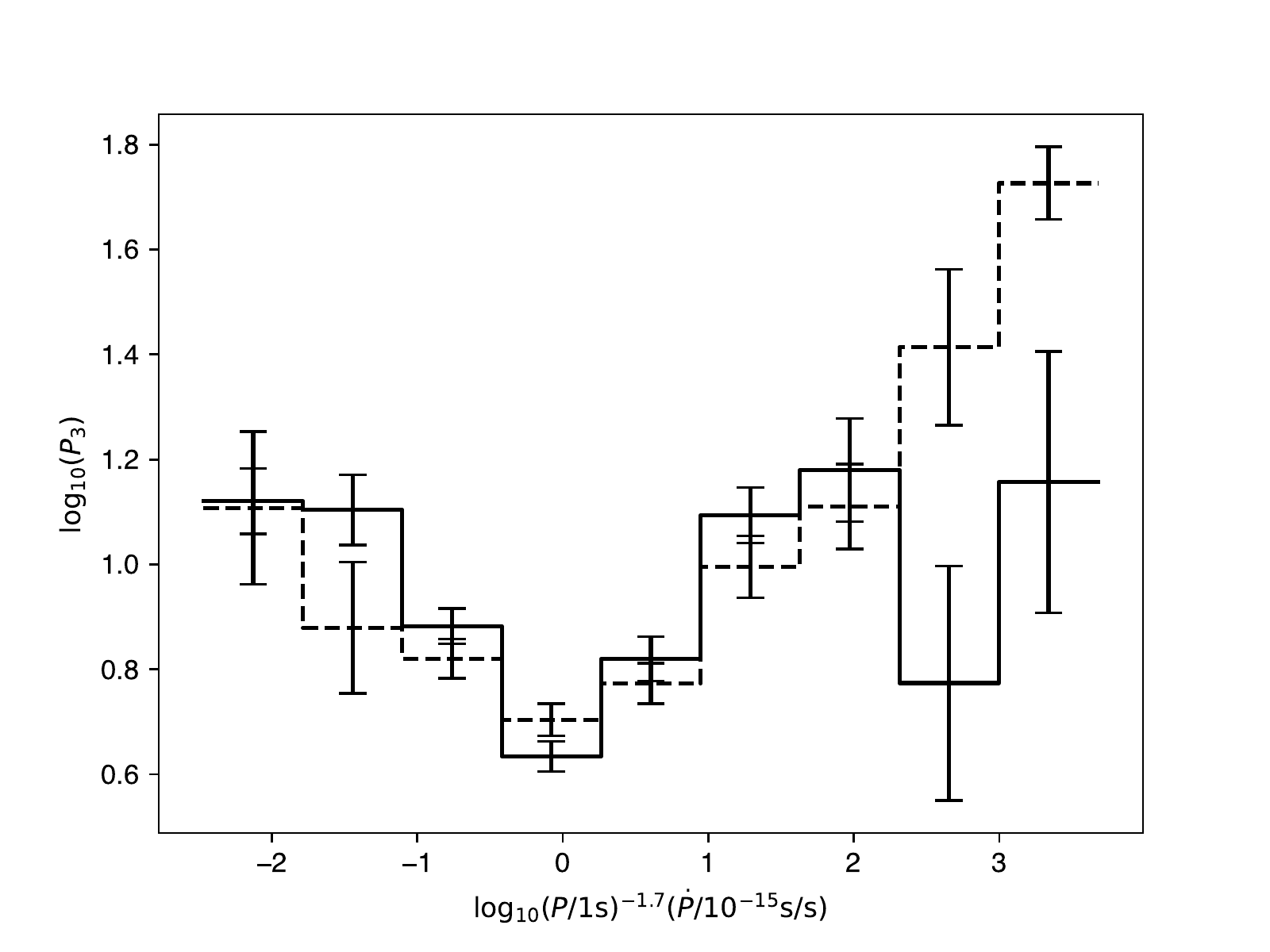}
    \caption{A histogram of the measured (solid) and predicted (dashed) weighted mean $P_3$ as a function of the combination in $P$ and \pdot\ corresponding to the direction of the strongest correlation in the $P$-\pdot\ diagram. The prediction is based on Eq.\,\ref{eq:mcmcfit} fitted to the measured distribution. The errorbars represent the standard deviation of values contributing to each bin, divided by the square root of the number of values.}
    \label{fig:histp3}
\end{figure}

The distribution of $P_3$ in the $P$-$\dot{P}$ diagram (Fig.\,\ref{fig:parappdotp3}) is V-shaped such that large $P_3$ values are found for the more energetic (large {\edot}) and young (small {\age}) pulsars, as well as the least energetic old pulsars. At around $\tau_c \simeq 10^{7.5}$ yr (or $\dot{E} \simeq 10^{31.5}$ erg/s), the smallest $P_3$ values dominate. A strong correlation (see also Table~\ref{tab:relation}) was found, such that the evolution is consistent with the \age\ direction and preferred over the \edot\ direction, consistent with a dependence on \age\ and  preferred over an evolution in the \edot\ direction. This V-shaped evolution of $P_3$  is further illustrated in Fig.\,\ref{fig:histp3} (solid line), which shows the evolution of the weighted mean $P_3$ in the direction in the $P$-$\dot{P}$ diagram suggested by the fit. The simulated distribution (dashed line) is computed by generating the same number of $P_3$ values as the original data from the fitted function (Eq.~\ref{eq:mcmcfit}) with the best-fit $P$-\pdot\ combination. Each $P_3$ value is generated from a Gaussian distribution with mean of the corresponding predicted value from the fitted function and a standard deviation as the combined error of the scatter of points $\sigma_e$ and the measurement error. This shows that the fit can reproduce the found evolution.

Near $\tau_c \simeq 10^{7.5}$ yr ($\dot{E}\simeq10^{31.5}$ erg/s) many pulsars have $P_3$ values relatively close to the alias border $P_3=2$. This suggests that the V-shaped evolution is a consequence of aliasing, arising from the emission pattern being observed once per stellar rotation. In the geometric framework of a carousel model, this could be caused by a monotonic evolution of the carousel rotation period. The oldest and least energetic pulsars have slowly rotating carousels resulting in large unaliased $P_3$ values. For somewhat younger pulsars the rotation is faster leading to a decrease in $P_3$. Even faster carousels for even younger pulsars then ultimately allow for aliasing to occur, such that the apparent $P_3$ can be large again despite that intrinsically now $P_3<2$ (this will be quantified in Sec.~\ref{sec:p3sim}). Although described in the framework of a carousel model, the aliasing effect could apply to any model which can produce periodic subpulse modulation.

Pulsars with multiple $P_3$ features in a single profile component are marked in open circles in Fig.\,\ref{fig:parappdotp3}, which could be a consequence of drift mode changes. The majority of these are relatively old low {\edot} pulsars. The number of these pulsars is too small to impact the conclusions related to $P_3$ evolution depending on which of the multiple $P_3$ measurements are used for a given pulsar, and the V-shaped evolution persists when the smallest $P_3$ is selected instead of the dominant feature. 

Fig.\,\ref{fig:parappdotp3} is for the \numdomdrift\ pulsars for which the dominant spectral feature is associated with drifting subpulses. The distribution for the \numdomlon\ pulsars for which the dominant spectral feature is a $P_3$-only feature is more uniform throughout the pulsar population without evidence for a similar evolution (see Appendix~Fig.~\ref{fig:parappdotp3only}). Very different distributions are to be expected, as at least some of the $P_3$-only features will be related to a timescale associated with nulling or mode changing, which can have a distinctly different origin (e.g.\,\citealt{Rankin1986,Basu2020}). This also explains the significantly different mean $P_3$ values for pulsars with drifting subpulses compared to $P_3$-only pulsars, as confirmed by Student's t-test giving a P-value of $4\times10^{-8}$.

\subsection{Spectral width evolution}
\label{sec:resultp3std}

\begin{figure}
    \centering
    \includegraphics[width=\linewidth]{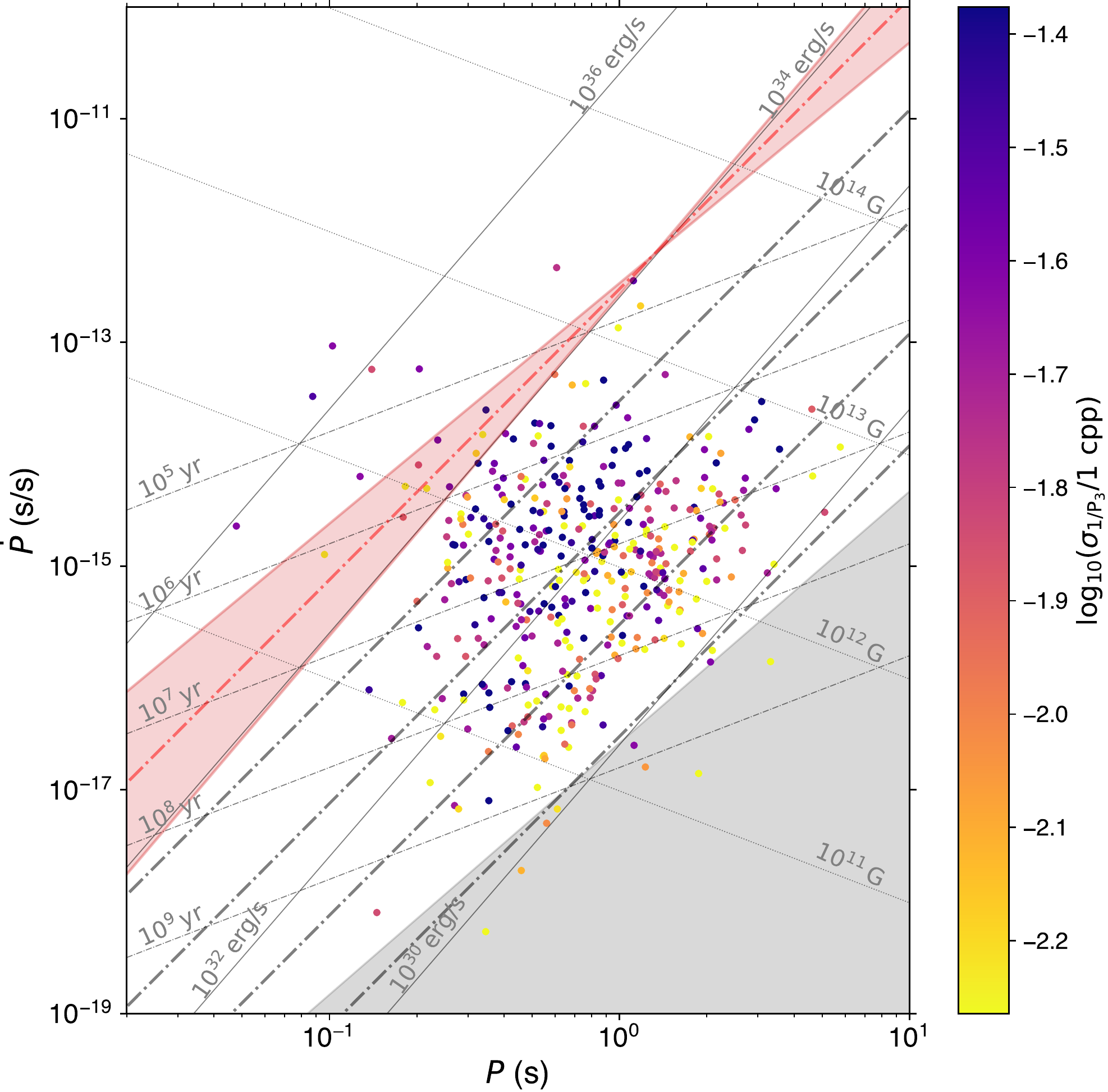}
    \caption{$P$-$\dot{P}$ diagram where the colour of the dots represent \pstd\ for each pulsar with a dominant drifting subpulse feature identified. The dot-dashed lines correspond to constant \pstd\ values according to a fit of Eq.\,\ref{eq:mcmcfit} (with $b=1$). The top shaded region indicates the $1\sigma$ uncertainty on the slope. See the caption of Fig.~\ref{fig:parappdotp3} for further details.}
    \label{fig:parappdotp3std}
\end{figure}

As discussed in \citet{WES2006}, older pulsars tend to exhibit more `coherent' drifting subpulses, corresponding to better defined $P_3$ values, while the spectral features of younger (or high \edot) pulsars are more `diffuse'. In our work, the spectral width \pstd\ in the $1/P_3$ direction quantifies the coherency of the drifting subpulses, thereby allowing a more detailed investigation compared to a binary coherent/diffuse classification in \citet{WES2006}. The $P$-$\dot{P}$ diagram in Fig.\,\ref{fig:parappdotp3std} shows \pstd\ for the dominant drifting subpulse feature, revealing that the most coherent subpulse modulation is indeed found for the oldest pulsars.

The found evolution (see Table~\ref{tab:relation}) is close to an evolution in the \edot\ direction, and is preferred over an evolution in $\tau_c$ direction. A histogram of the spectral width (Fig.\,\ref{fig:histp3std}) shows that its evolution starts to flatten for $\dot{E}\gtrsim10^{32}$ erg/s. This flattening is why a parabolic fit is preferred, rather than there being evidence for a turn-over in the evolution.

It is suggested that pulsars with smaller \age\ have faster intrinsic modulation (Sec.~\ref{sec:resultp3}), possibly because of faster rotating carousels. Therefore instabilities in their modulation period could have a larger effect on the observed spectral width. This is especially the case if their modulation is aliased, as this will amplify the apparent diffuseness of the spectral features compared to the intrinsic instability of the periodicities. The spectral width evolution therefore supports our suggestion that the youngest pulsars have aliased $P_3$, rather than the oldest.
The spectral width appears to be stronger linked to \edot\ than \age, while the opposite is true for the $P_3$ evolution. This implies that $P_3$ and \pstd\ are not both associated with only \age\ or \edot. Nevertheless, the $P$-\pdot\ evolution of $P_3$ and \pstd\ could be consistent with a common different combination of $P$ and \pdot.

In Fig.\,\ref{fig:parappdotp3std} the $P_3$-only pulsars are excluded, as that sample includes modulation caused by for example nulling (see Sec.~\ref{sec:resultp3}). Indeed the mean \pstd\ values are different for $P_3$-only and drifting subpulses pulsars as confirmed with a Student's t-test with a P-value of 0.0002. 

\subsection{Detectability and strength of drifting subpulses}
\label{sec:resultpasym}

\begin{figure}
    \centering
    \includegraphics[width=0.9\linewidth]{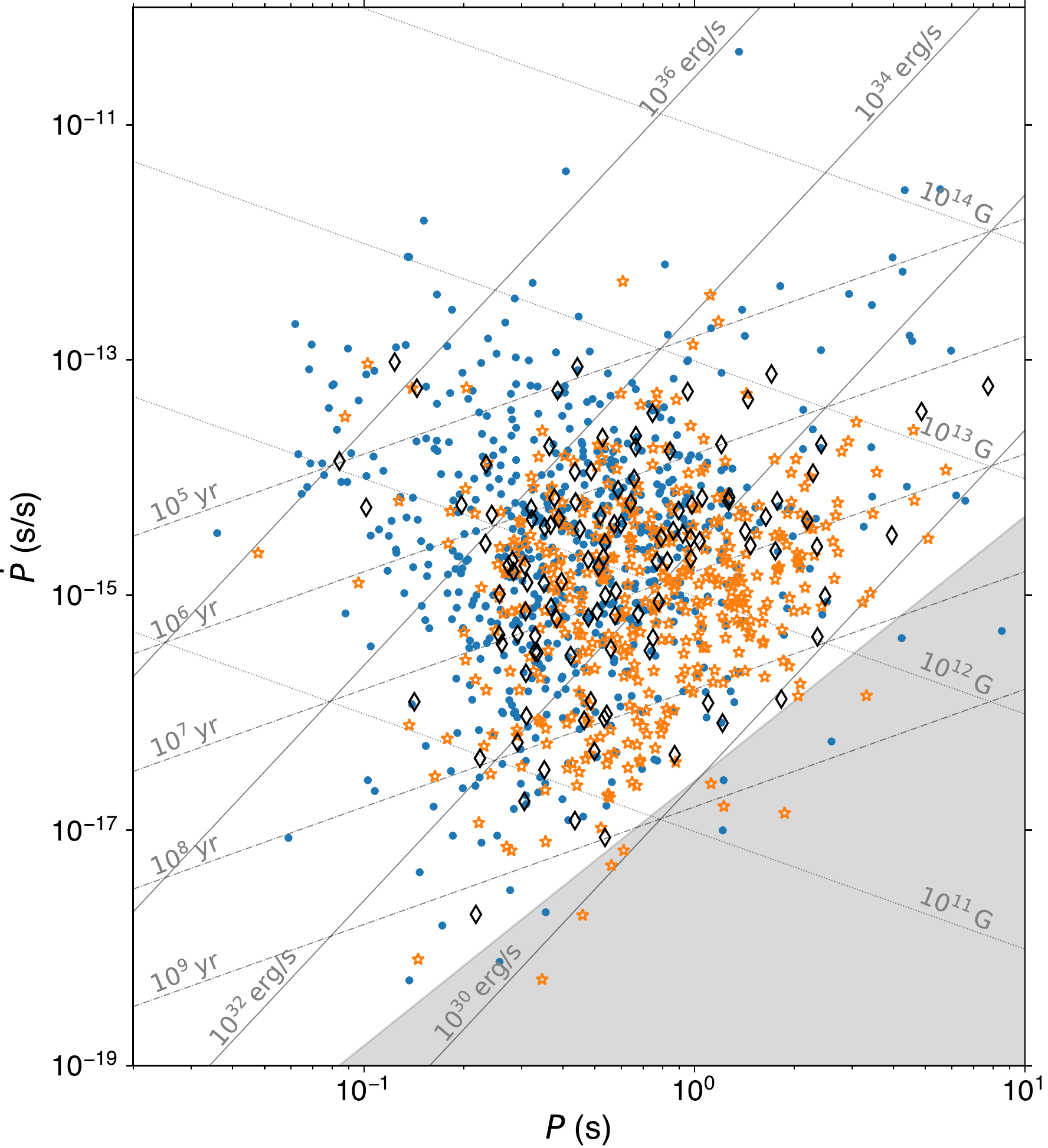}
    \caption{$P$-$\dot{P}$ diagram showing pulsars with detected drifting subpulses with stars, $P_3$-only pulsars with diamonds, and the other pulsars in the sample with the dots. See the caption of Fig.~\ref{fig:parappdotp3} for further details.}
    \label{fig:ppdotall}
\end{figure}

As shown in Fig.\,\ref{fig:ppdotall}, pulsars with detected drifting subpulses (stars) are close to the death line. They therefore typically have larger \age\ and lower \edot\ than pulsars without detectable periodic subpulse modulation (dots). The \age\ distributions of these two population are distinctly different, as the Kolmogorov-Smirnov test gives a P-value of $2\times10^{-19}$. This result confirms the finding of \citet{Ashworth1982,Rankin1986,WES2006,Basu2016,Basu2019} in our much larger sample. Pulsars with $P_3$-only features, shown with diamonds in Fig.\,\ref{fig:ppdotall}, are more spread out across the population. This is consistent with the findings in \citet{Basu2020} who found that pulsars with $P_3$-only features are found across a wide range of \edot. 

The preference for drifting subpulses to be detected in older and less energetic pulsars suggests that when the pulsars age their drifting subpulses become stronger. To test this, the evolution of \pasym\ (Eq.\,\ref{eq:AsymDriftPower}) is considered. This quantifies the asymmetry in the 2DFS arising when subpulses preferentially arrive earlier (or later) in successive pulses, and is independent of the spectral power being concentrated in a well defined $P_3$ feature. Especially because it is not significantly biased by the S/N until $\mathrm{S/N}<100$ (see Sec.\,\ref{sec:twodfsana}), unlike the detectability of drifting subpulses (see Sec.\,\ref{sec:fracdriters}), this is a useful quantity to consider. This quantity is determined for all pulsars, regardless if drifting subpulses are detected. For pulsars with multiple components, hence different \pasym\ measurements, the measurement with the smallest errorbar is selected. 

\begin{figure}
    \centering
    \includegraphics[width=\linewidth]{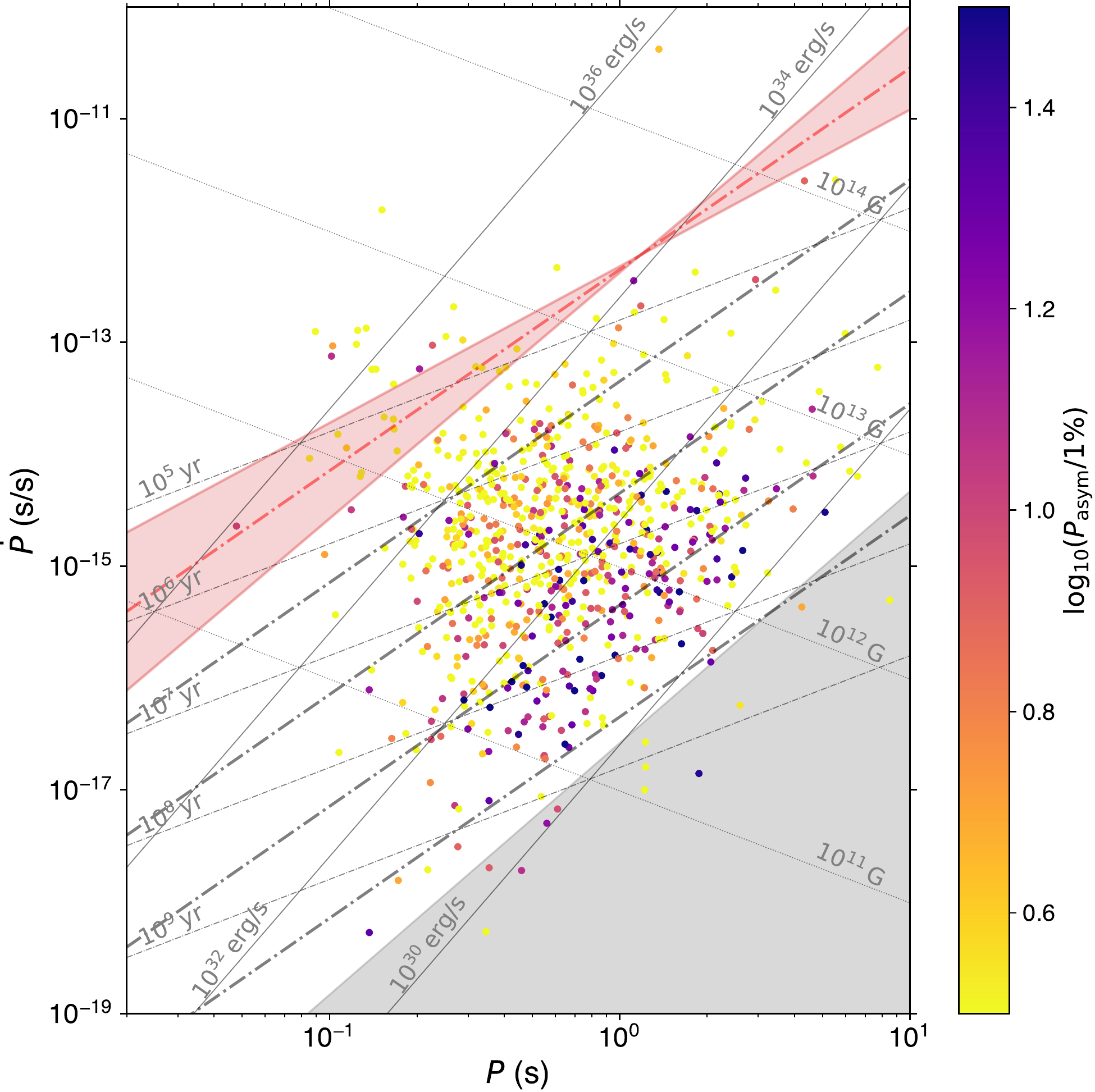}
    \caption{$P$-$\dot{P}$ diagram where the colour of the dots represent \pasym\ for the profile components for which \pasym\ has the smallest errorbar. Only pulsars for which $\mathrm{S/N}\geq100$ were considered. The dot-dashed lines correspond to constant \pasym\ values according to a fit of Eq.\,\ref{eq:mcmcfit} (with $b=1$ and $c=0$). The top shaded region indicates the $1\sigma$ uncertainty on the slope. See the caption of Fig.~\ref{fig:parappdotp3} for further details.}
    \label{fig:parappdotp2power}
\end{figure}

Fig.\,\ref{fig:parappdotp2power} shows the distribution of \pasym\ in the $P$-$\dot{P}$ diagram for pulsars with a $\mathrm{S/N}\geq100$. A strong and significant (see Table~\ref{tab:relation}) correlation is observed such that the older and less energetic pulsars tend to have the largest \pasym. This confirms that strong drifting subpulse signals are associated with pulsars closer to the death line.

The directions of evolution of \pasym\ and $P_3$ in the $P$-\pdot\ diagram are consistent (although the uncertainties are relatively large, see Table~\ref{tab:relation}), suggesting the same physical mechanism may be responsible for both. The transition between small and large values of \pasym\ occurs  at $\tau_{\mathrm{c}}\simeq10^7$ yr, which coincides with the transition from the smallest observed $P_3$ to larger values for the younger pulsars. The suggestion is made that younger pulsars are more likely to show aliased drifting subpulse patterns (Sec.\,\ref{sec:resultp3}), so the apparent drift direction of their subpulses is not necessarily constant throughout an observation. This would reduce \pasym, and together with the increasing diffuseness of the spectral features (see Sec.\,\ref{sec:resultp3std}) would reduce the detectability of drifting subpulses for younger pulsars. Although the decrease of \pasym\ for younger pulsars could in principle  point to the emergence of stronger stochastic disorganised subpulses in that part of the $P$-$\dot{P}$ diagram, there is no evidence for such an  evolutionary link from the analysis of modulation indices (Sec.\,\ref{sec:resultmod}). 

\subsection{Drift direction and $P_2$ evolution}
\label{sec:resultp2}

The number of pulsars with positive  and negative drift detected is \driftingpositive\ and \driftingnegative, respectively. Here positive drift corresponds to subpulses drifting from the leading to the trailing side of the pulse profile. So there is no preference for a drift direction in the pulsar population, thereby confirming the result of e.g. \citet{WES2006}. From a MCMC analysis no significant dependence on $P$ and \pdot\ is found. In the carousel model, the observed drift direction depends on whether there is an inner or outer line of sight cut through the emission beam. It can therefore be expected to be uncorrelated with the spin parameters. 

\begin{figure}
    \centering
    \includegraphics[width=\linewidth]{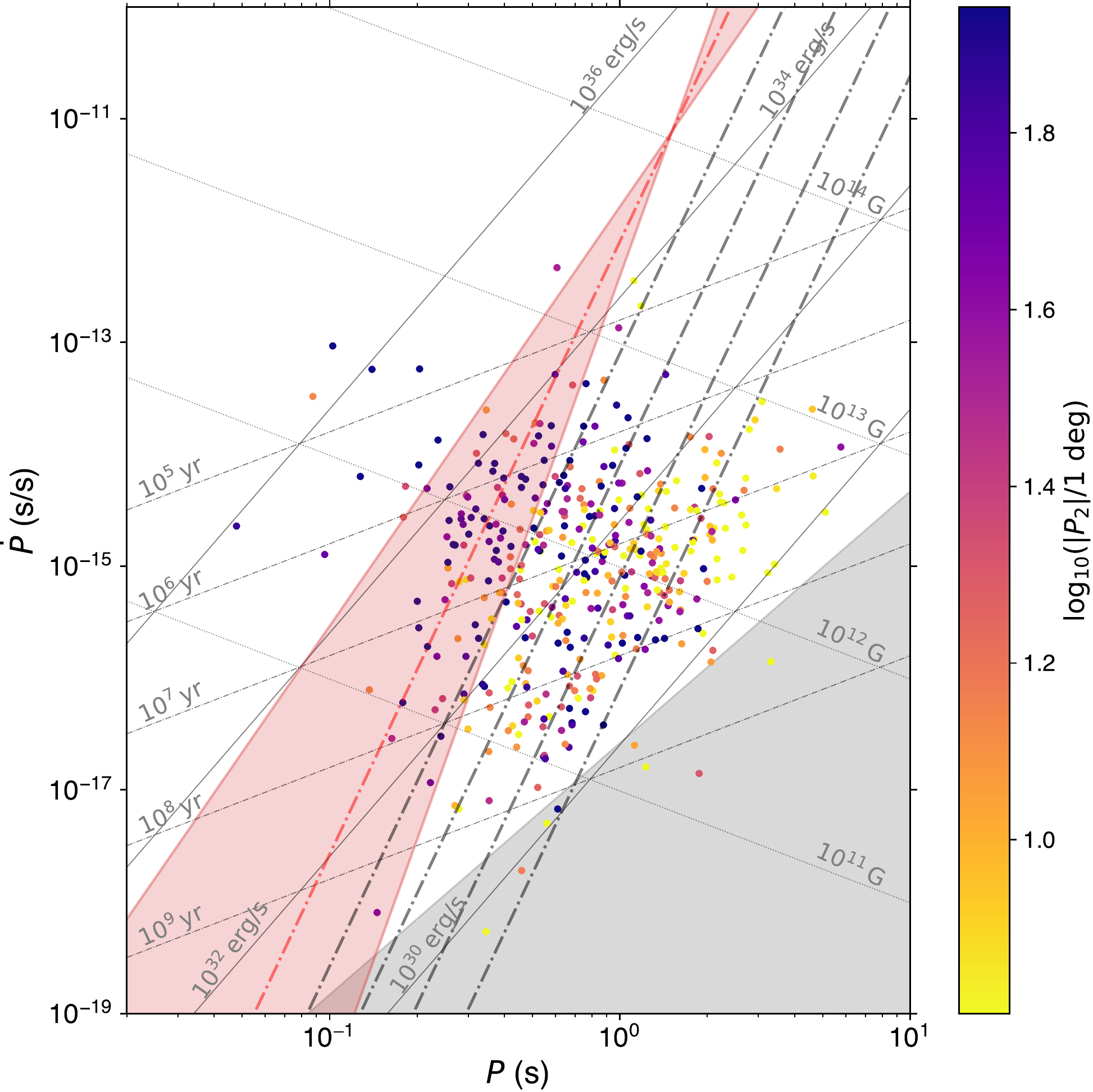}
    \caption{$P$-$\dot{P}$ diagram where the colour of the dots represent $|P_2|$ for each pulsar with a dominant drifting subpulse feature identified. The dot-dashed lines correspond to constant $|P_2|$ values according to a fit of Eq.\,\ref{eq:mcmcfit} (with $b=1$ and $c=0$). The top shaded region indicates the $1\sigma$ uncertainty on the slope. See the caption of Fig.~\ref{fig:parappdotp3} for further details.}
    \label{fig:linearppdotp2}
\end{figure}

Besides $P_3$, $P_2$ gives additional information about the appearance of the drifting subpulse patterns.  Fig.\,\ref{fig:linearppdotp2} shows the distribution of $|P_2|$ for those pulsars with detected drifting subpulses.
The population of pulsars with small $|P_2|$ values (lighter colour) is skewed towards the less energetic part of the population. Large $|P_2|$ are preferentially found for the higher \edot\ pulsars, and also for the less energetic pulsars. Although the correlation is strong (see Table~\ref{tab:relation}), the direction of evolution is relatively unconstrained. The Bayesian fitting (see Sec.~\ref{sec:mcmc}) assumes a symmetric error.  Since $P_2$ has asymmetric errorbars, the errorbar towards zero is chosen (i.e.\ the positive/negative errorbar is chosen for positive/negative $P_2$ value). This is usually the larger errorbar of the two, hence is the conservative choice.

In the geometric picture of a carousel producing the drifting subpulses, the larger $|P_2|$ for more energetic pulsars can be associated with a reduction of the number of beamlets in the carousel. The fact that less energetic pulsars tend to have small measured $|P_2|$ could result in spectral features in the 2DFS which are more clearly separated from the vertical axis. This will help to make drifting subpulses easier to detect for less energetic pulsars. 
An alternative reason for the larger $|P_2|$ values observed for more energetic pulsars would be additional disorganised subpulse modulation (with associated power along the vertical axis of the 2DFS) skewing the measured centroid position of the spectral features toward larger $|P_2|$.

It is suggested that the carousel should be rotating faster for more energetic pulsars to explain the $P_3$ evolution in the $P$-$\dot{P}$ diagram (Sec.\,\ref{sec:resultp3}), ignoring a possible evolution in the number of beamlets. Since $P_3$ quantifies the time for a subbeam to rotate over a subbeam separation, a reduction in the number of beamlets (increased separation) for energetic pulsars implies that an even faster carousel is required to compensate and explain the observed $P_3$ values. 

For fast rotating carousels, the beamlets could rotate appreciably during a time $P_2$, affecting the measured $P_2$ \citep{Gupta2004}. Depending on which hemisphere of the carousel is observed, its rotation is either with or against the rotation of the star, with opposite effects on the apparent $P_2$. Therefore, averaged over the population, this effect on $P_2$ is assumed to be cancelled out.
In addition, an evolution of the magnetic inclination angle $\alpha$ could affect the measured $P_2$, but it will be argued in Sec.\,\ref{sec:p2evo} that it would strengthen the statement that more energetic pulsars have faster carousels. In addition, it will be shown that the effect of pulse width evolution has no effect on this conclusion.

\subsection{Drift rate evolution}
\label{sec:resultdriftrate}

\begin{figure}
    \centering
    \includegraphics[width=\linewidth]{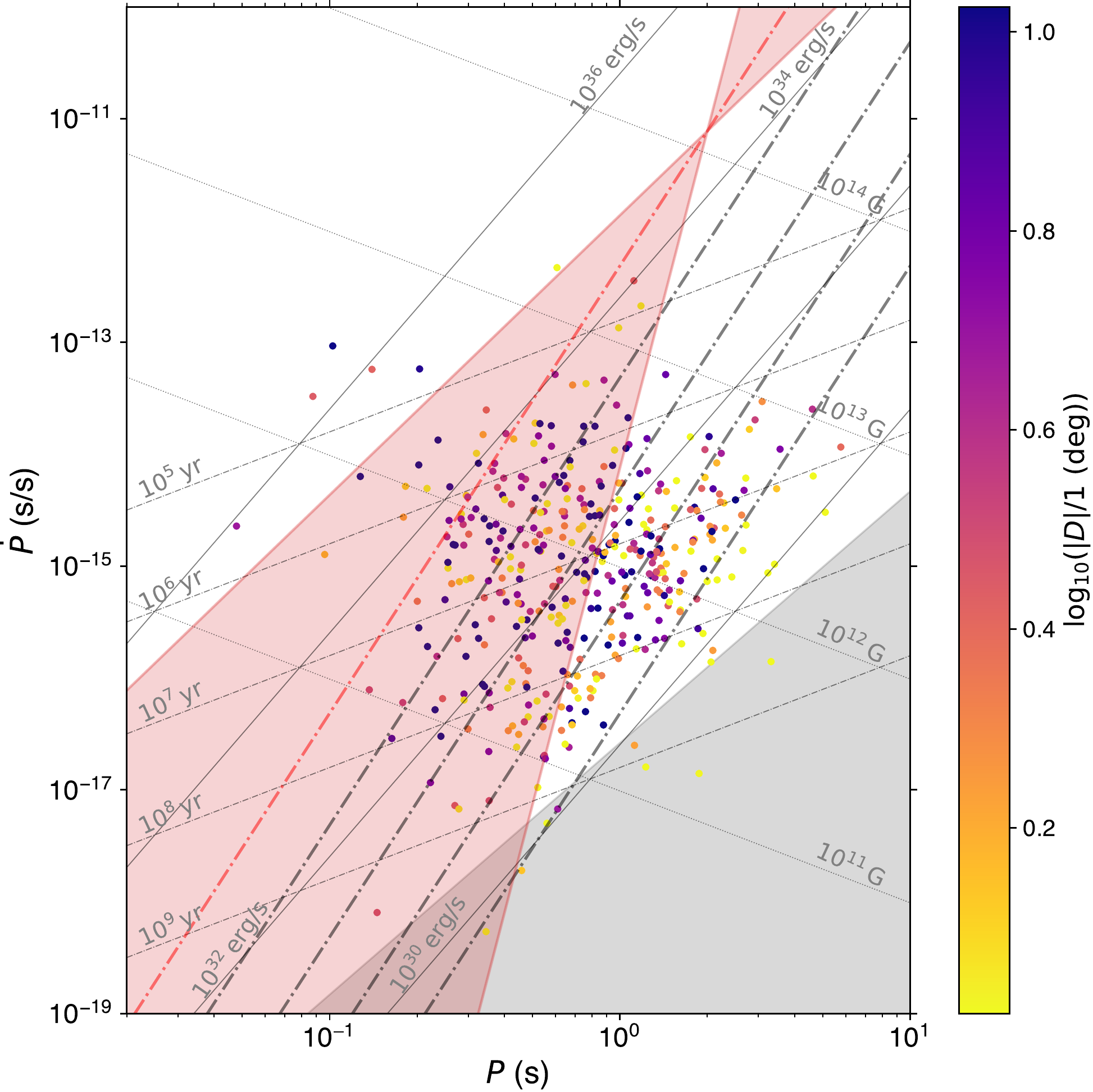}
    \caption{$P$-$\dot{P}$ diagram where the colour of the dots represent the drift rate for each pulsar with a dominant drifting subpulse feature identified. The dot-dashed lines correspond to constant $|D|$ values according to a fit of Eq.\,\ref{eq:mcmcfit} (with $a=1$ and $c=0$)}
    \label{fig:linearppdotdriftrate}
\end{figure}

The drift rate $|D|\equiv|P_2|/P_3$ does not provide independent information compared to what is discussed in Sec.~\ref{sec:resultp3} and \ref{sec:resultp2}, but it is a useful quantity as it relates to the gradient of the drift bands. The drift rate distribution shows a relatively weak correlation in the $P$-$\dot{P}$ diagram (Fig.\,\ref{fig:linearppdotdriftrate}) such that less energetic older pulsars typically have smaller $|D|$. A relatively modest correlation is found (see Table~\ref{tab:relation}), consistent with an absence of a correlation with $\dot{P}$.

The drift rate can be linked to how clear the drifting subpulse pattern is by comparing $P_2$ to the pulse width $W$. If a subpulse drifts over $W$ in $\sim$two stellar rotations or less ($2|D|\gtrsim W$), it is visible for such a short time that a stable pattern of multiple beamlets is required for the drifting subpulses to be easily recognisable in a pulse stack. The pattern can therefore be more susceptible to being distorted, making the subpulse pattern more difficult to recognise in a pulse stack. To quantify this, the drift rate is normalised by $W_{50}/2$ where  $W_{50}$ is the full width at half maximum of the pulse profile as determined by \citet{Posselt2021}. $W_{50}$ varies across the population, such that younger pulsars tend to have wider profiles (e.g.\,\citealt{Posselt2021}). 

\begin{figure}
    \centering
    \includegraphics[width=\linewidth]{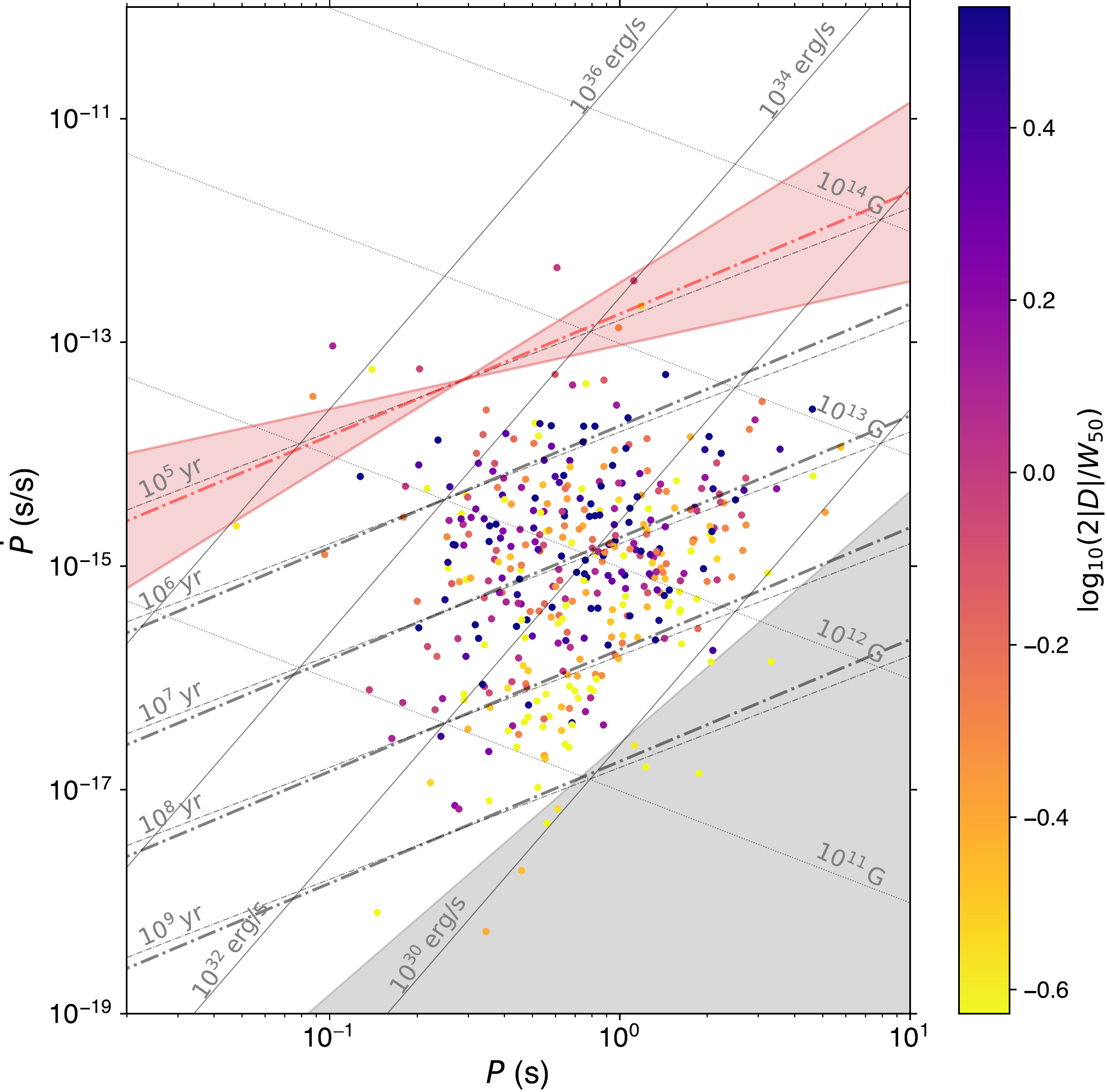}
    \caption{$P$-$\dot{P}$ diagram where the colour of the dots represent the drift rate normalised with $W_{50}/2$ for each pulsar with a dominant drifting subpulse feature identified. The dot-dashed lines correspond to constant $|D|$ values according to a fit of Eq.\,\ref{eq:mcmcfit} (with $b=1$ and $c=0$). The top shaded region indicates the $1\sigma$ uncertainty on the slope. See the caption of Fig.~\ref{fig:parappdotp3} for further details.}
    \label{fig:linearppdotdriftfrac}
\end{figure}

The normalised drift rate is shown in Fig.\,\ref{fig:linearppdotdriftfrac}, showing that a pulsar with a $\tau_c \gtrsim 5\times10^{7}$ yr ($\dot{E}\lesssim 5\times10^{32}$ erg/s) is more likely to have $|D|<W_{50}/2$. This corresponds well with the region in the $P$-$\dot{P}$ diagram for which drifting subpulses are strongest and most often detected (Sec.~\ref{sec:resultpasym}). Some energetic younger pulsars can have small drift rates, but many do not. The correlation is modest, as summarised in Table~\ref{tab:relation}.

\subsection{Modulation index evolution}
\label{sec:resultmod}

\begin{figure}
    \centering
    \includegraphics[width=\linewidth]{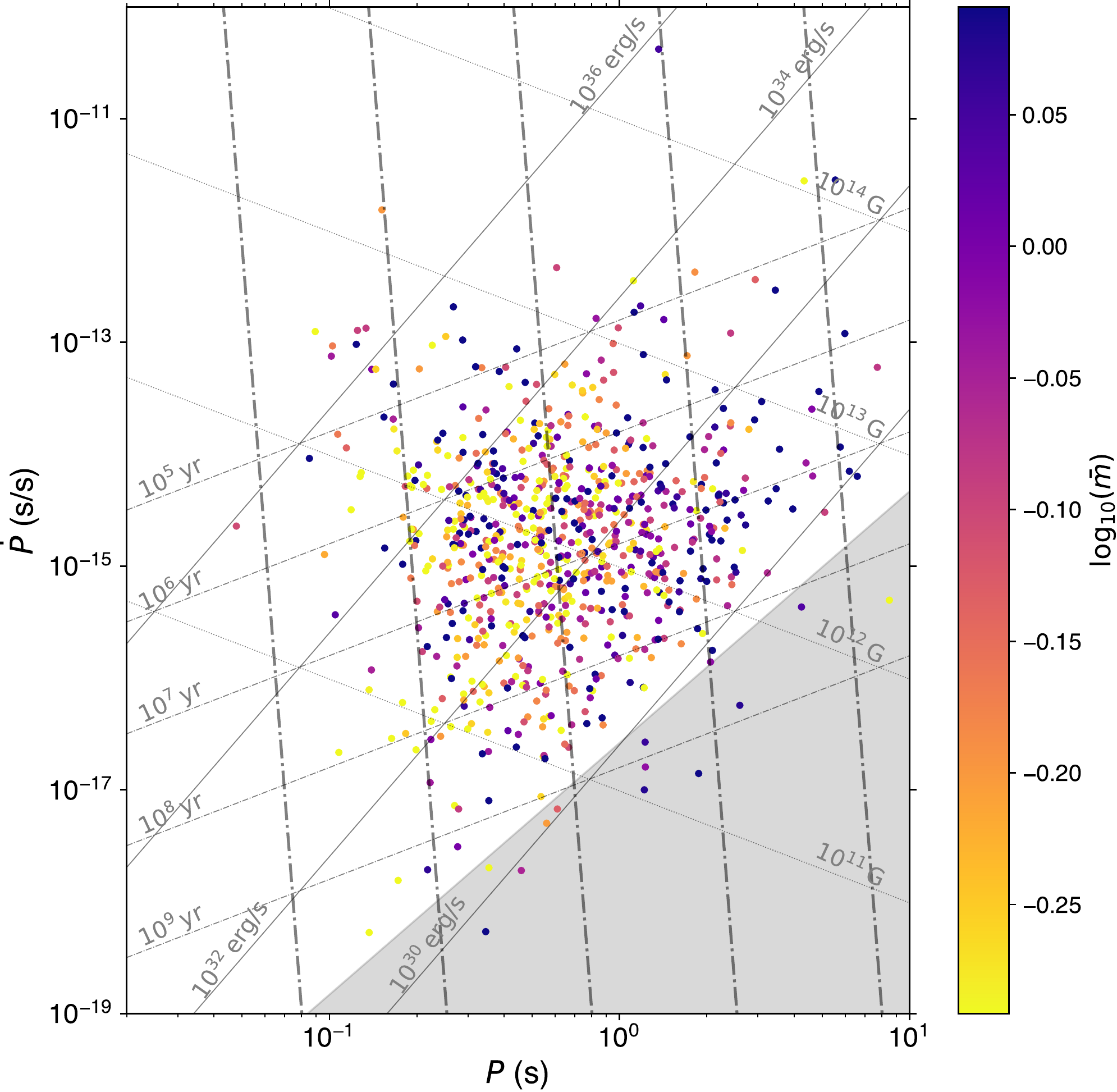}
    \caption{$P$-$\dot{P}$ diagram where the colour of the dots represent $\bar{m}$ of the MP for pulsars with a $\mathrm{S/N}> 100$. The dot-dashed lines correspond to constant $\bar{m}$ values according to a fit of Eq.\,\ref{eq:mcmcfit} (with $a=1$). See the caption of Fig.~\ref{fig:parappdotp3} for further details. Given the large errorbar, the direction of the correlation is not shown.}
    \label{fig:parappdotavmod}
\end{figure}

The average modulation index $\bar{m}$ quantifies the level of variability of single pulses (Eq.\,\ref{EqAvMod}). When an IP is detected, only the $\bar{m}$ corresponding to the MP is analysed here. Because $\bar{m}$ is biased such that it is lower at low S/N (see Sec.\,\ref{sec:method}), only observations with a $\mathrm{S/N}> 100$ are considered. In addition, pulsars with on-pulse regions that are so wide that no suitable off-pulse could be identified to subtract from the spectra are excluded as these $\bar{m}$ measurements will be biased high.

The $\bar{m}$ distribution in the $P$-$\dot{P}$ diagram (Fig.\,\ref{fig:parappdotavmod}) shows a skew such that pulsars with $P>1$ s have typically a large $\bar{m}$. The modest correlation is consistent with an absence of a correlation with $\dot{P}$. The significance of the quadratic term with $c=\AvModsnrparaan\pm\AvModsnrparaanerr$ in Eq.\,\ref{eq:mcmcfit} is less than $3\sigma$, but quantifies the flattening at small $P$ seen in the distribution in Fig.\,\ref{fig:histavmod}. The $\bar{m}$ evolution appears to be distinctly different in the $P$-$\dot{P}$ diagram compared to the detectability of drifting subpulses. 

Evidence of strong subpulse modulations in long period pulsars has been reported in the literature. For example, the radio magnetar XTE~J1810$-$197 (PSR~J1809$-$1943 in our sample) \citep{Serylak2009,Levin2019} and the recently discovered 75~s pulsar J0901$-$4046 \citep{Caleb2022} show clear and strong single pulse shape variations, resulting in large modulation indices. On the other hand, some long period pulsars and magnetars, for example the 23.5 s pulsar J0250+5854 \citep{Tan2018}, PSR J2144$-$3933 with a period of 8.5 s \citep{Young1999} and the magnetar J1622$-$4950 \citep{Levin2012} have modest modulation indices. Large modulation indices can also be found for younger short-period pulsars, especially when giant pulses are often observed (e.g.\,PSR~J1047$-$6709 has $\bar{m}=3.16\pm0.02$, caused by giant pulses as reported by \citealt{Sun2021}). So it is evident that there is a large scatter of $\bar{m}$ on the reported evolution in the $P$-$\dot P$ diagram, reflected in the modest correlation coefficient.

\citet{WES2006} found no evidence for the modulation indices being affected by whether the subpulses are organised or not (whether there are drifting subpulses or not), but marginal evidence for pulsars with coherent drifting subpulses to have a lower modulation index. We find both \pasym\ and \pstd\ to be uncorrelated with $\bar{m}$ (see online material \ref{ApendixModIndexEvolution}). 
Also no evidence is found for $\bar{m}$ being different for pulsars for which drifting subpulses or $P_3$-only features are detected (see online material \ref{ApendixModIndexEvolution}). 

A reason for drifting subpulses to be more difficult to detect for less energetic pulsars could be the appearance of additional strong stochastic modulation, which would imply a higher modulation index. However, the different evolution of $\bar{m}$ and the probability of detecting drifting subpulses in the $P$-$\dot{P}$ diagram, as well as the non-correlation between $\bar{m}$ and drifting subpulses related quantities, do not support this. Therefore, the main reason for drifting subpulses to be more prominent in some pulsars appears to be the stability of the produced subpulse patterns, rather than additional stochastic variability. 

\section{Discussion}
\label{sec:discussion}

In this section all the results related to subpulse modulation across the pulsar population are discussed. 
First this is done mathematically with minimum physical interpretation, 
progressing to a more explicit comparison to the carousel model.

\subsection{Drifting subpulse population}
\label{sec:overviewdis}

As shown in Sec.\,\ref{sec:fracdriters}, drifting subpulses are common and are detectable in about 60\% of the pulsars if sufficient S/N is available. This suggests that the physical conditions required for the drifting subpulse phenomenon cannot be distinctly different from those needed for the radio emission mechanism itself. However, some pulsars do not show drifting subpulses despite a high S/N observation being available, suggesting that the non-detection is intrinsic to the pulsar. This could be caused by highly irregular drifting subpulse patterns, potentially amplified by aliasing, generating erratic changes in apparent drift direction. Furthermore, disruption of the drifting subpulse pattern by nulling and/or mode changing might cause pulsars only to show drifting subpulses for a fraction of the time.

Some pulsars which do not show drifting subpulses do exhibit other modulations, such as amplitude modulation (a $P_3$-only feature). In at least some pulsars, periodic amplitude modulation may well be linked to the same physical phenomenon as drifting subpulses, with the line of sight determining whether this manifests itself as subpulse drift or not.

We found that the observed $P_3$ distribution for pulsars with drifting subpulses shows V-shaped evolution such that at around $\tau_c \simeq 10^{7.5}$ yr (or $\dot{E} \simeq 10^{31.5}$ erg/s), the smallest $P_3$ values are observed, and larger $P_3$ values are associated with both more and less energetic pulsars (Sec.~\ref{sec:resultp3}). The minimum corresponds to 
$P_3\simeq2$, the Nyquist period. This strongly suggests alias plays an important role. Aliasing is a sampling effect such that a beat between the rotation period and the intrinsic modulation period is observed. 
So a picture is emerging where the turning point in the $P_3$ distribution is interpreted as a change in alias order. The fact that at the high \edot\ end of the distribution broader spectral features are observed (Sec.~\ref{sec:resultp3std}), and the drifting subpulse features are less strong (Sec.~\ref{sec:resultpasym}) suggests a transition from high \edot\ aliased modulation to unaliased lower \edot\ pulsars.
This is further explored in Sec.~\ref{sec:p3sim}. 

\citet{Basu2019} report an anti-correlation between $P_3$ and \edot\ for pulsars with drifting subpulses, which they associate with low \edot\ pulsars ($\dot{E}<10^{32.5}$ erg/s). This anti-correlation is confirmed by the subset of our sample corresponding to a similar low \edot\ range below the turning point in the V-shaped evolution. However, our analysis of a larger sample suggests that the correlation is stronger in a $P$ and \pdot\ combination closer to \age\ than \edot. Unlike for drifting subpulses, \citet{Basu2020} found periodic subpulse modulation in the form of amplitude modulation (modulation without evidence for phase drift) to occur throughout the full \edot\ range such that for $\dot{E}>10^{32.5}$ erg/s it is the only form of periodic subpulse modulation.

They report that amplitude modulation typically has a $P_3 \sim 10 - 200$, larger than those associated with drifting subpulses and uncorrelated with \edot.
We also concluded that $P_3$-only pulsars are seen in the full \edot\ range, and the mean $P_3$ for these pulsars is larger than that of drifting subpulses, and that no clear evolution of their $P_3$ values is seen across the $P$-\pdot\ diagram (Sec.~\ref{sec:resultp3}).

\citet{Basu2020} suggest that amplitude modulation forms a distinct class, with a different physical origin compared to drifting subpulses. 
It can be noted that other pulsar properties also show a transition at a comparable \edot, such as profile morphology (e.g \citealt{Rankin2022}) and linear polarisation (e.g. \citealt{Weltevrede2008,Posselt2022}). However, our sample includes pulsars above $\dot{E} > 10^{32.5}$ erg/s for which clear drifting subpulses are identified (examples include PSRs J1453$-$6413, J1645$-$0317, and J1918+1444). This suggests that there is no sharp transition between pulsars with drifting subpulses and those with amplitude modulations. This conclusion is further strengthened by the fact that the evolution of the drift rate is gradual, such that high \edot\ pulsars show less strong pulse phase modulation (see Fig.\,\ref{fig:linearppdotdriftrate}). 
The larger drift rate at high \edot\ could be interpreted to mean either that there is more additional erratic emission biasing the $P_2$ measurements, or an actual evolution in drift rate. In any case, this strongly suggests that many pulsars in the $P_3$-only class will have weak subpulse phase modulation that is hard to detect. 
However, this does not explain the fact that $P_3$-only pulsars have typically larger $P_3$ values, suggesting that for some $P_3$-only pulsars the modulation is governed by different physics. For example, some of the $P_3$-only features may be caused by periodic nulling.

\subsection{Modelling the $P_3$ evolution}
\label{sec:p3sim}

\begin{figure}
    \centering
    \includegraphics[width=8cm]{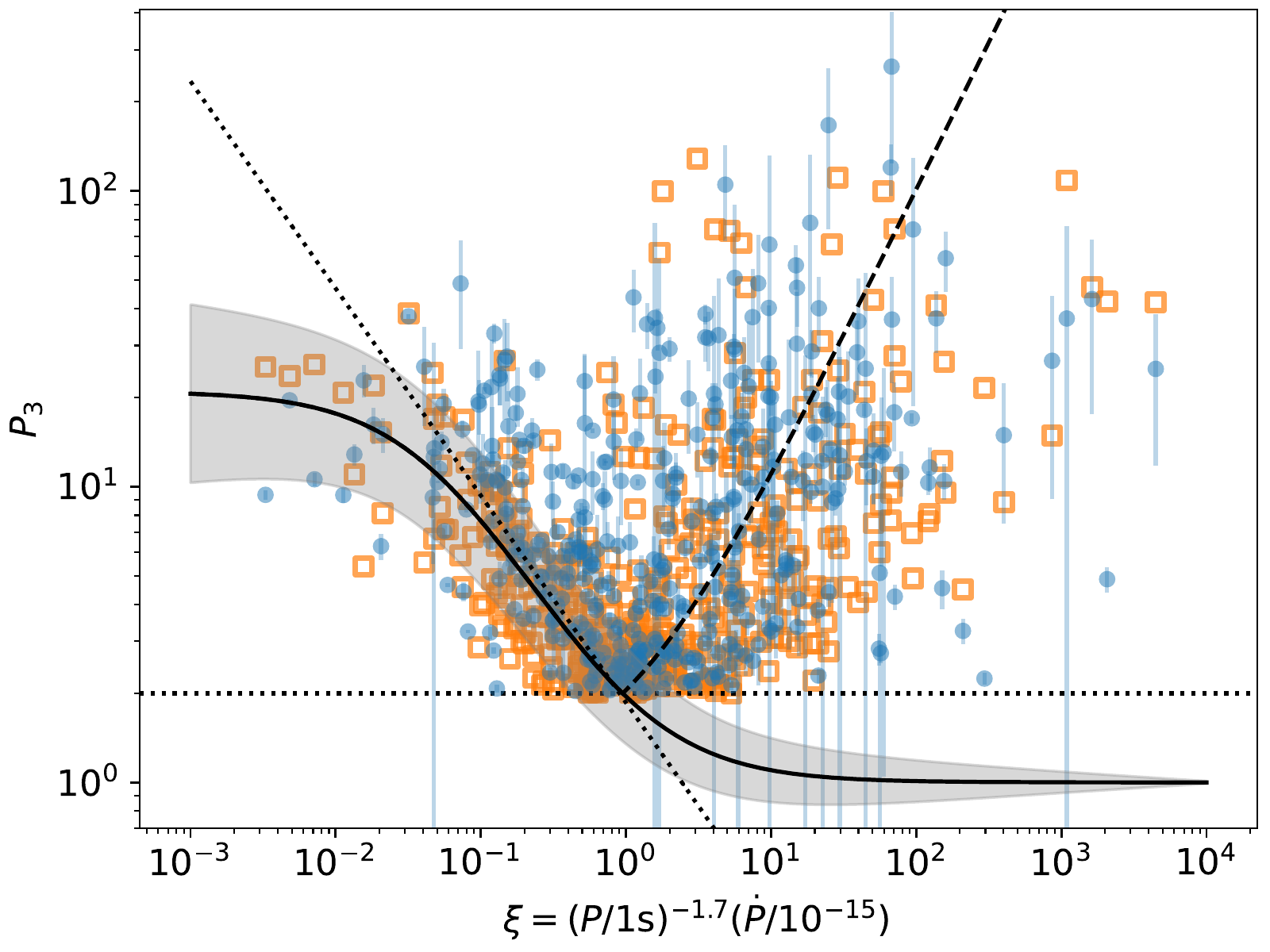}
    \caption{A comparison between the observed $P_3$ values (blue points) with a simulation (orange open squares) as function of $\xi$ (the $P$-$\dot P$ combination for which $P_3$ is observed to evolve strongest). The simulation assumes that the intrinsic $P_3$ decreases with $\xi$ (solid line). When the intrinsic $P_3<2$ (horizontal dotted line), the observed $P_3$ will be aliased. The dashed line (Eq.~\ref{eq:aliasedModel}) is the aliased version of the solid line. Intrinsic $P_3$ values are drawn at random from the model, where the grey shaded region indicates the width of the distribution. The intrinsic $P_3$ values are transformed to the observed $P_3$ after taking alias into account (Eq.~\ref{eq:aliasformula}).
    The diagonal dotted line represents a power law relationship with an exponent equal to $-0.7$.}
    \label{fig:p3_xi_new}
\end{figure}

In Sec.~\ref{sec:overviewdis} it was demonstrated that the distribution of the observed $P_3$ is V-shaped. We consider the direction of evolution of $P_3$ in the $P$-$\dot{P}$ diagram as found in Sec. \ref{sec:resultp3}, i.e. $a/b=\Pverdriftparag$.
Hereafter, we use symbol \mbox{$\xi=(P/{\rm s})^{\Pverdriftparag}(\dot{P}/10^{-15})$} to describe the $P_3$ evolution, and in effect regard it as a proxy both for the characteristic and the true age of the pulsar.
The observed evolution is shown in Fig.~\ref{fig:p3_xi_new} with blue points, and is characterised by large $P_3$ values for low and high $\xi$, and the smallest $P_3$ values dominate the region with $\xi \sim 0.8$. 
The fact that the turning point of the distribution occurs at $P_3\sim 2$ strongly suggests a transition between (first-order) aliased and non-aliased subpulse modulation. Furthermore, in this interpretation, the non-aliased drift in older pulsars is suggested by drifting subpulses being stronger and more stable (larger {\pasym}, and smaller {\pstd}) compared with those of younger pulsars. The purpose of this section is to demonstrate that it is possible to find a monotonic relationship between $P_3^\mathrm{int}$, the intrinsic $P_3$, and $\xi$ which can plausibly reproduce both the unaliased and aliased observed values of $P_3$. This will be done without considering a specific physical origin for the drifting subpulses.

To test if a model where the intrinsic $P_3$ monotonically decreases with $\xi$ can reproduce the data, a synthetic sample of pulsars is generated by considering a description of $P_3^\mathrm{int}(\xi)$. When $P_3^\mathrm{int} < 2$, alias occurs and we take its lowest order aliased value (see, e.g., \citealt{ES2002})
\begin{align}
\label{eq:aliasformula}
\frac{1}{P_3} = \left|\frac{1}{P_3^\mathrm{int}} - 1\right|.
\end{align}
Note that this approach is independent of any physical model, as it is a mathematical description of alias and it relies only on the fact that the emission is observed once per stellar rotation.

We first considered $P_3^\mathrm{int}(\xi)$ to follow the diagonal dotted line in Fig.~\ref{fig:p3_xi_new}. By design, this would describe the data in the low-$\xi$ (typically older pulsars) region well.
At high $\xi$ (younger pulsars), high alias orders (not included in eq.~\ref{eq:aliasformula}) are implied by the $1/P_3^\mathrm{int}\gg 0.5$ (very fast modulation) reached. As a consequence, the model fails to reproduce the systematic increase of the observed $P_3$ for $\xi \gtrsim 0.8$ because a near uniform distribution in apparent $P_3$ would be expected. To avoid this, we insist on a model where the alias order remains small, such that either an unaliased $P_3$ is observed, or the lowest possible order alias described by Eq.~\ref{eq:aliasformula}. This requires a relation for $P_3^\mathrm{int}(\xi)$ which flattens out at large $\xi$ compared to the diagonal dotted line in Fig.~\ref{fig:p3_xi_new}.

The dashed line in Fig.~\ref{fig:p3_xi_new} represents the aliased $P_3$ observed at large $\xi$. This line reproduces the observed increase of $P_3$ with $\xi$, and we define this line as 
\begin{equation}
\label{eq:aliasedModel}
P_3(\xi)= (\xi-q)^k/w.
\end{equation}
The unaliased $P_3^\mathrm{int}$ (obtained with Eq.~\ref{eq:aliasformula}), extended over the full $\xi$ range, is represented as the solid line.
The specific choice of $q=-1.05$, $w=1$, $k=1$ is used. These parameters ensure that the Nyquist point ($\xi=q+2w$) is close to 1, and that $P_3(\xi)$ flattens
at low $\xi$ ($q+w < 0$). Other functional forms may fit the data equally well, but it is important to realise that, however contrived the relation between the dashed and solid curves appear in a $P_3$ plot, the chosen forms are simply related via Eq.~\ref{eq:aliasformula}.

Overall the solid line in Fig.~\ref{fig:p3_xi_new} represents a monotonic gradual increase of the intrinsic modulation period as the pulsar ages (evolves towards lower $\xi$). If interpreted as a carousel of beamlets, this would be an evolution from being initially fast circulation relative to the neutron star surface to near stationary as the pulsar `dies'. Young pulsars are then observed to have a large single-aliased $P_3$ given by the dashed line. 
It should also be noted that since the dashed and solid lines are aliases of each other, a monotonic increasing and decreasing $P_3^\mathrm{int}(\xi)$ could explain the observed $P_3$ evolution equally well, although we have argued in Sec.~\ref{sec:overviewdis} that an increasing $P_3^\mathrm{int}$ with \age\ is preferred by the observations.

The model for $P_3$ (solid line for small $\xi$, dashed line for large $\xi$) fits the observed distribution well (blue points). Fig.~\ref{fig:p3_xi_new} also shows the consequence of allowing a moderate degree of scatter (grey shaded region) around the underlying model for $P_3^\mathrm{int}$ (solid line). The synthetically generated orange squares are drawn from this distribution, and transformed to the observed $P_3$. The distribution of orange squares matches that of the blue points well, including the observed scatter, thereby demonstrating that the observed non-monotonic evolution of $P_3$ can be reproduced with a monotonic evolution of $P_3^\mathrm{int}$.

In the model it is assumed that the fractional uncertainty in $P_3^\mathrm{int}$ decreases from $0.5$ at $\xi=10^{-3}$ to $0.01$ at $\xi=10^{4}$. This indicates that the increased scatter in observed $P_3$ values for young (high $\xi$) pulsars is a consequence of aliasing boosting the scatter in $P_3^\mathrm{int}$, while according to the model $P_3^\mathrm{int}$ is better defined for these pulsars. This boost of variability in $P_3^\mathrm{int}$ is argued in Sec.~\ref{sec:resultp3std} to be responsible for the evolution in spectral width as well. A reason for an increased scatter in $P_3^\mathrm{int}$ for older pulsars is the observation that drift mode changes are more likely for these pulsars (see Sec.~\ref{sec:resultp3}).

\subsection{$P_2$ evolution}
\label{sec:p2evo}

So far, the results have been discussed independent of a specific model for drifting subpulses. Here, some geometric effects are discussed in the context of the rotating carousel model where the drifting subpulses are produced by a number of subbeams rotating around the magnetic axis. 

It was found that pulsars with a higher \edot\ tend to have larger $P_2$ (Sec.\,\ref{sec:resultp2}). Two possible interpretations were suggested. It could relate to an increased stochastic subpulse modulation, which skews the measured $P_2$. Alternatively, it can geometrically be interpreted as a reduction of the number of beamlets in the carousel. Here we discuss implications of the latter interpretation. But additional geometric effects related to this interpretation will be considered first.

The evolution in pulse width $W$ is expected to contribute to the $P_2$ evolution. A larger $W$ means the subbeams can be expected to be further separated in pulse longitude, and for typical geometries $P_2\propto W$. Pulsars with larger \edot\ tend to have larger $W$ (e.g.\,\citealt{Johnston2019,Posselt2021}). Therefore, from this effect, one would expect that $P_2$ is larger for high \edot\ pulsars. One factor affecting the $W$ evolution is the evolution of the magnetic inclination angle $\alpha$, which is aligning over the lifetime of a pulsar (e.g.\,\citealt{Lyne1988,Tauris1998,Weltevrede2008,Johnston2017}). However, alignment would predict $W$ to typically be larger for older pulsars, opposite to what is observed (e.g.\,\citealt{Johnston2019,Posselt2021}). This suggests that other factors (especially the shrinking of the polar cap over time) dominate the observed increase of $W$ towards high \edot\ pulsars. Nevertheless, to test if our conclusions related to $P_2$ evolution are affected by $W$ evolution, the evolution of $P_2/W_{50}$ is assessed (see Appendix~\ref{app:ppdothist}). This shows that also $P_2/W_{50}$ tends to be higher for high \edot\ pulsars, hence the conclusion that they tend to have a smaller number of subbeams remains valid.

$P_2$ is related to the so called complexity parameter (e.g.\,\citealp{Gil2000}), defined as $C=2r_p/\Delta$, where $2r_p$ is 
the polar cap diameter and $\Delta$ the typical separation between beamlets. 
The pulse width relates to $2r_p$ for a central line of sight cut. Therefore, for a packed polar cap with subbeams, $C$ represents the number of subbeams crossed by the line of sight, hence relates to the expected complexity in the pulse structure. Since $\Delta$ would correspond to $P_2$ in pulse longitude, $C$ can also be expressed as $W/P_2$ for a central cut. More generally one can expect $C\propto (P_2/W)^{-1}$. As discussed, pulsars with a larger \edot\ tend to have a larger $P_2/W$, hence $C$ can be expected to be small.

Observationally, $C$ has been linked to modulation indices, which measure the amount of variability of single pulses. \cite{Jenet2003} argue that a high $C$ implies overlapping subpulses, hence a reduced modulation index. Therefore, $\bar{m}$ and $C$ are thought to be anti-correlated. 
In Sec.\,\ref{sec:resultmod} it was found that large $P$ pulsars typically have a larger $\bar{m}$. This should correspond to a small $C$, hence a large $P_2/W$. However, large $P_2/W$ are found to associate to larger \edot\ pulsars which tend to have small $P$. This suggests that the evolution of $\bar{m}$ is governed by more than $C$, such as an evolution in the variability of the individual subbeams.

Specific physical models for subpulse generation can predict the scaling of $C$ with $P$ and {\pdot} \citep{Jenet2003}. These include the \citet{RS1975} sparking gap model (see also \citealt{Gil2000}), and three other models, which relate to continuous current outflow instabilities \citep{Arons1979,Hibschman2001}, surface magnetohydrodynamic wave instabilities \citep{Lou2001}, and outer magnetospheric instabilities \citep{Jenet2003}. However, none of the $P$ and \pdot\ dependence of these four predictions for $C$ (see \citealt{Jenet2003}) correlate with the found evolution of $\bar{m}$ with $P$ and {\pdot}. A linear fit of Eq.\,\ref{eq:mcmcfit} reveals that $b/a=\AvModsnrlng\pm\AvModsnrlngerr$, inconsistent with the predicted $b/a$ of $-4/9$, $-1/3$, 1 and $-1/5$ respectively.
This again might suggest that the physics governing $C$ is not directly related to $\bar{m}$, rather than that the models for $C$ are incorrect. 

\subsection{Comparison with theoretical models}
\label{sec:comptheo}

\begin{figure}
    \centering
    \includegraphics[width=0.9\linewidth]{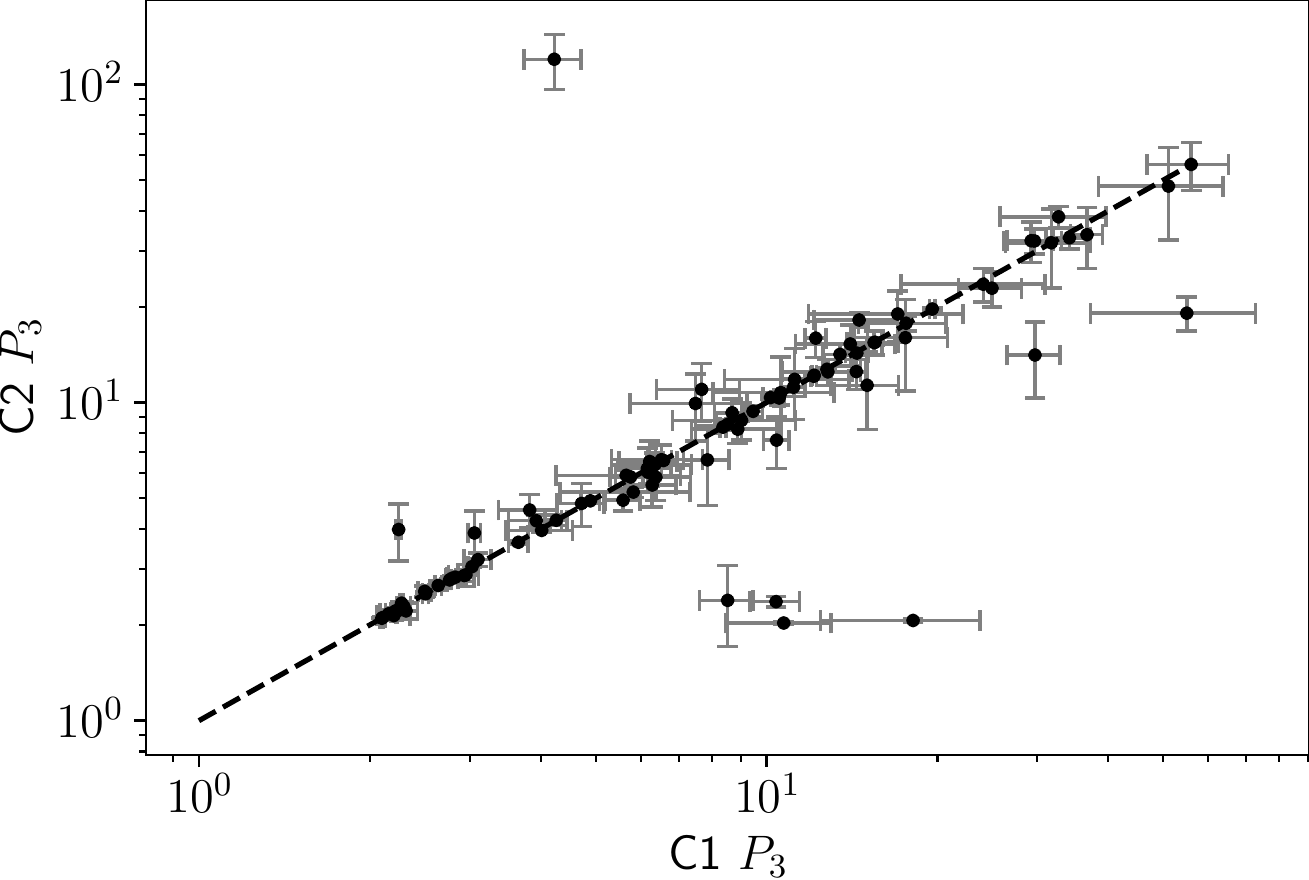}
    \caption{Comparison of $P_3$ values as measured in two profile components of the same pulsar. For clarity, only relatively well defined spectral features with $P_3<3\sigma_{1/P_3}$ are shown. The dashed line corresponds to equal $P_3$ values for the two components.}
    \label{fig:p3comp}
\end{figure}

We demonstrated in Sec.~\ref{sec:p3sim} that the origin for the apparent V-shaped $P_3$ evolution is suggestive of aliasing. The model employed (Eq. \ref{eq:aliasedModel})  is purely mathematical, without imposing specific physics to the drifting subpulse phenomenon. Here we discuss the results in light of the geometric and physical notions of emission carousels and $\mathbf{E}\times\mathbf{B}$ drift, concepts first introduced by \cite{RS1975} (hereafter RS) and since developed by many authors (e.g.\,\citealt{Gil2003alias}).

A prediction of the carousel model is that the measured $P_3$ is determined by the circulation time of the carousel and the number of subbeams within the polar cap. As a consequence, in general different profile components should have identical $P_3$ values. Fig.\,\ref{fig:p3comp} shows a scatter plot for pulsars for which relatively well defined $P_3$ values are measured in separate pulse profile components. A tight correlation is observed, with most pulsars having consistent $P_3$ values for different components. Some exceptions exist (PSRs J1224$-$6407, J1345$-$6115, J1427$-$4158, J1430$-$6623 and J1722$-$3207 in the figure). Exceptions are to be expected in the presence of drift mode changes (see also Sec.~\ref{sec:resultp3}). In addition, so-called core components can have separate modulations.
The correlation therefore powerfully demonstrates that drifting subpulse patterns in separate profile components are linked to a single physical origin, as is expected in a carousel.

Accepting that carousels are present on or just above polar caps (whose minimum size is defined by the last open fieldline touching the light cylinder), it is commonly assumed that their drift over the surface, so called $\mathbf{E}\times\mathbf{B}$ drift, directly reflects a particle flow and emission on the open fieldlines. 

Although exceptional circumstances may apply in specific pulsars, the general nature of $\mathbf{E}\times\mathbf{B}$ drift is to be weak (i.e.\,near-corotation) in young fast pulsars and speed up rapidly as a pulsar ages and spins down (being dependent on the acceleration parameter $\sim B/P^2$). This is true both in its original formulation in RS and later modifications \citep{Gil2003model}. Such behaviour is exactly the opposite of what our observations suggest. 

It can be pointed out that we do not directly observe the circulation time, but $P_3$, which depends on the number of sparks in the carousel. However, to reproduce the observed $P_3$ evolution, the spark number would have to increase dramatically when applied to younger pulsars -- and
again our measurements show that if the number of sparks changes, it is in exactly the opposite sense (Sec.~\ref{sec:p2evo}). Thus $\mathbf{E}\times\mathbf{B}$ physics cannot plausibly yield the solid line of Fig.~\ref{fig:p3_xi_new}, which we argue represents the intrinsic subpulse drift.
 
We therefore face a paradox. Drift explained by the acceleration parameter would result in values of the circulation time (and hence of $P_3$ for any likely number of sparks) which decreases as the pulsar ages from right to left in Fig.~\ref{fig:p3_xi_new}. In fact such measured values do appear in the figure, but we have interpreted them as lowest order aliased with their true values close to $P_3\sim 1$.
 
One way out would be to suppose that the left and right halves of Fig.~\ref{fig:p3_xi_new} are governed by different physics \citep{Basu2016,Basu2020}, an assumption weakened by our results  (Sec.~\ref{sec:overviewdis}).  In any case, such a model would mean that by coincidence the transition at $P_3\sim2$ has both a physical significance and plays its role as an alias boundary dependent on our sampling rate $P$. An important feature of our results is the continuity of the drift phenomenon throughout pulsar evolution.

Alternatively we can abandon the assumption that the solid black curve in Fig.~\ref{fig:p3_xi_new} represents direct observation of $\mathbf{E}\times\mathbf{B}$-driven drift, while still insisting that it describes observations of real spark motion. However, this line is mathematically related to the dashed alias line -- and it is the dashed line which is approximately consistent with the magnitude and variation of RS-predicted $\mathbf{E}\times\mathbf{B}$ drift. As stressed above, each line is the alias of the other. The $P_3$ system we observe can therefore be seen as our periodic sampling of the beat between an internal $\mathbf{E}\times\mathbf{B}$ system within the pulsar and the pulsar period itself (see \citealt{Wright2022}).

One advantage of this interpretation is that the apparent drift reversals found in some older pulsars (e.g.\,J1750$-$3035, \citealp{Szary2022}) can naturally be understood as a slight variation in $\mathbf{E}\times\mathbf{B}$ without appeal to observational aliasing, and the pulsar's $\mathbf{E}\times\mathbf{B}$ does not change sign. Sudden changes in drift rates can also be seen as loss or gain of a subbeam in a near-fixed $\mathbf{E}\times\mathbf{B}$ environment. These and related points are further discussed in \citet{Wright2022}.     

A particularly interesting feature of this approach (to be pursued in more detail elsewhere) is the formulation of the expected observable $P_3$ represented by the dashed line in Fig.~\ref{fig:p3_xi_new}. This now can be seen as representing the underlying $\mathbf{E}\times\mathbf{B}$ system of the pulsar which our periodic sampling reveals.
It is now possible to apply the acceleration parameter to pulsars on this line and estimate the spacing and number of polar cap sparks within the pulsars' beat system. If the polar cap has a radius $r_p\propto s/P^{1/2}$ so that the $N$ sparks have a separation $\Delta\approx 2\pi r_p/N$ then it can be shown that 
\begin{equation}
\label{eq:GeoffPrediction}
P_3 \simeq s\left(\frac{\Delta}{95\;\mathrm{m}}\right)\left(\frac{P\dot E}{3\times10^{31}\;\mathrm{erg}}\right)^{\frac{1}{2}},
\end{equation}
where $P_3$ represents the values on the dashed alias line (see equation (9) of \citealt{Wright2022}), and $s$ is the ratio of the actual $r_p$ to that defined by the last closed field line touching the light cylinder. Divergences from the line $P_3\propto({P\dot E})^{\frac{1}{2}}$ are a measure of the product $s\times \Delta$. The significance here is that the dependence of $P_3$ on $P\dot E$ in Eq.~\ref{eq:GeoffPrediction} is derived on a purely theoretical basis and implies $a/b= -2$, a near-coincidence with the observationally-derived result of Sec.~\ref{sec:resultp3} ($a/b=\Pverdriftparag\pm\Pverdriftparagerr$), and would imply a value $k=1/2$ in Eq.~\ref{eq:aliasedModel}. Application of this equation both to pulsars individually and collectively could give realistic estimates of $s\times \Delta$ to be compared with profile and polarisation data.  

\section{Conclusions}

A catalogue of subpulse modulation properties at $1280$~MHz is presented, reporting on the findings of a systematic analysis of the subpulse modulations for \censuspsr\ pulsars in the TPA single-pulse legacy data. The wealth of single-pulse data, with 1.6 million pulses analysed and
covering a significant fraction of the known pulsars, allows a detailed investigation into the evolution of various drifting subpulse related quantities across the pulsar population. By design, the obtained dataset is suitable with sufficient quality and length to identify drifting subpulses in many sources. A semi-automated process and new techniques are developed, with shuffle-normalised spectra aiding identification of weak (possibly broad) spectral features and the assessment of their significance. In the course of the analysis, we corrected the periods and spin-down rates for 12 pulsars, for which harmonically related values were reported in the literature.

The strength of subpulse modulation across the population of pulsars is quantified with a flux density weighted and pulse-longitude averaged modulation index $\bar{m}$. Our results reveal that $\bar{m}$ evolves most strongly with $P$ such that large $P$ pulsars tend to have larger $\bar{m}$. 
Although $\bar{m}$ evolution has been linked to the evolution of a `complexity parameter' in the past, our observations suggest that it is not the dominant relevant physical parameter. Instead, the intrinsic variability of subbeams might play a more important role.

Although the focus of this paper is drifting subpulses, we also identify a class of pulsars where amplitude modulations with a period $P_3$ but without subpulse phase modulations (so-called $P_3$-only pulsars). These are found to be distributed roughly uniformly throughout the pulsar population. Some modulations of this nature may be generated by nulling or mode changing. However, there is no evidence of a sharp transition between such pulsars and those with drifting subpulses. This therefore strongly suggests that many $P_3$-only  pulsars have weak drift that is hard to detect.

Drifting subpulses are detected across a wide range of the $P$-\pdot\ space. For \driftfrac\ of pulsars in our sample ({\driftnum} in total) drifting subpulses are identified. 
This implies that 60\% of the overall population of pulsars would be found to have detectable drifting subpulses if data with sufficient S/N were available. Unlike the $P_3$-only pulsars, drifting subpulses are more likely to be detected for the older and less energetic pulsar population, and their $P_3$ values are typically smaller. 

The modulation periods $P_3$ for pulsars with drifting subpulses are found to follow a V-shaped evolution in the $P$-\pdot\ diagram such that at a characteristic age $\tau_c \simeq 10^{7.5}$ yr the smallest $P_3$ values are observed. Both older and younger pulsars typically have larger $P_3$, with the minimum occurring at the Nyquist period $P_3\simeq2$. This strongly points to alias being a fundamental ingredient required for the understanding of drifting subpulse evolution. The V-shaped evolution of $P_3$ can be reproduced, in a way which is independent of a specific physical model, by a monotonically evolving intrinsic underlying $P_3^\mathrm{int}$. In this interpretation, pulsars evolve from a fast $P_3^\mathrm{int}$ with observed aliased $P_3>P_3^\mathrm{int}$ (typically of the lowest order for pulsars with small \age) to unaliased $P_3=P_3^\mathrm{int}$ (for pulsars with large \age).

This mathematical description of drifting subpulses offers an explanation for various observed trends. The small \age\ (large \edot) pulsars are naturally expected to have more erratic subpulse modulation patterns because the alias effect will enhance the observed effect of any irregularities in the underlying subpulse modulation. So this interpretation not only describes the observed $P_3$ evolution, but also explains why drifting subpulses are more often detected in large \age\ pulsars. Furthermore, for small \age\ pulsars with drifting subpulses, it explains their broader (less well defined $P_3$) and weaker (less systematic drift) spectral features, and their more erratic drifting subpulse patterns may generate bias in the observed evolution of $P_2$.

A natural geometric representation of drifting subpulses is to see them modelled as a carousel of discrete subbeams rotating around the magnetic axis. 
Since we find evidence that the number of subbeams does not decrease for larger \age, our interpretation of the observed $P_3$ implies that the carousel must slow down as it evolves with $P_3^\mathrm{int}$ increasing.

This creates an apparent contradiction to the conventional view that carousel rotation is driven by the $\mathbf{E}\times\mathbf{B}$ effect.
Measured by the original \citet{RS1975} formula the carousel drift should begin as near-corotation and gradually accelerate as a pulsar ages. Here we find exactly the opposite.
Later changes introduced by partial screening of the polar cap potential (e.g.\,\citealt{Gil2003model,Basu2016}) have successfully modified this view. However, this model in its present form cannot be extended to young pulsars.

To resolve this, we suggest that the observed subpulse modulation represents a sampling of a beat system, internal to the pulsar, between its $\mathbf{E}\times\mathbf{B}$ timescale and its rotation period (\citealt{Wright2022}). The polar carousel of subbeams are then seen as generated by time-delayed interaction between separate regions of the magnetosphere. Such a carousel would initially drift strongly relative to the neutron star and gradually slow as the pulsar ages, which is what our observations suggest.

\section*{Acknowledgements}

The MeerKAT telescope is operated by the South African Radio Astronomy Observatory, which is a facility of the National Research Foundation, an agency of the Department of Science and Innovation. Pulsar research at Jodrell Bank Centre for Astrophysics and Jodrell Bank Observatory is supported by a consolidated grant from the UK Science and Technology Facilities Council (STFC). XS acknowledges financial support from the President's Doctoral Scholar Award. GW thanks the University of Manchester for Visitor status. MeerTime data is housed and processed on the OzSTAR supercomputer at Swinburne University of Technology. 
MB acknowledges support from
ARC grant CE170100004.

\section*{Data Availability}

The data underlying this article will be shared on reasonable request
to the corresponding author. See also Appendix~\ref{app:online}.



\bibliographystyle{mnras}
\bibliography{TPAsubpulsemodulation} 




\appendix

\section{Online tables and figures}
\label{app:online}
The full Table~\ref{tab:p3s} and all original and shuffle-normalised spectra can be found here\footnote{\url{10.5281/zenodo.6900582}}. In the same place Table A.1 can be found, which contains additional information related to the observations. Both tables can also be found in the supplementary material associated with this publication.

\section{Spectral analysis}
\label{app:spectra}

The on-pulse regions are algorithmically defined as that part of the template where the amplitude exceeds 1 per-cent of the maximum amplitude. This either forms a single interval which is labelled as the MP, or a MP and an IP interval if the amplitude of the template is below 1 per-cent roughly in between the MP and IP. In this work, that part of the profile where the maximum amplitude of the template occurs is labelled as the MP. In some cases, the analysis benefits from splitting the profile into at most two components.
Unless the components are defined manually, the split is applied at the first minimum following the first (possibly local) maximum amplitude in the template
in the relevant on-pulse region. Further details about the MP/IP definition, and profile splitting can be found in the online material \ref{sec:AppendixManualOnpulse}. 

The FFT length used for the spectra is at most 512 pulses (or the largest power of two that fits in the length of the pulse sequence), and the minimum is limited to 64 pulses. The number of pulses that contributes to the LRFS is a multiple of the FFT length, so reducing the FFT length could lead to the inclusion of more pulses in the analysis. In addition, a lower spectral resolution is preferred if the S/N is relatively low. The FFT length used for the calculation of the longitude resolved modulation index is always 16 pulses since spectral resolution is not required, and short FFT lengths are preferred as it ensures most pulses can be used in the analysis. See online material \ref{sec:AutoFFTlength} 
for further details.

The dynamic range used for the colour range for the LRFS (see e.g.\, colourbar on the right of panel a ii) of Fig.~\ref{fig:allspectra}) is such that 98 per-cent of the spectral values in the pulse longitude range shown occupy at least 50 per-cent of the dynamic range. This could require clipping to be applied: samples exceeding a threshold value corresponding to the darkest colour are set to the threshold. The level of clipping applied can be read from the numerical scale next to the colour scale. If the maximum value is less than that indicated with `Max.' next to this, clipping has been applied. So in the case of the LRFS for PSR J1539$-$6322 in panel (a ii) of Fig.~\ref{fig:allspectra}, the largest spectral sample is about $1.00/0.02=50$ times stronger than the color scale suggested. This helps to emphasise the weaker spectral features, such as in this case the first harmonic.

As for the LRFSs, the dynamic range of the 2DFSs is adjusted to ensure that weak features are visible by clipping the brightest features. The applied algorithm is identical to that used in the LRFS. Only samples in the shown $1/P_2$ range are considered, and either the full $1/P_3$ range or that highlighted by the dotted lines in the 2DFS are used. All spectral powers are normalised in the produced plots. 

For the 2DFS, the full available $1/P_2$ range is not always shown, since the spectral power tends to be concentrated towards $1/P_2=0$. The shown range is algorithmically determined by ensuring that at least 90 per-cent of the spectral power is shown. The shown range is always symmetric with respect to $1/P_2=0$, and no fewer than two spectral bins at either side of the shown range are excluded when the full range is not shown. In {\nrManualVrange} cases the shown range is defined manually to ensure that all defined features are fully visible. The $1/P_2$ ranges of the shuffle-normalised 2DFSs always match those of the regular 2DFSs for easier comparison.

The shuffle-normalised spectra suppress the stochastic pulse-to-pulse variability, which often manifests itself as a vertical band of excess power in the middle of the 2DFS. However, it can be reduced too much resulting in negative powers to appear. This is most noticeable in the shuffle-normalised LRFS (panel d ii of of Fig.~\ref{fig:allspectra}) which has distinctly negative spectral power away from the $\sim0.08$ cpp feature. This is because the shuffling process randomises (smears out in the $1/P_3$ direction) all variability, including the power associated with the organised drifting subpulses. So the average spectral power that gets subtracted will over-estimate the spectral power associated with stochastic pulse-to-pulse variability. Despite this artifact, the shuffle-normalised spectra have proven to be an effective tool to identify weak spectral features of interest.

The vertically integrated power of the visually most dominant drifting subpulse feature, otherwise the clearest $P_3$-only feature, is shown in the bottom panel of the 2DFS. The approximate range in $1/P_3$ occupied by the feature is indicated by the two horizontal lines in the 2DFS. In the case of column (e) in Fig.~\ref{fig:allspectra}, this implies that the weaker drift feature is shown in the bottom panel attached to the 2DFS. The spectral feature in the bottom panel always corresponds to the orange feature in the left side panel of the 2DFS. The dotted lines are also shown in the left side panel of the 2DFS. The thicker dotted lines indicate the feature that is visually dominant when there are multiple profile components with spectral features. So in the case of PSR J1539$-$6332 in column (a) of Fig.~\ref{fig:allspectra}, the dominant feature is the one in the first 2DFS. The dominant feature is used in the statistical analysis when spectral properties of different pulsars are compared.

To assess the significance of the spectral features, a number of diagnostics are utilised. The pulse sequence is split into two shorter sequences with half the number of pulses to verify that the feature is persistent. Where multiple observations of the same pulsar are available, these could similarly be used. The shuffle-normalised plots provide a quick indication of the significance of features in the presence of stochastic pulse shape variability. Following \cite{WES2006,wse07}, the $P_3$ and $P_2$ of the identified spectral features are measured by determining the centroid of the spectral power within rectangular regions in the 2DFS. An error on the centroid follows from the rms determined from the off-pulse spectra. However, in most cases the systematic error arising from the subjectivity in the selection of the rectangular region dominates the uncertainty. To quantify this, multiple somewhat different rectangular regions are considered. The quoted errors correspond to the range of centroid locations (and their errors). In order for a feature to be classed as a drifting subpulse feature, the sign of the derived $P_2$ should be constrained within the quoted error.

Unlike $P_3$, $P_2$ depends on observing frequency as well as pulse longitude (e.g. \citealt{WES2006,wse07}). This makes $P_2$ a somewhat ill defined quantity. An additional complication in measuring $P_2$ from the 2DFS is that in general there is stochastic pulse shape variability leading to power concentrated along the vertical axis. This introduces a bias in $P_2$ such that it is measured to be further away from zero than the actual $P_2$ corresponding to the drifting subpulses. This bias also makes the error on $P_2$ asymmetric, hence separate errors in both directions are quoted. The 2DFS of PSR J1539$-$4828 (panel b iii of Fig.~\ref{fig:allspectra}) shows that changes in the alias order is another reason for the determined centroid of a spectral feature to deviate from what corresponds to the typical subpulse separation $P_2$.

\begin{figure}
    \centering
    \includegraphics[width=\linewidth]{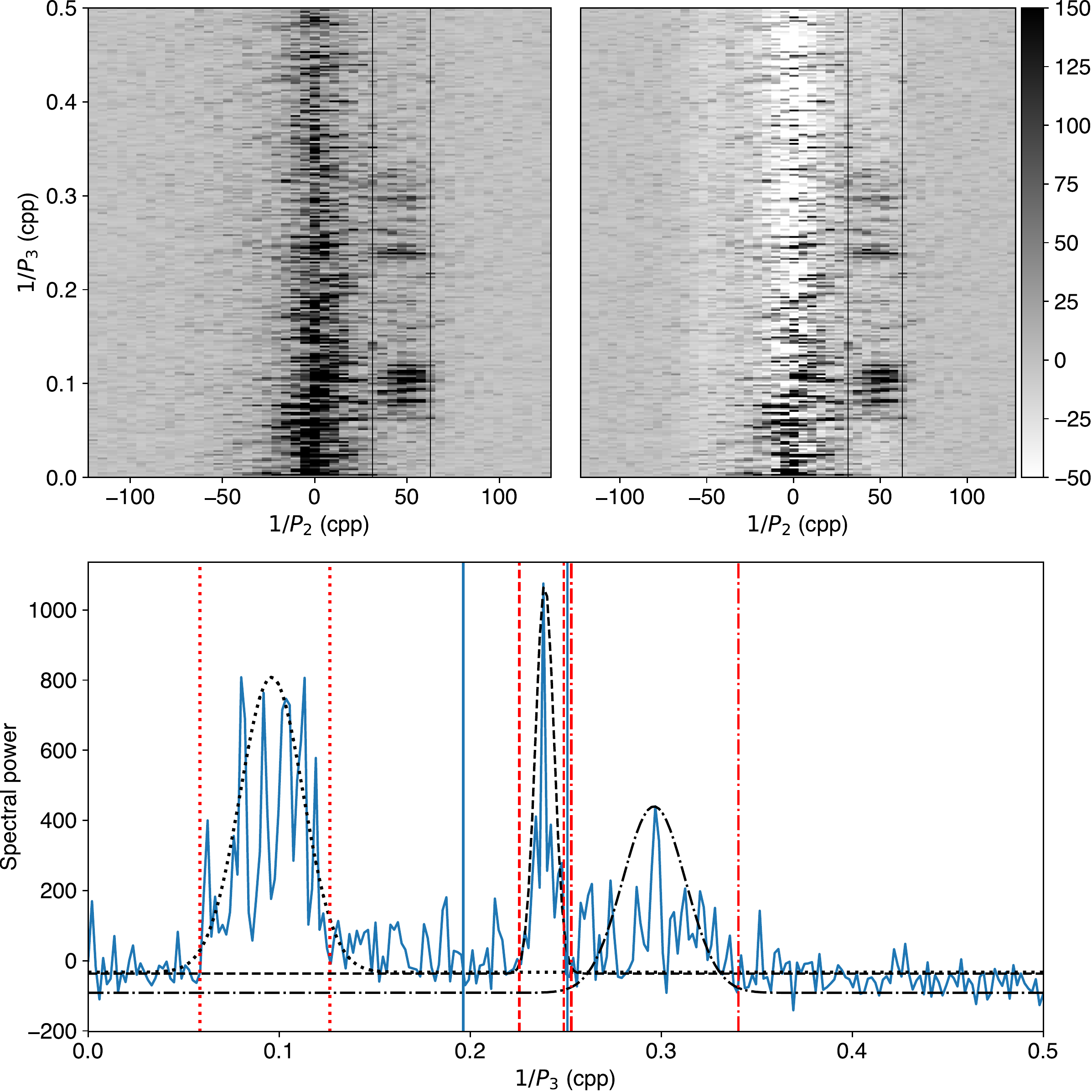}
    \caption{The top left and right panels show respectively the original and shuffle-subtracted 2DFS for PSR J0108$-$1431. The bottom panel shows the 2DFS power integrated between the vertical solid lines as shown in the top panels. There are three spectral features, covering the ranges indicated with dotted, dashed and dash-dotted lines. The three Gaussians are defined by the determined $\mu_{1/P_3}$ and $\sigma_{1/P_3}$, with a baseline/amplitude corresponding to the lowest/largest spectral power within the $1/P_3$ range covering the feature.}
    \label{fig:shuffp3spec}
\end{figure}

The spectral width $\sigma_{1/P_3}$ is determined using an approach similar to that used to produce the shuffle-normalised spectra (See Sec.~\ref{sec:shuffle_norm}). An example of the shuffle-subtracted 2DFS for PSR~J0108$-$1431 is shown in the top right panel of Fig.\,\ref{fig:shuffp3spec}, with the original spectrum showing in top left and the resulting integrated spectral power in the bottom panel. As discussed in the context of the shuffle-normalised spectra, the baseline subtraction can overcompensate, resulting in negative spectral powers. To avoid this affecting the measurement of the spectral widths, the baseline is raised such that no negative spectral powers are left in the integrated spectrum within the $1/P_3$ range covered by the spectral feature.

The spectrum in Fig.\,\ref{fig:shuffp3spec} has three drift modes identified as indicated by the vertical lines in the bottom panel. The three Gaussian curves show the derived $\mu_{1/P_3}$ and $\sigma_{1/P_3}$, as well as the applied baseline which is different for the three features. The calculated $\sigma_{1/P_3}$ represents the widths of the spectral features well.

\section{Additional population analysis}
\label{app:ppdothist}

\begin{figure}
    \centering
    \includegraphics[width=\linewidth]{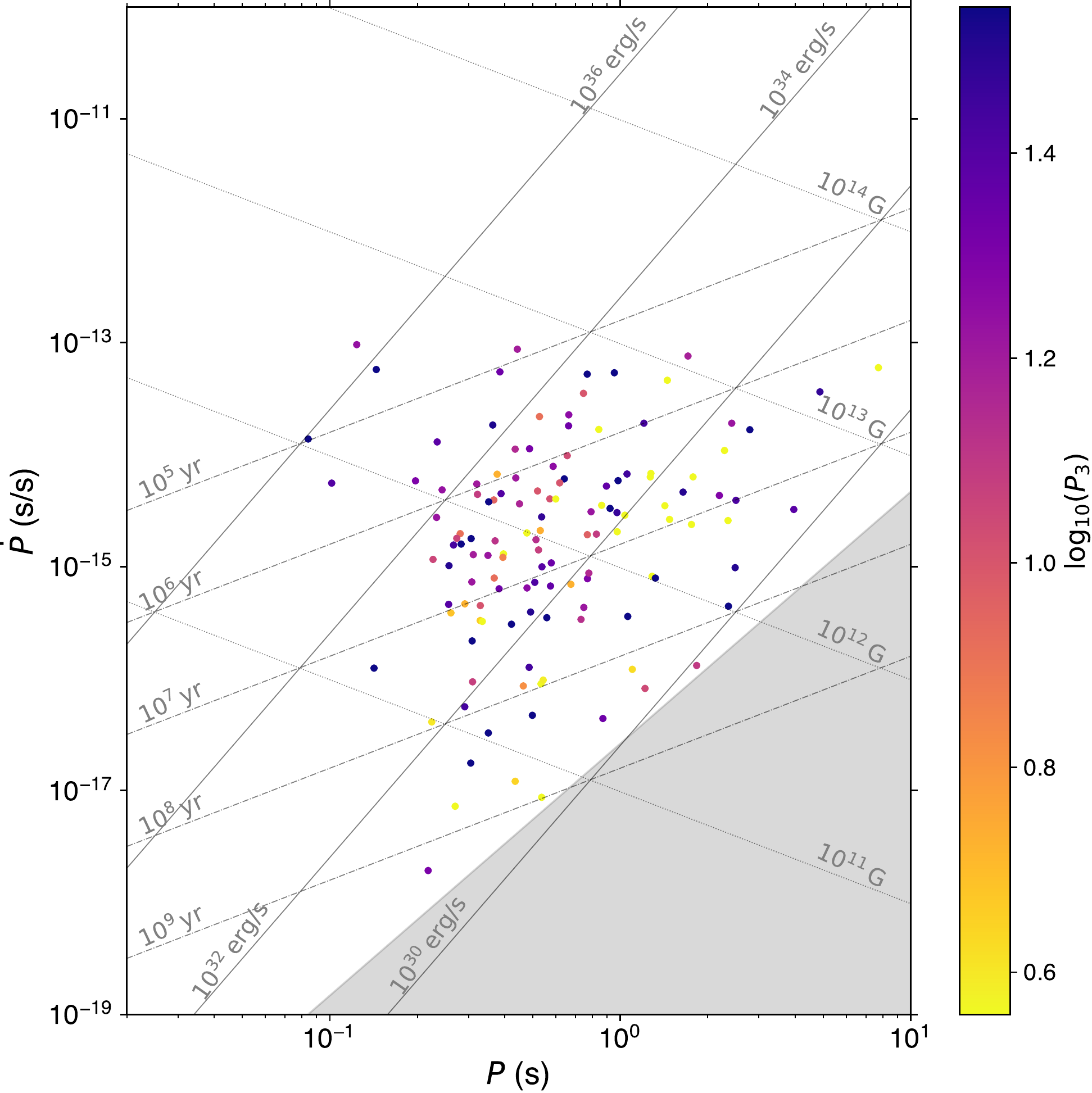}
    \caption{$P$-$\dot{P}$ diagram where the colour of the dots represent $\log_{10}(P_3)$ for each pulsar with a dominant $P_3$-only spectral feature. See the caption of Fig.~\ref{fig:parappdotp3} for a description of the various lines.}
    \label{fig:parappdotp3only}
\end{figure}

Fig.~\ref{fig:parappdotp3only} shows the distribution of $P_3$ in the $P$-$\dot{P}$ diagram for pulsars with their dominant spectral feature being $P_3$-only. Unlike Fig.~\ref{fig:parappdotp3} (for pulsars for which the dominant spectral feature is associated with drifting subpulses), the distributions of the two classes of $P_3$ values are very different, with no evidence for a V-shaped evolution in Fig.~\ref{fig:parappdotp3only}.

\begin{figure}
    \centering
    \includegraphics[width=\linewidth,trim= 0 0 0 1.3cm, clip]{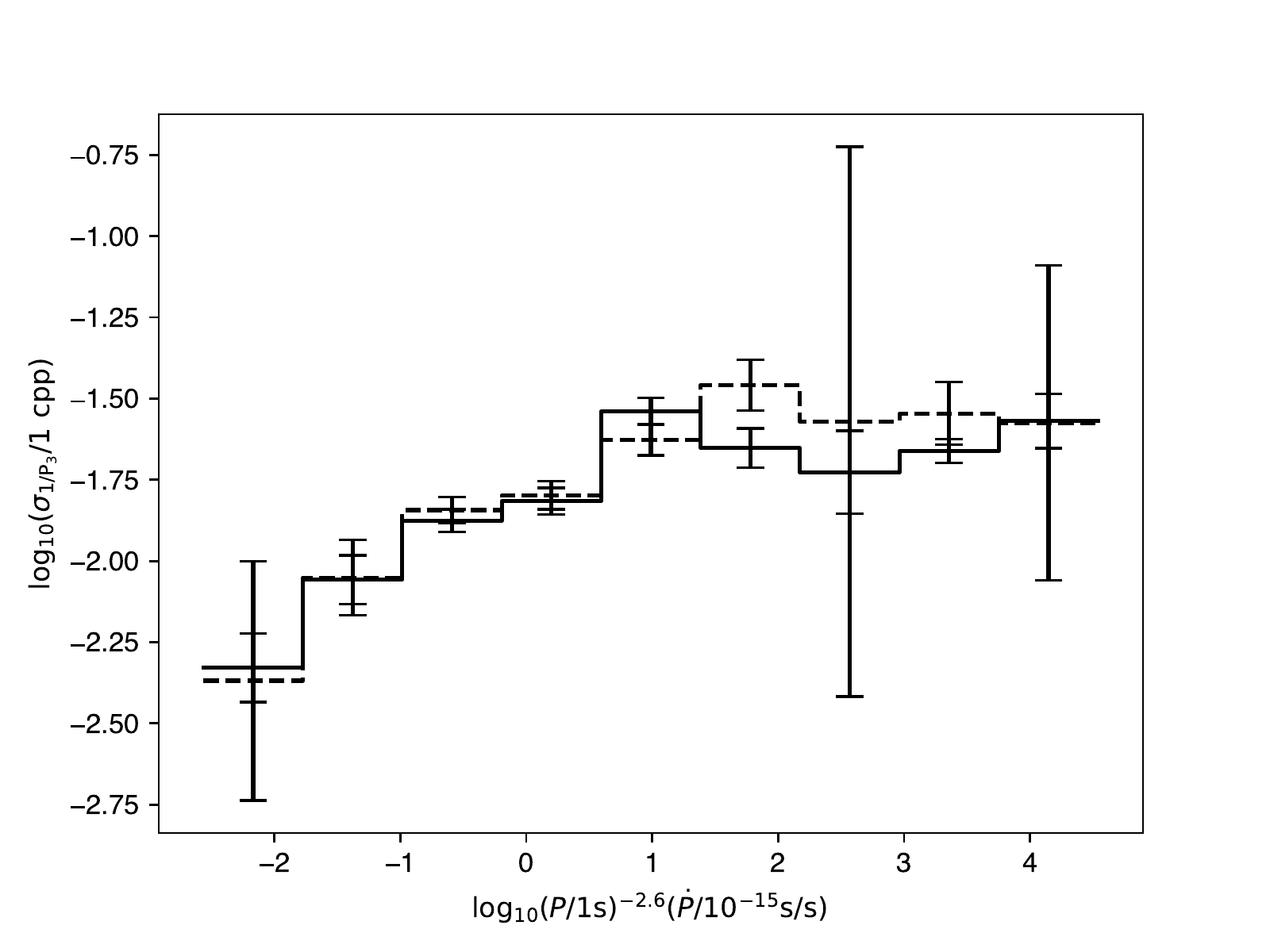}
    \caption{A histogram of the measured (solid) and predicted (dashed) weighted mean $\log_{10} (\sigma_{1/P_3})$ as a function of the combination in $P$ and \pdot\ corresponding to the direction of the strongest correlation in the $P$-$\dot{P}$ diagram. The prediction is based on Eq.\,\ref{eq:mcmcfit} fitted to the measured distribution. The errorbars represent the standard deviation of values contributing to each bin, divided by the square root of the number of values.}
    \label{fig:histp3std}
\end{figure}

As demonstrated in Sec.~\ref{sec:resultp3std}, the spectral width varies across the $P$-$\dot{P}$ diagram. Fig.~\ref{fig:histp3std} shows that the spectral width shows a significant flattening towards high $\dot{E}$ pulsars. 

\begin{figure}
    \centering
    \includegraphics[width=\linewidth,trim= 0 0 0 1.3cm, clip]{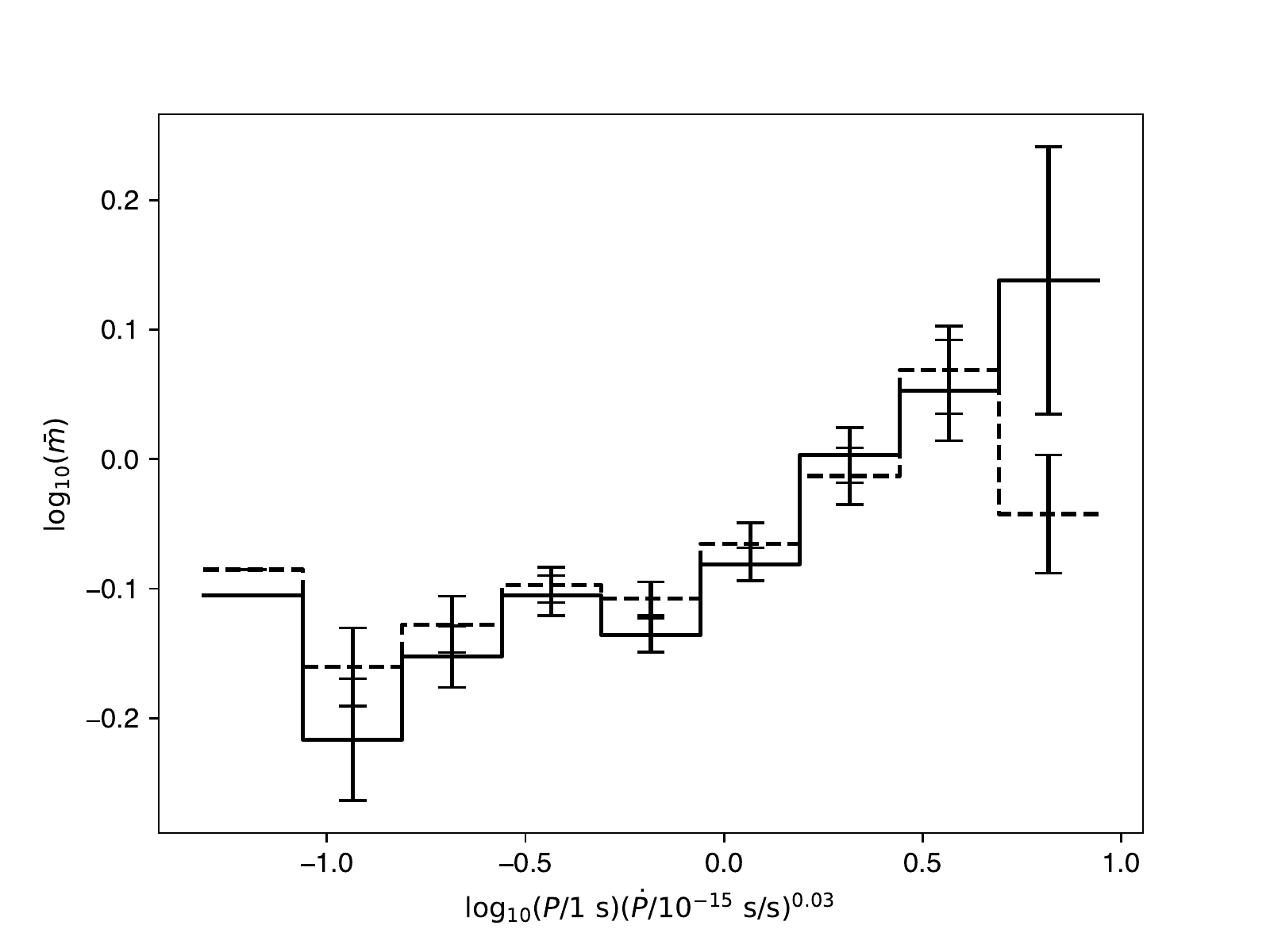}
    \caption{A histogram of the measured (solid) and predicted (dashed) weighted mean $\log_{10} (\bar{m})$ as a function of the combination in $P$ and \pdot\ corresponding to the direction of the strongest correlation in the $P$-$\dot{P}$ diagram. The prediction is based on Eq.\,\ref{eq:mcmcfit} fitted to the measured distribution. The errorbars represent the standard deviation of values contributing to each bin, divided by the square root of the number of values.}
    \label{fig:histavmod}
\end{figure}

As shown in Fig.~\ref{fig:histavmod}, the average modulation index $\bar{m}$ evolution is non-linear in the $P$-$\dot{P}$ diagram, such that there is flattening of $\bar{m}$ for small $P$ pulsars. This illustrates why a quadratic form was preferred when performing the fit in Sec.~\ref{sec:resultmod}. The predicted distributions in Figs.~\ref{fig:histp3std} and \ref{fig:histavmod} are derived as for Fig.~\ref{fig:histp3}.

\begin{figure}
    \centering
    \includegraphics[width=\linewidth]{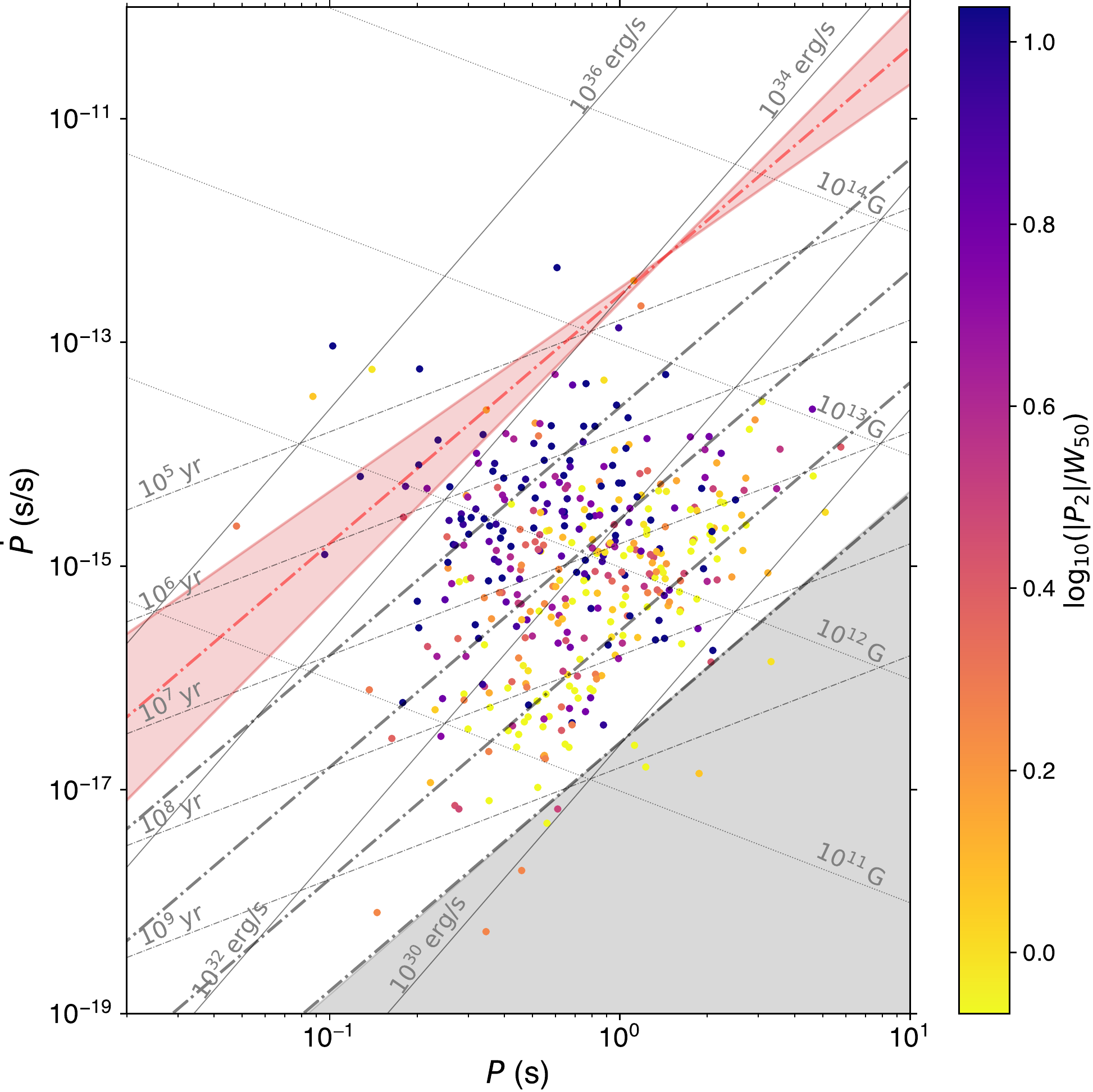}
    \caption{$P$-$\dot{P}$ diagram where the colour of the dots represent $|P_2|/W_{50}$ for each pulsar with a dominant drifting subpulse feature identified. The dot-dashed lines correspond to constant $|P_2|/W_{50}$ values according to a fit of Eq.\,\ref{eq:mcmcfit} (with $b=1$ and $c=0$). The top shaded region indicates the $1\sigma$ uncertainty on the slope. See the caption of Fig.~\ref{fig:parappdotp3} for further details.}
    \label{fig:linearppdotp2frac}
\end{figure}

Fig.\,\ref{fig:linearppdotp2frac} shows the distribution of $|P_2|/W_{50}$ for those pulsars with detected drifting subpulses. Large $|P_2|/W_{50}$ are found for higher \edot\ pulsars, and the values decrease towards lower \edot\ pulsars. A linear fit of Eq.\,\ref{eq:mcmcfit} (with $c=0$ and $b=1$) reveals that the correlation is strong ($\corrPhorabsfracdriftln\pm\correrrPhorabsfracdriftln$) with $a/b=\Phorabsfracdriftlng\pm\Phorabsfracdriftlngerr$. The direction of evolution of $|P_2|/W_{50}$ is consistent with both \edot\ ($a/b=-3$) and \age\ direction ($a/b=-1$).

\appendix
\setcounter{section}{3}

\section{Further description of the subpulse modulation analysis and the sample}
\label{app:pip}

\subsection{Single pulse stack}
\label{app:pulse_stack}

\begin{figure}
    \centering
    \includegraphics[width=\linewidth]{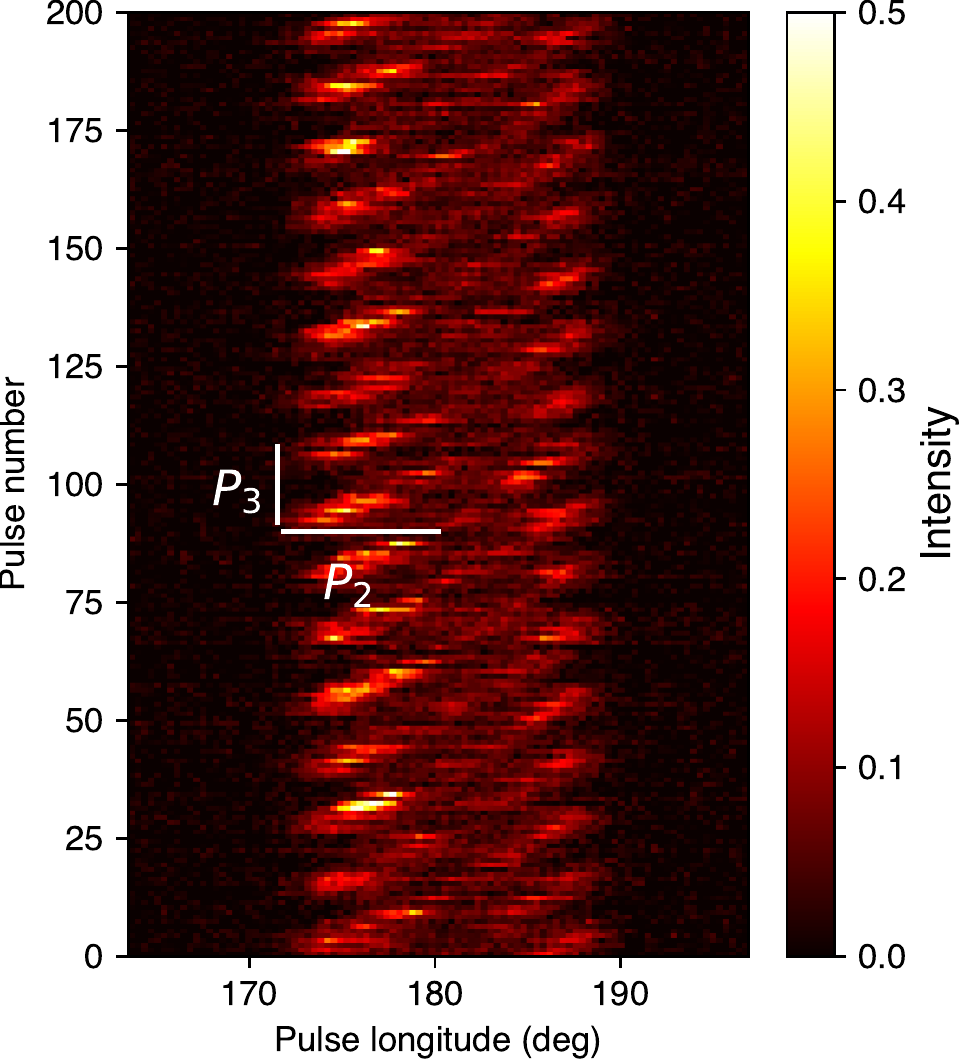}
    \caption{A sequence of 200 pulses of PSR J1539$-$6322. The $P_2$ and $P_3$ spacings are indicated.}
    \label{fig:J1539_stack}
\end{figure}

An example of a pulse stack is shown in Fig.\,\ref{fig:J1539_stack}. This is part of the full pulse sequence of PSR~J1539$-$6322. Clear drifting subpulses are seen from the leading to trailing edge of the pulse profile. The spacings $P_2$ and $P_3$ are marked for one drift band. The corresponding spectra of the full sequence are shown in panels (a) of Fig.~\ref{fig:allspectra}.

\subsection{Summary of MP/IP on-pulse and off-pulse definitions}
\label{sec:AppendixManualOnpulse}

For {\nrpulsarswithIPautodetect} pulsars the IP was correctly identified with the automated algorithm described in Sect.~\ref{sec:prepsrsalsa} 
and for {\nrpulsarswithIPmanual} 
pulsar manual intervention was required. For PSR J1828$-$1101 this is because severe scattering in the interstellar medium blends the MP and IP. There are also {\nrpulsarswithnoIP} pulsars with the IP falsely identified, this is due to the large separation among different pulse profile components. 

In {\nrMPorIPManualDouble} of the {\nrMPdoublePlusNrIPdouble} cases where either the on-pulse regions of the MP or IP were split, the on-pulse regions were specified manually. First of all, manual splitting is required if profile components (i.e.\ parts of the profile with distinct subpulse modulation properties) are not separated by a local minimum. In other cases, the pulse profile has more than two distinct components, hence a choice is made where two components are of most interest for displaying in the plots with spectral results. There are {\nrMPorIPManualOnpulseNoDouble} cases where the on-pulse regions are adjusted without defining two components. These account for the imperfections in the template matching with the observed profile, and also for excluding the long scattering tails of some pulsars.

Off-pulse regions are required for baseline subtraction and analysis of the spectra (see Sect.~\ref{sec:lrfsmod} and \ref{sec:twodfs}). 
To avoid any pulsar signal to be unintentionally included, the size of the off-pulse region excludes margins at both side of the MP and IP with a width of 30 per-cent of the determined on-pulse widths. When too little off-pulse interval remained\footnote{If the total off-pulse region occupies less than 2 per-cent of the rotation.}, no margins are excluded. This was the case for {\ComponentWidthpsrs} which have either intrinsically wide pulses, or due to scattering.
Although this automatic process works well for the majority of the pulsars, there are {\nrPulsarsManualOffpulse} pulsars for which an off-pulse region is defined manually.
This is to avoid unnecessarily wide off-pulse regions or to compensate for imperfections in the template compared to the observed pulse profile.

For {\nrPulsarsNoOffSubtrTwodfs} pulsars ({\NoOffSubtrTwodfspsrs}) the profile is too wide so that no suitable regions can be identified for off-pulse subtraction, which affects some of the further processing (see section \ref{sec:psrsalsa}). 
These are excluded from the analysis where relevant as explained in Sec.~\ref{sec:ppdot}. 
Moreover for {\nrNoBaselineSlopeSub} pulsars ({\NoBaselineSlopeSubpsrs}) the off-pulse regions were too small to meaningfully subtract a first order polynomial, and a constant was subtracted instead.

\subsection{RFI rejection}
\label{sec:app_RFI}

For {\nrPulsarsNskipOrNread} pulsars the analysis was improved by further rejecting some initial and/or final pulses in the observation beyond the mitigation applied by the processing described in Sec~\ref{sec:singlepulsepipeline} 
thereby shortening the dataset analysed.
The advantage of this approach is that the remaining pulses form a continuous sequence of pulses. However, for {\nrPulsarsAutoClean} pulsars no sufficiently long relatively RFI free sequence pulses could be identified. In those cases the worst by RFI affected pulses were excluded. This was done by iteratively finding outliers in off-pulse root-mean-square (rms)\footnote{Defined to have a rms exceeding the median rms by $4\sigma$.} The affected pulses are removed from the analysis in two different ways. First of all, the pulses are removed from the sequence resulting in a shorter sequence. This sequence is used for those types of analysis for which the order of the pulses is not relevant (e.g. the calculation of modulation index, see section \ref{sec:lrfsmod}). 
Another sequence is created by setting all samples in the affected pulses to zero. This sequence is created for those parts of the analysis which depend on the order of the pulses, e.g. the calculation of the fluctuation spectra (see sections \ref{sec:lrfsmod} and \ref{sec:twodfs}).
For these {\nrPulsarsAutoClean} pulsars, only a modest {\percentzap} of pulses were removed. 

Periodic RFI, affecting a narrow range of fluctuation frequencies, can be effectively excluded from the analysis in the spectral domain. This has been done for {\zapspec} pulsars (\zapspecpsrs).

\subsection{Pulse phase resolution determination}
\label{sec:AutoRebinning}
The used pulse longitude resolution is determined automatically by progressively reducing the native resolution of the data with factors of 2. The algorithm aims for the detection of a significant modulation index for at least half of the on-pulse bins, but avoiding the number of on-pulse bins to become less than 8. In addition, if the number of significant modulation index points in the on-pulse region exceeds 50, the resolution is lowered to avoid an excessive density of modulation index points in the produced plots.

The analysis of the modulation index is done separately for the MP and IP, and the pulse longitude resolution is determined separately by considering the full MP and IP on-pulse regions respectively. Plots are generated separately for the MP and IP, and the resolution of the pulse profile reflects the resolution used in this analysis. All spectra are based on data using the resolution determined for the MP. Since the pulse stacks used for the determination of spectra and modulation indices can be different (because of RFI being removed either by removing pulses or setting intensities to zero), the resolution of the pulse profile might be different compared to that used for the spectra. This is made consistent between pulse profile and spectral resolution for two pulsars (J1908+0916 and J2048+2255). There are 14 other cases where the resolution is defined manually. Seven pulsars (PSRs~J0729$-$1448, J0905$-$5127, J1331$-$5245, J1700$-$4422, J1702$-$4128, J1842+0358 and J1856+0245) have reduced resolution to decrease the number of modulation index points shown on the profile or to increase the significance of spectral bins. Seven cases (PSRs J0111$-$7131, J0133$-$6957, J1632$-$1013, J1737$-$3102, J1851+0418, J1901+0234 and J1901+1306) have too small automatically defined resolution, where the pulse profile is not resolved properly, so the resolution is increased. 

\subsection{Automatic FFT length determination}
\label{sec:AutoFFTlength}

The optimum FFT length is determined by progressively reducing the FFT length by factors of two.
The number of significant (3$\sigma$) spectral points in the full MP on-pulse region of the LFRS is assessed for each pulse longitude bin. If there are no pulse longitude bins for which at least a quarter of the spectral points is significant, the spectral resolution is reduced further. For {\nrPulsarsManualFFTsize} pulsars the FFT length was manually set. This has been done, for example, to avoid very sharp spectral features to become unresolved.

\subsection{Spectral off-pulse subtraction in 2DFS}
\label{sec:app_offpulse_sub_2dfs}
The off-pulse spectrum is calculated by using an off-pulse region with the same width as that of the on-pulse region. The off-pulse spectrum is averaged in the $1/P_2$ direction before subtraction. This averaging reduces the rms fluctuations in the power of the off-pulse spectrum, and therefore the final spectrum will be less noisy. This averaging is justified when the off-pulse noise is Gaussian.
Non-Gaussian (correlated) noise, such as periodic baseline variations with a timescale~$\lesssim P$ would affect different $1/P_2$ spectral bins differently, potentially making the noise subtraction less effective if the noise spectrum is averaged before subtraction.
However, this type of variability is in general not found to be an issue in the analysed data. In addition, because this type of variability tends to be non-stationary, even without averaging the subtraction of the off-pulse spectrum from the on-pulse spectrum will not suppress the variability accurately.
Besides improving the sensitivity of the obtained spectra, the averaging of the off-pulse spectra also implies that it is not necessary for a continuous off-pulse region with a width equal to that of the on-pulse region to be available. This did benefit the analysis of {\nrcatOffpulse} pulsars: {\psrcatoffpulsepsrs}.

\subsection{Correlation analysis}
\label{sec:app_Bayesian}

The posterior distributions of the fitted parameters are sampled from a likelihood function and prior distributions. We use a Gaussian likelihood function to account for both {\sig} and the measurement uncertainties. Uniform priors on all parameters are taken, except for \sig\ which was forced to be positive by setting the prior probability zero when $\sigma_{e}<0.01$.
The posterior distributions are determined using a Markov Chain Monte-Carlo sampler, \textsc{emcee} \citep{emcee} in python. We use 50 walkers and 20,000 steps, and the first 5000 are removed for the burn-in phase. Each chain is thinned with half of the median correlation length of all parameters, and all walkers are added together to form a flattened chain. In the calculation, the values of $P$ and \pdot\ are normalised by 1~s and $10^{-15}$~s/s respectively to help convergence. The initial guesses of parameters (except for \sig) are taken to be the optimised parameters of a linear least square fit without considering the errorbars on the measurements using \textsc{scipy.curvefit} \citep{scipy}.

\subsection{Additional information related to the sample}
\label{app:sample}

\begin{figure}
    \centering
    \includegraphics[width=\linewidth]{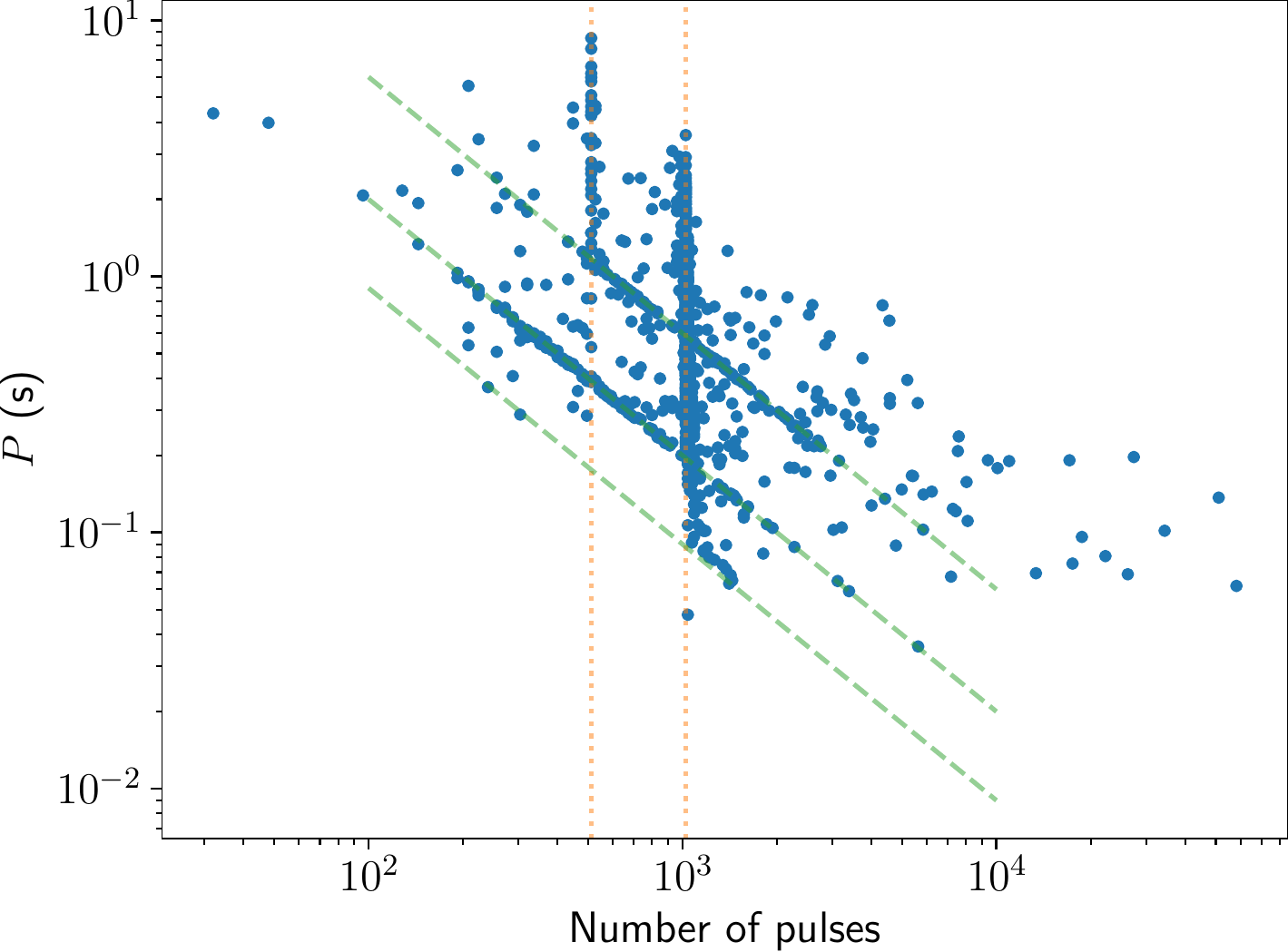}
    \caption{The pulse period as a function of the number of recorded pulses, with 512 and 1024 pulses indicated by the vertical dotted lines. The diagonal dashed lines correspond to observations of 90, 200 and 600 seconds (from bottom to top).}
    \label{fig:nrpulses}
\end{figure}

Fig.\,\ref{fig:nrpulses} shows that the majority of pulsars have at least 1024 pulses recorded (points along the right dotted line). Observations with around 512 recorded pulses (left dotted line) correspond mostly to pulsars with relatively long pulse period ($> 2$ s) which were added to the campaign to extend the coverage of the pulsar population. The diagonal dashed lines correspond to fixed observing durations. These lines correspond to the integration times corresponding to a lower limit applied to all observations (90 s), that adopted for the earliest observations (200 s) and that used for a later extension of the campaign (600 s) for pulsars with uncertain flux densities or sky positions.

\begin{figure}
    \centering
    \includegraphics[width=\linewidth]{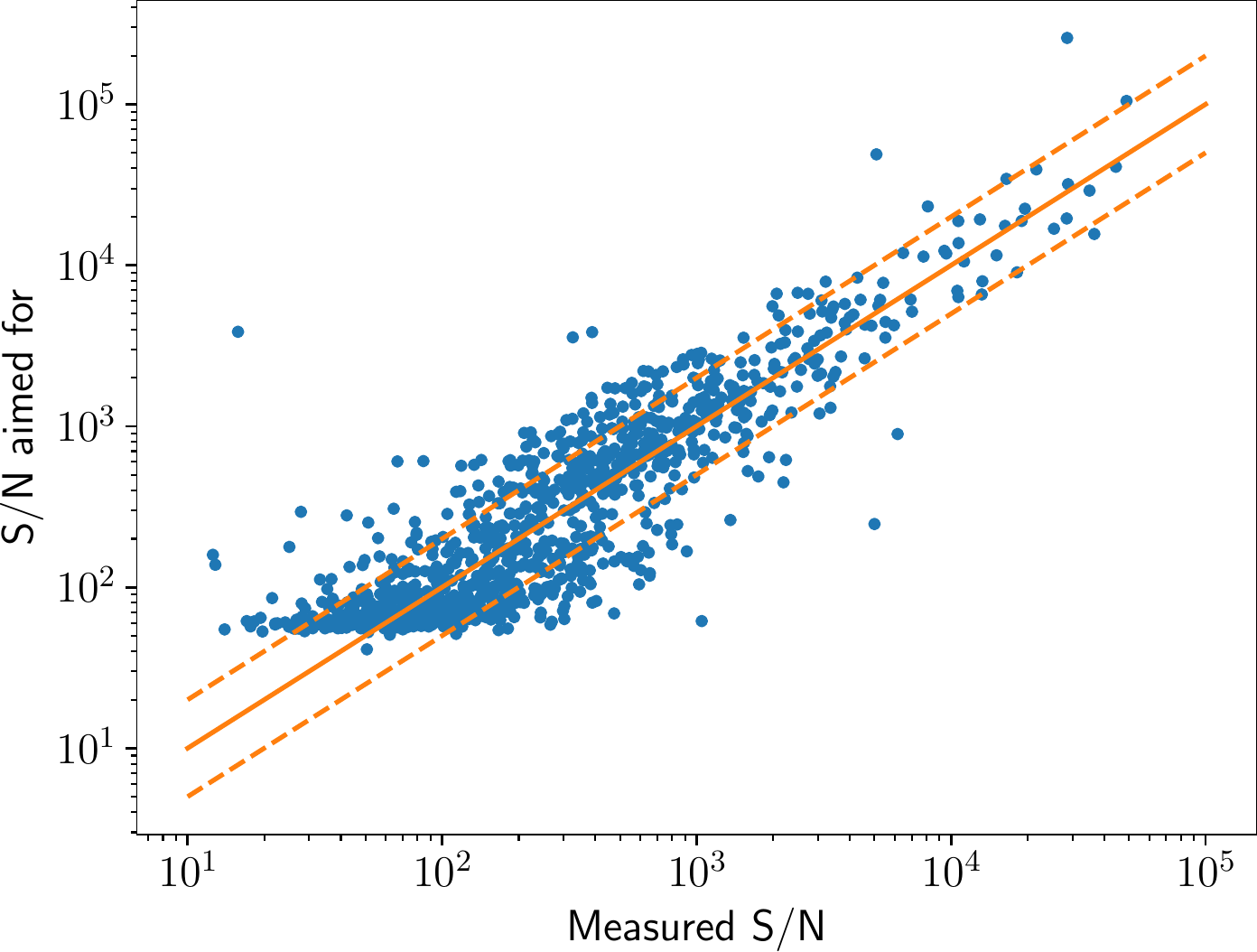}
    \caption{Comparison of the S/N aimed for by the methodology of \citet{Song2021} for the observations analysed in this work with the measured S/N (dots). The one-to-one relation corresponds to the solid line, and the two dashed lines correspond to the measured S/N being a factor of 2 higher and lower, respectively.
    }
    \label{fig:snrcompdata}
\end{figure}

Fig.\,\ref{fig:snrcompdata} shows a comparison of the measured S/N of the integrated pulse profiles for our sample\footnote{After mitigation of RFI (see Sect.~\ref{sec:method}).} 
compared to that aimed for using the methodology of \citet{Song2021}.
For many pulsars the S/N of the data is consistent with the expected S/N within a factor of two (i.e. within the range indicated by the dashed lines). All observations were aimed to have a S/N of at least $\sim40$, but there is a tail in the measured S/N distribution towards lower S/N.  

\begin{figure}
    \centering
    \includegraphics[width=\linewidth]{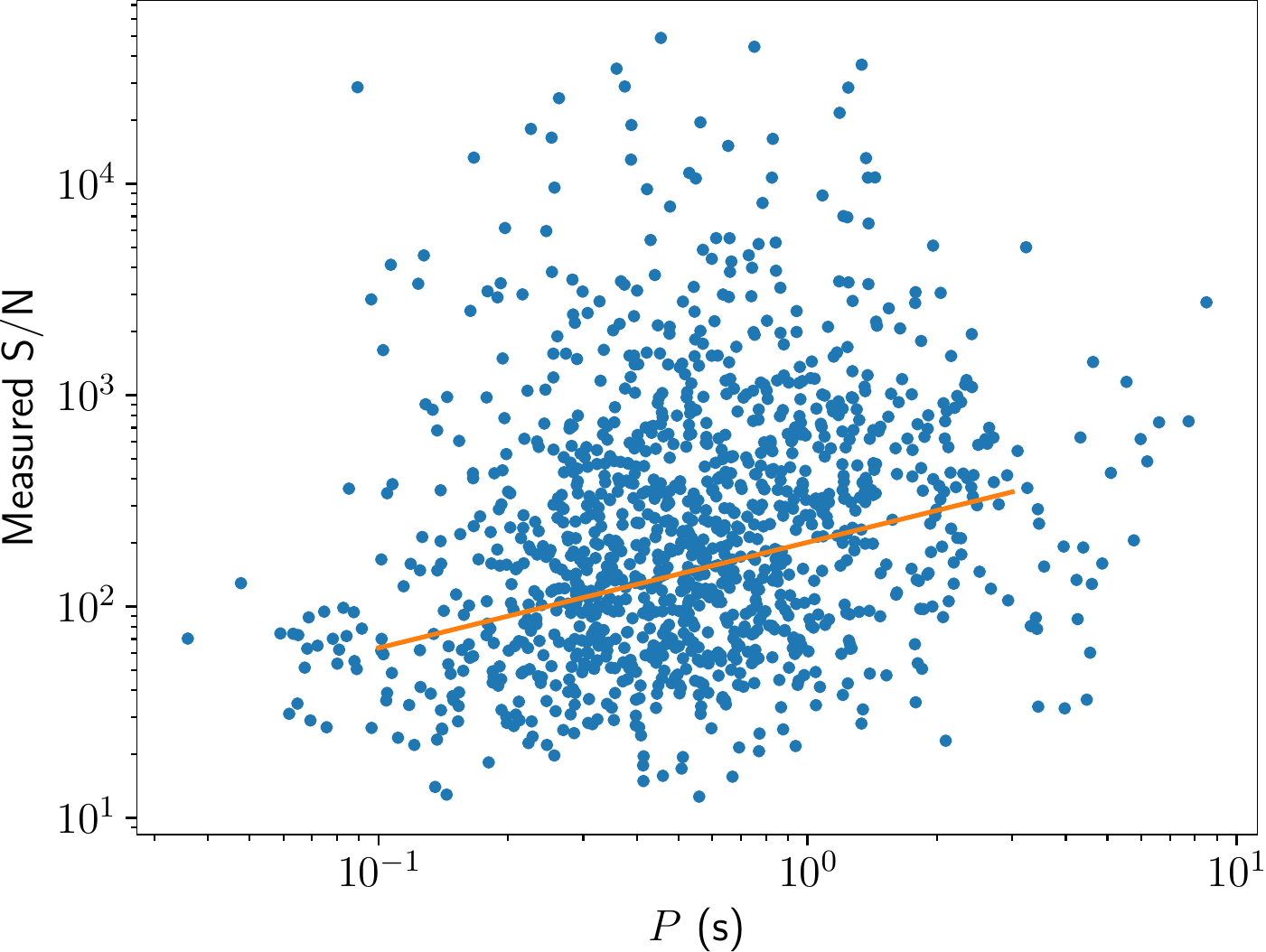}
    \caption{Measured profile S/N as a function of pulse period of the sample (dots). The solid line shows a relation $\mathrm{S/N} \propto \sqrt{P}$.}
    \label{fig:snrvsp}
\end{figure}

Fig.\,\ref{fig:snrvsp} shows the S/N as a function of pulse period $P$. The S/N is larger for longer period pulsars, because they have typically a longer observing length $t_\mathrm{obs}$, as indicated by the solid line corresponding and noting that $\mathrm{S/N} \propto \sqrt{t_\mathrm{obs}}$.

\subsection{Comparison with other drifting subpulse surveys}
\label{sec:app_survey_comparison}

\subsubsection{Comparison with \citet{WES2006}}
\label{sec:wescomp}
\citet{WES2006} used similar methods to those used in our study, including the use of the 2DFS to identify drifting subpulses. In addition, their observations were taken at a similar center frequency, although using a narrower bandwidth. There are \tpawes\ pulsars included in both surveys, of which \tpaweshasp\ were reported to have periodic modulation (drifting subpulses or $P_3$-only) by \citet{WES2006}. In this sample of overlapping pulsars, the majority (\tpawespconsist) have consistent $P_3$ measurements. So the majority of measurements are in agreement, and the following explains the differences in the obtained results.

There are 7 pulsars for which inconsistent $P_3$ were found. For five pulsars (PSRs J1136+1551, J1913$-$0440, J1919+0021, J1935+1616 and J1941$-$2602), no drifting subpulses are identified in our data, while \citet{WES2006} observed a broad feature with a spectral width larger than 0.2 cpp with a relatively weak evidence of drifting subpulses. Our non-detection could be due to irregular or sporadically occurring drifting subpulses. For PSR~J1607$-$0032, our data reveals a periodicity close to $1/P_3\sim0$, which was also seen in the spectra in \citet{WES2006}, but they also found evidence for a broad feature at higher $1/P_3$ frequency extended to the alias border at 0.5 cpp. Hence a longer pulse sequence might be required to confirm their reported behaviour, especially if it occurs sporadically. Features similarly broad compared to the ones seen in the above pulsars in \citet{WES2006} are identified in our data for other sources, so we are sensitive for this type of features (see panel (e) of Fig.~\ref{fig:allspectra} 
for an example of a broad and weak spectral feature). One pulsar (PSR J0034$-$0721) has inconsistent $P_3$ measured because it has drift mode changes (e.g.\, \citealt{Huguenin1970,Ilie2020}).

Of the \tpawespconsist\ pulsars for which $P_3$ is consistent between \citet{WES2006} and our measurements, \tpawesptwo\ pulsars have consistent $P_2$ measurements.
For the remaining four pulsars (PSRs J0624$-$0424, J0837+0610, J1848$-$0123 and J2346$-$0609) the measured $P_3$ are close to the alias border, hence apparent changes in subpulse drift direction could affect the measurement of $P_2$ (see panel (b) in Fig.~\ref{fig:allspectra}). 

In this work, we identified \tpanowes\ pulsars with periodic subpulse modulations, while no features were reported by \citet{WES2006}. This is mainly because of the higher S/N in our observations. This is particularly beneficial for broader, less well defined spectral drifting subpulse features.

\subsubsection{Comparison with \citet{Basu2016,Basu2019}}
\label{sec:basucomp}

\citet{Basu2019} reported 61 pulsars with drifting subpulses, where \tpabl\ pulsars are covered in our sample. Within this sample, \tpablconsist\ pulsars have consistent $P_3$ values, although we classify two pulsars as having $P_3$-only subpulse modulation. One of them, PSR J0837+0610, was discussed in Sec.\,\ref{sec:wescomp}. For the other, PSR J1625$-$4048, weak subpulse phase variation with pulse longitude was reported in \citet{Basu2019}, which was not confirmed in our relatively short observation.

Of the three remaining pulsars, two have no significant periodic subpulse modulation in our data (PSRs J0815+0939 and J2006$-$0807). PSR J0815+0939 is known to be bi-drifting (different profile components have subpulses which drift in opposite directions) at lower observing frequencies \citep{Champion2005,Szary2017} and the S/N of our data is relatively low. PSR J2006$-$0807 is known to show nulling at lower frequencies, drifting subpulses and mode changes \citep{Basu2019psr}. Nulling is also observed in our data, which can disrupt the pattern of drifting subpulses.
Also the pulse profile shape is distinctly different at our observing frequency, so the drifting subpulses might be clearer at lower frequencies. 
For PSR J1921+1948, the measured $P_3$ is inconsistent with \citet{Basu2019}, a likely consequence of the known drift mode changes \citep{Basu2019}.

Besides pulsars for which drifting subpulses were identified, \citet{Basu2019} lists 69 pulsars without drifting subpulses. Of these, \tpanoblall\ are in our sample, with drifting subpulses detected in \tpanobldrift\ of them. Of these \tpanobldrift, \cite{Basu2016} identified $P_3$-only features with consistent $P_3$ values for \intpanoblinbe\ pulsars. Reasons why drifting subpulses could be identified in our data, but not in the analysis of \citet{Basu2019}, includes better data quality and the different observing frequency. For some of these pulsars a spectral feature which is broad in the $1/P_3$ direction was identified in our data, which are harder to detect.

In addition to the pulsars discussed above, \citet{Basu2016} reported on a few pulsars with $P_3$-only features, of which  \tpabe\ were observed by the TPA as well. Of these, \tpabepconsist\ pulsars have consistent $P_3$ measurements and some phase drift was detected for \tpabepdrift\ of them. The higher S/N of the TPA data for these pulsars is likely the reason why the phase drift could be identified. Of the \tpabe\ pulsars observed by both the TPA and \citet{Basu2016}, \inbenotpa\ did not show evidence for periodic subpulse modulations in our data.
Because the periods of the modulation reported by \citet{Basu2016} are long (larger than 10$P$), longer pulse sequence may be required to confirm these results.

\section{Supplementary Population Analysis}
\label{ApendixModIndexEvolution}

\begin{figure}
    \centering
    \includegraphics[width=\linewidth]{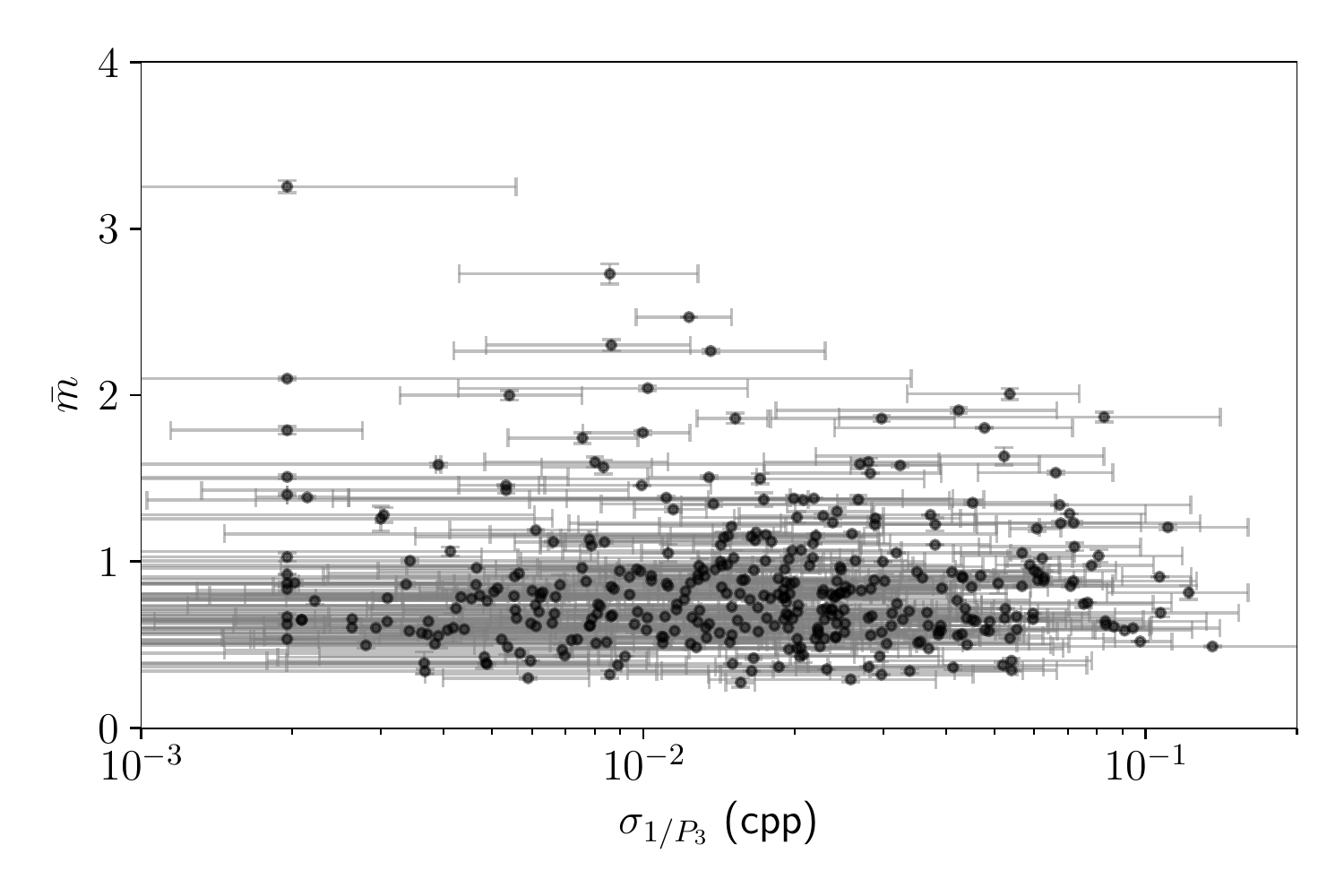}
    \caption{Scatter plot of the average modulation indices $\bar{m}$ and spectral widths \pstd\ with the corresponding errorbars.}
    \label{fig:avmodpstd}
\end{figure}

\begin{figure}
    \centering
    \includegraphics[width=\linewidth]{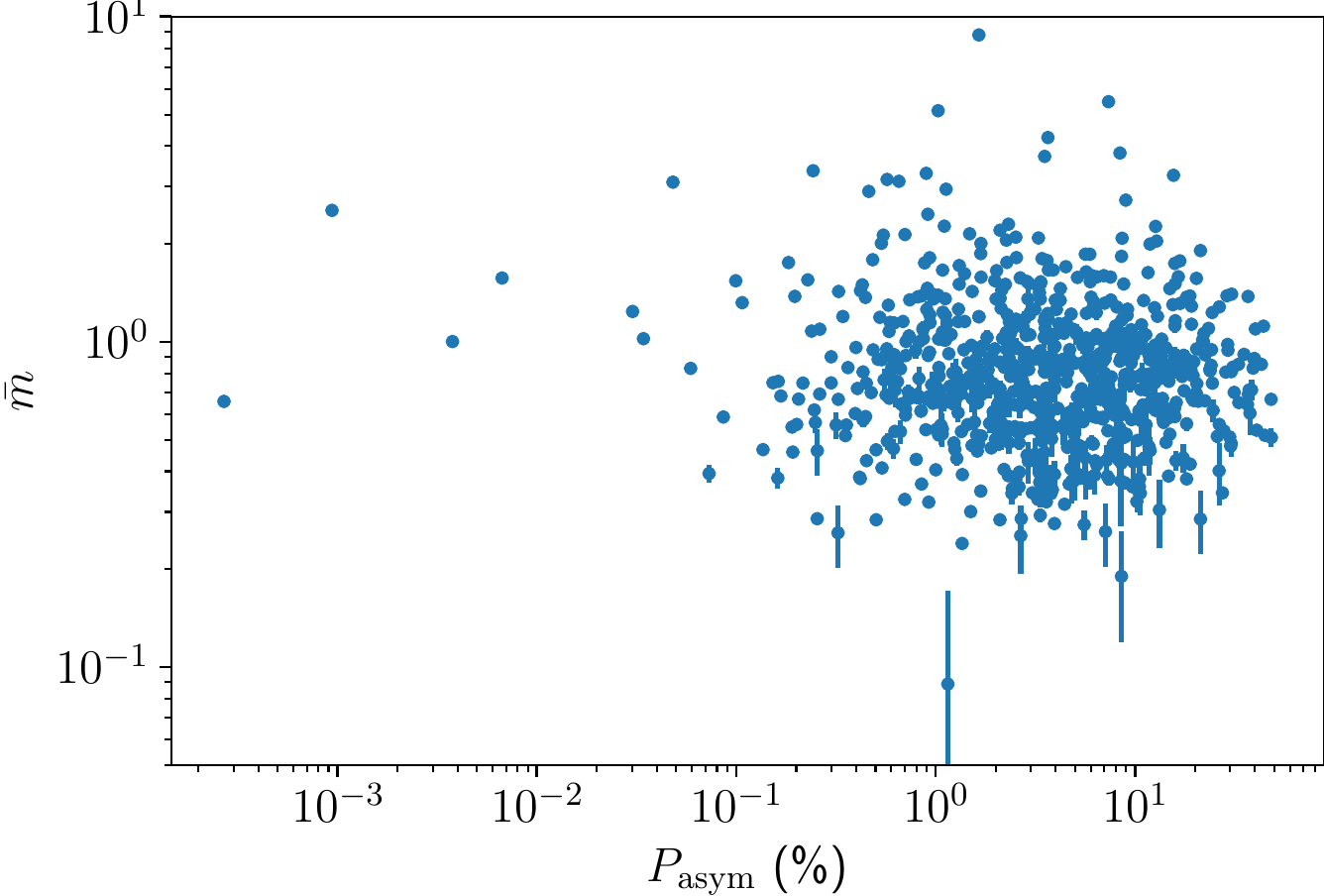}
    \caption{Scatter plot of average modulation indices $\bar{m}$ and \pasym. The errorbars on  \pasym\ are typically large, so for clarity they are not shown.}
    \label{fig:avmodp2power}
\end{figure}

The spectral width \pstd\ is uncorrelated with $\bar{m}$, as shown in Fig.~\ref{fig:avmodpstd}. To quantify the strength of the correlation, a Spearman ranked order correlation coefficient $\rho$ is calculated, and the errorbar is found by using bootstrapping. Each measurement of $\bar{m}$ and \pstd\ are replaced by a value generated from a Gaussian distribution with a mean being the measured value and standard deviation the corresponding errorbar. The generated values are used to re-calculate $\rho$, and the process is repeated 100 times. The standard deviation of obtained $\rho$ values defines the errorbar on $\rho$. The resulting $\rho=\corrlogpstdavm\pm\correrrlogpstdavm$ confirms there is no significant correlation between $\bar{m}$ and \pstd. Also no correlation between $\bar{m}$ and \pasym\ is seen with $\rho = \corrlogpoweravm\pm\correrrlogpoweravm$ (see Fig.\,\ref{fig:avmodp2power}). 

\begin{figure}
    \centering
    \includegraphics[width=\linewidth]{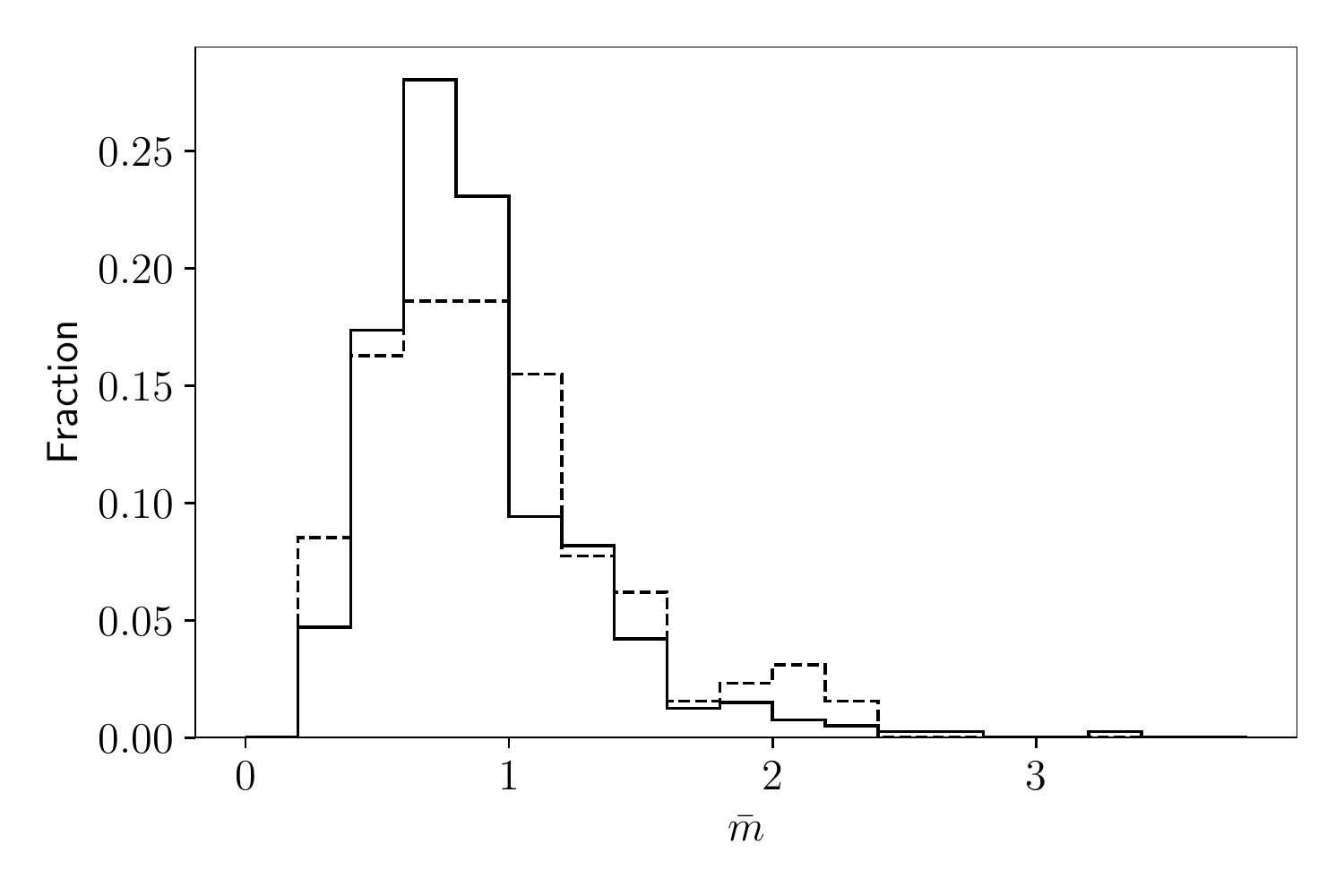}
    \caption{Distribution of average modulation indices for pulsars with drifting subpulses (solid line) and with $P_3$-only features (dashed line).}
    \label{fig:avmoddriftp3}
\end{figure}

In addition, $\bar{m}$ for pulsars with drifting subpulses and $P_3$-only features are compared (see Fig.\,\ref{fig:avmoddriftp3}). The Kolmogorov-Smirnov (KS) test suggests that the null hypothesis of the two distributions being identical  cannot be rejected at a 10\% probability.

\newpage
\onecolumn
\appendix 

\setcounter{section}{1}

\begin{longtable}{llccrrrrrr}
\caption{A summary of observation related parameters. For each row the pulsar name, the observation date and time, and the number of participating MeerKAT dishes $N_d$ used are reported. The number of pulse longitude bins $N_b$ for the MP and IP can be different and affects the calculation of the modulation indices. The spectra of the IP are always generated with the $N_b$ corresponding to the MP. All spectra are calculated with a FFT length $N_\mathrm{fft}$. The number of pulses included in the pulse profile is shown as $N_p$ (profile), and is always a multiples of 16. The same number of pulses is used for the calculation of modulation indices. The number of pulses discarded because of for example RFI is $N_p$ (zapped) thereby interrupting the sequence of pulses analysed. In addition, $N_p$ (ignored) corresponds to the total number of pulses removed from the start and the end of the observation, thereby avoiding the analysed sequence of pulses to be disrupted. These are only shown for a small number of pulsars, as for most pulsars the full sequence was analysed. Pulsar names marked with a $*$ were excluded from the population analysis because of severe scattering. Those marked with a $\dag$ showed some visual evidence for scattering, but not severe enough to exclude these pulsars.}\\ \hline\hline 
\hspace{5mm}PSRJ & \hspace{3mm}PSRB & Observation & \multicolumn{1}{c}{$N_d$} & \multicolumn{1}{c}{$N_b$}  & \multicolumn{1}{c}{$N_b$} & \multicolumn{1}{c}{$N_\mathrm{fft}$} & \multicolumn{1}{c}{$N_p$} & \multicolumn{1}{c}{$N_p$}  & \multicolumn{1}{c}{$N_p$}  \\ & & & & \multicolumn{1}{c}{(MP)} & \multicolumn{1}{c}{(IP)} & & \multicolumn{1}{c}{(profile)} & \multicolumn{1}{c}{(zapped)} & \multicolumn{1}{c}{(ignored)}\endfirsthead
 \hline \hline 
\hspace{5mm}PSRJ & \hspace{3mm}PSRB & Observation & \multicolumn{1}{c}{$N_d$} & \multicolumn{1}{c}{$N_b$}  & \multicolumn{1}{c}{$N_b$} & \multicolumn{1}{c}{$N_\mathrm{fft}$} & \multicolumn{1}{c}{$N_p$} & \multicolumn{1}{c}{$N_p$}  & \multicolumn{1}{c}{$N_p$}  \\ & & & & \multicolumn{1}{c}{(MP)} & \multicolumn{1}{c}{(IP)} & & \multicolumn{1}{c}{(profile)} & \multicolumn{1}{c}{(zapped)} & \multicolumn{1}{c}{(ignored)} \endhead

 J0034$-$0721 & B0031$-$07 & 2019-10-18 22:29:51 & 58 & 256 &  & 512 & 1040 &  &  \\
 J0038$-$2501 &  & 2019-12-06 18:43:17 & 58 & 256 &  & 512 & 2336 &  &  \\
 J0045$-$7042 &  & 2019-11-12 15:55:50 & 57 & 128 &  & 512 & 1632 &  &  \\
 J0045$-$7319 &  & 2021-02-12 12:19:26 & 28 & 256 &  & 256 & 368 &  &  \\
 J0108$-$1431 &  & 2019-11-15 19:58:27 & 61 & 256 &  & 512 & 1040 &  &  \\
 J0111$-$7131 &  & 2019-12-06 18:54:39 & 58 & 128 &  & 512 & 1472 &  &  \\
 J0113$-$7220 &  & 2019-10-27 21:22:54 & 58 & 512 &  & 256 & 880 & 32 &  \\
 J0131$-$7310 &  & 2020-06-26 03:01:16 & 58 & 256 &  & 512 & 3440 &  &  \\
\end{longtable}

\begin{landscape}
\setcounter{table}{1}
\small 
\setlength\LTleft{0pt}
\setlength\LTright{0pt} 
\sisetup{detect-weight=true,detect-inline-weight=math}
\begin{longtable}{llS[table-format=1.4]SS[table-format=5.3]S[table-format=5.3]|rrS[table-format=1.5]S|ccS[table-format=2.3]S[table-format=2.4]r@{\;}l}
\caption{A summary of the spectral subpulse modulation measurements. 
For each pulsar the name, spin parameters $P$ and $\dot{P}$, as well as the characteristic age and spin-down energy loss rate are reported. This is followed by the number of pulses  (a multiples of 16) in the integrated pulse profile and as used for the modulation index analysis, the corresponding S/N, the average modulation index of the MP, and a measure of the asymmetry in the 2DFS (for the profile component with the smallest error on $P_{\mathrm{asym}}$). Each pulsar can have multiple profile components for which spectral analysis is performed, which are listed as subsequent rows. The component is identified as being for the MP/IP followed by the component number. This corresponds to the 2DFS number as shown in the online appendix with figures. A profile component can have multiple spectral features. Each feature is assigned a class (drift/$P_3$-only/$P_2$-only correspond to drifting subpulses, $P_3$-only or $P_2$-only features as defined in the main text. 
The measured $P_2$, $P_3$, and spectral width are listed. The $P_3$ of the dominant spectral feature of the pulsar (as used in population analysis) is indicated in bold. Possibly scattered pulsars are marked with a $*$ if the scattering prevented inclusion in population analysis, or with a $\dag$ when partially scattered but included.
} \\ \hline\hline 
\hspace{5mm}PSRJ & \hspace{3mm}PSRB & \mc{$P$} & \mc{$\dot P$} & \mc{$\tau_c$} & \mcn{$\dot{E}$} & Pulses & S/N & \mc{$\bar{m}$} & \mcn{$P_{\mathrm{asym}}$} & Class & Comp & \mc{$P_3$} & \mc{\hspace{4mm}$\sigma_{1/P_3}$} & \multicolumn{2}{c}{$P_2$} \\ 
 & & \mc{(s)} & \mc{($10^{-15}$ s/s)} & \mc{($10^{5}$ yr)} & \mcn{($10^{33}$ erg/s)} & & & & \mcn{(\%)} & & & & \mc{\hspace{4mm}(cpp)} & \multicolumn{2}{c}{(deg)} \\ 
 \hline \endfirsthead
 \hline \hline 
\hspace{5mm}PSRJ & \hspace{3mm}PSRB & \mc{$P$} & \mc{$\dot P$} & \mc{$\tau_c$} & \mcn{$\dot{E}$} & Pulses & S/N & \mc{$\bar{m}$} & \mcn{$P_{\mathrm{asym}}$} & Class & Comp & \mc{$P_3$} & \mc{$\sigma_{1/P_3}$} & \multicolumn{2}{c}{$P_2$} \\ 
 & & \mc{(s)} & \mc{($10^{-15}$ s/s)} & \mc{($10^{5}$ yr)} & \mcn{($10^{33}$ erg/s)} & & & & \mcn{(\%)} & & & & \mc{(cpp)} & \multicolumn{2}{c}{(deg)} \\ 
 \hline  \endhead

J0034$-$0721 & B0031$-$07 & 0.943 & 0.408 & 366.0 & 0.019 & 1040 & 1985 & 1.458(1) & -15.0(7) & drift & MP C1 & \bfseries 6.6(2) & 0.010(4) & $-23\!$ & $^{+3}_{-11}$  \\ 
 
 & & & & & & & & & & drift & MP C1 & 9.9(7) & 0.015(6) & $-23\!$ & $^{+2}_{-64}$  \\ 
 
J0038$-$2501 &  & 0.2569 & 0.001 & 53600.0 & 0.002 & 2336 & 73 & 2.01(9) & 38(16)  & & & & & &  \\ 
 
J0045$-$7042 &  & 0.6323 & 2.49 & 40.2 & 0.389 & 1632 & 36 & 1.6(2) & -13(15)  & & & & & &  \\ 
 
J0045$-$7319 &  & 0.9263 & 4.46 & 32.9 & 0.222 & 368 & 70 & 0.90(6) & -5(3)  & & & & & &  \\ 
 
J0108$-$1431 &  & 0.8076 & 0.077 & 1660.0 & 0.006 & 1040 & 957 & 0.421(8) & 18.9(9) & drift & MP C1 & \bfseries 10.5(9) & 0.017(2) & $7.4\!$ & $^{+3}_{-0.4}$  \\ 
 
 & & & & & & & & & & drift & MP C1 & 4.2(1) & 0.005(7) & $7.7\!$ & $^{+2}_{-0.4}$  \\ 
 
 & & & & & & & & & & drift & MP C1 & 3.4(1) & 0.023(9) & $8\!$ & $^{+3}_{-1}$  \\
\end{longtable}
\end{landscape}

\bsp	
\label{lastpage}
\end{document}